\documentclass{jfm}
\usepackage[singlelinecheck=false]{caption}                                         % Justified captions
\usepackage[pdfstartview=FitH,bookmarks=true,bookmarksopenlevel=1]{hyperref}        % Hyper-references
\usepackage{xcolor}
\usepackage{tabularx}

\input tcilatex
\newcolumntype{Y}{>{\centering\arraybackslash}X}

\begin{document}

\title[The multicomponent diffuse-interface model]{The multicomponent diffuse-interface model\\and its application to water/air interfaces}
\author{E.~S.~Benilov\footnote{Email address for correspondence: Eugene.Benilov@ul.ie}}
\affiliation{Department of Mathematics and Statistics, University of Limerick,\\Limerick V94 T9PX, Ireland}
\maketitle
\date{}

\begin{abstract}
Fundamental properties of the multicomponent diffuse-interface model (DIM),
such as the maximum entropy principle and conservation laws, are used to
explore the basic interfacial dynamics and phase transitions in fluids. Flat
interfaces with monotonically-changing densities of the components are proved
to be stable. A liquid layer in contact with oversaturated but stable vapour is
shown to either fully evaporate or eternally expand (depending on the initial
perturbation), whereas a liquid in contact with saturated vapour always
evaporates. If vapour is bounded by a solid wall with a sufficiently large
contact angle, spontaneous condensation occurs in the vapour. The external
parameters of the multicomponent DIM -- e.g., the Korteweg matrix describing
the long-range intermolecular forces -- are determined for the water--air
combination. The Soret and Dufour effects are shown to be negligible in this
case, and the interfacial flow, close to isothermal.

\end{abstract}
\maketitle

\section{Introduction}

The diffuse-interface model (DIM) was proposed by \cite{Korteweg01} as an
attempt to avoid the abrupt change of parameters in the models of
liquid/vapour interfaces existing at the time. It is based on two assumption:

(i) The long-range attractive intermolecular force (the van der Waals force)
can be modelled by a \emph{pair-wise} potential, so that the force affecting a
molecule is the algebraic sum of those exerted on it by other molecules.

(ii) The characteristic distance over which the van der Waals force acts is
much smaller than the interfacial thickness.

The resulting representation of the molecular interaction has been
incorporated into models of various phenomena, such as phase transitions in
ferroelectric materials \citep{Ginzburg60}, spinodal decomposition
\citep{Cahn61,LowengrubTruskinovsky98}, growth of, and oscillations in,
crystals
\citep{CollinsLevine85,TangCarterCannon06,HeinonenAchimKosterlitzYingEtAl16,
PhilippeHenryPlapp20}, solidification of alloys
\citep{StinnerNestlerGarcke04,NestlerGarckeStinner05}, phase separation in
polymer blends \citep{ThieleMadrugaFrastia07,MadrugaThiele09}, electrowetting
\citep{LuGlasnerBertozziKim07}, contact lines
\citep{Jacqmin00,PismenPomeau00,DingSpelt07,YueZhouFeng10,YueFeng11,SibleyNoldSavvaKalliadasis14,DingZhuGaoLu17,BorciaBorciaBestehornVarlamovaHoefnerReif19},
contact lines in liquids with surfactants
\citep{ZhuKouYaoWuYaoSun19,ZhuKouYaoLiSun20}, Faraday instability
\citep{BorciaBestehorn14,BestehornSharmaBorciaAmiroudine21}, Rayleigh--Taylor
instability
\citep{ZanellaTegzeLetellierHenry20,ZanellaLetellierPlappTegzeHenry21},
cavitation \citep{PetitpasMassoniSaurelLapebieMunier09}, nucleation and
collapse of bubbles
\citep{MagalettiMarinoCasciola15,MagalettiGalloMarinoCasciola16,GalloMagalettiCasciola18,GalloMagalettiCoccoCasciola20},
capillary condensation \citep{Pomeau86,Benilov22b} liquid films
\citep{Benilov20d,Benilov22b}, tumor growth
\citep{FrigeriGrasselliRocca15,RoccaScala17,DaiFeireislRoccaSchimpernaSchonbek17},
classification of high-dimensional data
\citep{BertozziFlenner12,BertozziFlenner16}, \emph{etc.}

The present paper is mainly concerned with application of the DIM to
evaporation of drops and condensation of vapour on a solid. These phenomena
have been examined using the single-component version of the DIM
\citep{Benilov22a,Benilov22b}, where the fluid/gas interface is modelled by
that between the liquid and vapour phases of the same fluid. The evaporation
in this case was shown to be due to a flow caused by a weak imbalance of
chemical potentials of the liquid and vapour phases.

It is unclear, however, how the results obtained via the single-component DIM
are modified by the effect of air, whose density (at, say, $25^{\circ
}\mathrm{C}$) exceeds that of water vapour by a factor of more than $50$.
Furthermore, there are three additional physical effects in multicomponent
fluids: diffusion (of water vapour in air), the Soret effect
(thermodiffusion), and the Dufour effect (heat flux due to density gradient).
Only one of these, the diffusion, has been examined before
\citep[e.g., ][]{DeeganBakajinDupontHuberEtal00,DunnWilsonDuffyDavidSefiane09,EggersPismen10,ColinetRednikov11,RednikovColinet13,Morris14,StauberWilsonDuffySefiane14,StauberWilsonDuffySefiane15,JanecekDoumencGuerrierNikolayev15,SaxtonWhiteleyVellaOliver16,SaxtonVellaWhiteleyOliver17,RednikovColinet19,WrayDuffyWilson19},
but the models employed in these papers do not include the flow-induced
evaporative flux discovered via the DIM. Thus, the multicomponent
diffuse-interface model appears to be a tool describing all the mechanisms at
work in evaporation/condensation of liquids into/from air.

The same can hopefully be said about the dynamics of contact lines, as most of
the existing models work for some fluids (including water) only if the
so-called slip length -- effectively, the interfacial thickness -- is set to
be unrealistically small \citep{PodgorskiFlessellesLimat01,WinkelsPetersEvangelistaRiepenEtal11,PuthenveettilSenthilkumarHopfinger13,Limat14,BenilovBenilov15}.

Before using a new model, one generally needs to examine its basic properties,
test it on problems with well-understood physics (to ensure that the
mathematics captures it), and parameterise this model for the intended
applications. These are the three aims of the present work in the context of
the multicomponent DIM.

The following results are reported:

\begin{enumerate}
\item Multicomponent flat interfaces with monotonically-changing densities of
the species (components) are all stable. This conclusion follows from the
entropy principle and conservation laws.

\item Several new physical effects are described, the most interesting of
which is evaporation of a flat liquid layer in contact with saturated vapour.
This phenomenon appears to be similar to evaporation of drops surrounded by
saturated vapour
\citep{DeeganBakajinDupontHuberEtal00,EggersPismen10,Benilov20c,
Benilov21b,Benilov22a}, but with one important difference: the drops evaporate
because the curvature of their boundary increases the effective saturation
pressure (the so-called Kelvin effect), making the saturated vapour
effectively \emph{under}saturated and, thus, encouraging evaporation. This
explanation is clearly inapplicable to layers with flat boundaries -- yet they
evaporate anyway. The actual mechanism is based on the long-range interaction
of the liquid/vapour interface with the substrate -- which implies that, for
macroscopic liquid films, this effect is weak. It should be important,
however, for nano-films whose thickness is comparable to the interfacial thickness.

\item The multicomponent DIM is parameterised for water/air interfaces at
normal conditions, which can be used in the future for modelling evaporation
and condensation of water in the Earth's atmosphere or the dynamics of contact
lines of water drops.
\end{enumerate}

The paper has the following structure. In section \ref{Sec 2}, the problem is
formulated mathematically. Section \ref{Sec 3} examines the entropy principle
and conservation laws. In section \ref{Sec 4}, the governing equations are
nondimensionalised and the main nondimensional parameters are identified.
Sections \ref{Sec 5}--\ref{Sec 7} examine basic solutions of the DIM, and in
section \ref{Sec 8}, the DIM is parameterised for water/air interfaces.
Section \ref{Sec 9} summarises the results obtained, including the effects
partly mentioned in item 2 of the above list.

\section{Formulation\label{Sec 2}}

\subsection{Thermodynamics\label{Sec 2.1}}

When studying hydrodynamics of a compressible fluid, one has to deal with its
thermodynamic properties. In this section, they are described briefly and in a
self-contained form (for the benefit of readers specializing is incompressible hydrodynamics).

Consider an $N$-component compressible non-ideal fluid, characterised by the
temperature $T$ and partial mass densities $\rho_{i}$ where $i=1...N$. The
fluid's thermodynamic properties are fully described by two functions: the
internal energy $e(\rho_{1}...\rho_{N},T)$ and entropy $s(\rho_{1}...\rho
_{N},T)$ -- both specific, i.e., per unit mass. The dependence of the fluid
pressure $p$ on $(\rho_{1}...\rho_{N},T)$, or the equation of state, is
defined by%
\begin{equation}
p=\rho\sum_{i}\rho_{i}\left(  \frac{\partial e}{\partial\rho_{i}}%
-T\frac{\partial s}{\partial\rho_{i}}\right)  , \label{2.1}%
\end{equation}
\citep[e.g.][]{GiovangigliMatuszewski13}, where%
\begin{equation}
\rho=\sum_{i}\rho_{i} \label{2.2}%
\end{equation}
is the total density (here and hereinafter, the summation is implied to be
from $1$ to $N$ unless stated otherwise). The partial chemical potentials, in
turn, are given by%
\begin{equation}
G_{i}=\frac{\partial\left(  \rho e\right)  }{\partial\rho_{i}}-T\frac
{\partial\left(  \rho s\right)  }{\partial\rho_{i}}. \label{2.3}%
\end{equation}
Note that $e(\rho_{1}...\rho_{N},T)$ and $s(\rho_{1}...\rho_{N},T)$ are not
fully arbitrary, but should satisfy the fundamental thermodynamic relation, or
the Gibbs relation, which can be written in the form%
\begin{equation}
\frac{\partial e}{\partial T}=T\frac{\partial s}{\partial T}. \label{2.4}%
\end{equation}
The equivalence of this equality and the standard form of the Gibbs relation
is shown in Appendix \ref{Appendix A}.

Using equations (\ref{2.1})--(\ref{2.4}), one can derive the following
identities:%
\begin{equation}
\frac{\partial p}{\partial T}=\sum_{i}\rho_{i}\frac{\partial G_{i}}{\partial
T}+\rho s, \label{2.5}%
\end{equation}%
\begin{equation}
\frac{\partial p}{\partial\rho_{j}}=\sum_{i}\rho_{i}\frac{\partial G_{i}%
}{\partial\rho_{j}}, \label{2.6}%
\end{equation}
then rewrite (\ref{2.6}) in the form%
\begin{equation}
G_{i}=\frac{\partial}{\partial\rho_{i}}\left(  \sum_{j}\rho_{j}G_{j}-p\right)
. \label{2.7}%
\end{equation}
In what follows, one also needs the specific heat capacity at constant
volume,
\begin{equation}
c=\frac{\partial e}{\partial T}, \label{2.8}%
\end{equation}
where the traditional subscript $_{V}$ is omitted. In this paper, $c$ is
assumed to be positive, which is indeed the case for neutral fluids \citep{Lyndenbell99}.

Define also%
\begin{equation}
a_{i}=-\frac{\partial e}{\partial\rho_{i}}, \label{2.9}%
\end{equation}
which will be referred to as the generalised van der Waals parameter (of the
$i$-th species), and%
\begin{equation}
B=-p-\rho\sum_{i}\rho_{i}a_{i}. \label{2.10}%
\end{equation}
$B$ is not one of the standard thermodynamic functions, but is convenient to
use when thermodynamics is coupled to hydrodynamics: as seen later, $B$
characterises the release (consumption) of heat due to the fluid's mechanical
compression (expansion). Using equation of state (\ref{2.1}) and definition
(\ref{2.9}) of the van der Waals parameter, one can rearrange expression
(\ref{2.10}) in the form%
\begin{equation}
B=\rho T\sum_{i}\rho_{i}\frac{\partial s}{\partial\rho_{i}}. \label{2.11}%
\end{equation}

\subsection{Examples: the Enskog--Vlasov and van der Waals
fluids\label{Sec 2.2}}

In the low-density limit, the specific internal energy and entropy of any
fluid should tend to those of ideal gas, i.e.,%
\[
e\sim\frac{T}{\rho}\sum_{i}c_{i}\rho_{i},\qquad s\sim\frac{\ln T}{\rho}%
\sum_{i}c_{i}\rho_{i}-\frac{1}{\rho}\sum_{i}R_{i}\rho_{i}\ln\rho_{i}%
\qquad\text{as}\qquad\rho_{i}\rightarrow0,
\]
where $c_{i}$ is the specific partial heat capacity, and $R_{i}=k_{B}/m_{i}$
is the specific gas constant ($m_{i}$ is the molecular mass of the $i$-th
species, and%
\[
k_{B}=1.380649\times10^{-23}\mathrm{m}^{2}\mathrm{s}^{-2}\mathrm{kg\,K}^{-1}%
\]
is the Boltzmann constant). Note also that one can replace $\ln T\rightarrow
\ln T/\bar{T}$ and $\ln\rho\rightarrow\ln\rho/\bar{\rho}$, where are $\bar{T}$
and $\bar{\rho}$ are suitable dimensional scales. This would make the
arguments of the logarithms nondimensional (with none of physically measurable
parameters depending on $\bar{T}$ and $\bar{\rho}$).

A simple description of non-ideal fluids is delivered by the Enskog--Vlasov
(EV) model, according to which%
\begin{equation}
e=\frac{T}{\rho}\sum_{i}c_{i}\rho_{i}-\frac{1}{\rho}\sum_{ij}a_{ij}\rho
_{i}\rho_{j}, \label{2.12}%
\end{equation}%
\begin{equation}
s=\frac{\ln T}{\rho}\sum_{i}c_{i}\rho_{i}-\frac{1}{\rho}\sum_{i}R_{i}\rho
_{i}\ln\rho_{i}-\Theta(\rho_{1}...\rho_{N}), \label{2.13}%
\end{equation}
where $\Theta(\rho_{1}...\rho_{N})$ is an arbitrary analytic function, and
$a_{ij}$ characterises the long-range attraction between molecules of the
$i$-th and $j$-th species (with the implication that $a_{ij}=a_{ji}$). For a
pure fluid, $a_{11}$ coincides with the generalised van der Waals parameter
defined by (\ref{2.9}) -- hence, the notation.

One can readily verify that expressions (\ref{2.12})--(\ref{2.13}) satisfy the
Gibbs relation (\ref{2.4}). They can be viewed as two-term expansions in small
$T$, under an extra assumption that the heat capacity $c$ and the generalised
van der Waals parameter $a_{i}$ are independent of $T$ and are zeroth-degree
homogeneous functions of $\rho_{i}$ (for a pure fluid, $c$ and $a_{1}$ do not
depend on $\rho_{1}$ at all).

The EV fluid model originates from the Enskog--Vlasov kinetic theory
\citep{Desobrino67,Grmela71} and, as such, is naturally suited for the use
with the DIM which can be viewed as a hydrodynamic approximation of the EV
kinetic equation \citep{Giovangigli20,Giovangigli21}. Equations (\ref{2.12}%
)--(\ref{2.13}) work very well for inert gases \citep{BenilovBenilov19a}, they
will be shown to work reasonably well for water, nitrogen, and oxygen (section
\ref{Sec 8} of this paper).

The EV involves too many parameters to be used as an illustration of general
results -- so, in such cases, the simpler van der Waals model will be
employed. It is a particular case of equations (\ref{2.12})--(\ref{2.13}) with%
\begin{equation}
\Theta=-\frac{1}{\rho}\sum_{i}R_{i}\rho_{i}\ln\left(  1-%
%TCIMACRO{\dsum \limits_{j}}%
%BeginExpansion
{\displaystyle\sum\limits_{j}}
%EndExpansion
b_{j}\rho_{j}\right)  , \label{2.14}%
\end{equation}
where $b_{i}$ is the reciprocal of the maximum density of the $i$-th species.
Physically, $1/b_{i}$ can be interpreted as density of the closest packing.

Given expressions (\ref{2.12})--(\ref{2.14}), definition (\ref{2.1}) of the
pressure and definition (\ref{2.3}) of the chemical potential yield%
\begin{equation}
p=\frac{T}{1-%
%TCIMACRO{\dsum \limits_{j}}%
%BeginExpansion
{\displaystyle\sum\limits_{j}}
%EndExpansion
b_{j}\rho_{j}}\sum_{i}R_{i}\rho_{i}-\sum_{ij}a_{ij}\rho_{i}\rho_{j},
\label{2.15}%
\end{equation}%
\begin{equation}
G_{i}=TR_{i}\ln\frac{\rho_{i}}{1-%
%TCIMACRO{\dsum \limits_{j}}%
%BeginExpansion
{\displaystyle\sum\limits_{j}}
%EndExpansion
b_{j}\rho_{j}}+\frac{Tb_{i}%
%TCIMACRO{\dsum \limits_{j}}%
%BeginExpansion
{\displaystyle\sum\limits_{j}}
%EndExpansion
R_{j}\rho_{j}}{1-%
%TCIMACRO{\dsum \limits_{j}}%
%BeginExpansion
{\displaystyle\sum\limits_{j}}
%EndExpansion
b_{j}\rho_{j}}-2\sum_{j}a_{ij}\rho_{j}+T\left(  R_{i}+c_{i}-c_{i}\ln T\right)
. \label{2.16}%
\end{equation}
For a pure fluid ($N=1$), equation (\ref{2.15}) reduces to the classical van
der Waals equation of state \citep{Vanderwaals93}.

\subsection{Hydrodynamics\label{Sec 2.3}}

Consider a fluid flow characterised by the species densities $\rho
_{i}(\mathbf{r},t)$, mass-averaged velocity $\mathbf{v}(\mathbf{r},t)$, and
temperature $T(\mathbf{r},t)$, where $\mathbf{r}=\left[  x,y,z\right]  $ is
the position vector and $t$, the time. Let the species be affected by forces
$\mathbf{F}_{i}$ (which will be later identified with the van der Waals
forces), and the fluid as a whole, affected by viscosity. The shear viscosity
$\mu_{s}$ and bulk viscosity $\mu_{b}$ depend generally on $\rho_{i}$ and $T$.

Let the flow be governed by following equations:%
\begin{equation}
\frac{\partial\rho_{i}}{\partial t}+\mathbf{\nabla}\cdot\left(  \rho
_{i}\mathbf{v}+\mathbf{J}_{i}\right)  =0, \label{2.17}%
\end{equation}%
\begin{equation}
\frac{\partial\left(  \rho\mathbf{v}\right)  }{\partial t}+\mathbf{\nabla
}\cdot\left(  \rho\mathbf{vv}\right)  =\mathbf{\nabla}\cdot\boldsymbol{\Pi
}+\sum_{i}\rho_{i}\mathbf{F}_{i}-\mathbf{\nabla}p, \label{2.18}%
\end{equation}%
\begin{multline}
\frac{\partial}{\partial t}\left(  \rho e+\frac{1}{2}\rho\left\vert
\mathbf{v}\right\vert ^{2}\right)  +\mathbf{\nabla}\cdot\left[  \left(  \rho
e+\frac{1}{2}\rho\left\vert \mathbf{v}\right\vert ^{2}+p\right)
\mathbf{v}-\boldsymbol{\Pi}\cdot\mathbf{v}+\mathbf{Q}\right] \\
=\sum_{i}\mathbf{F}_{i}\cdot\left(  \rho_{i}\mathbf{v}+\mathbf{J}_{i}\right)
. \label{2.19}%
\end{multline}
Here, the viscous stress tensor is%
\begin{equation}
\boldsymbol{\Pi}=\mu_{s}\left[  \boldsymbol{\boldsymbol{\nabla}}%
\mathbf{v}+\left(  \boldsymbol{\boldsymbol{\nabla}}\mathbf{v}\right)
^{tr}-\frac{2}{3}\mathbf{I}\left(  \boldsymbol{\boldsymbol{\nabla}}%
\cdot\mathbf{v}\right)  \right]  +\mu_{b}\,\mathbf{I}\left(
\boldsymbol{\boldsymbol{\nabla}}\cdot\mathbf{v}\right)  , \label{2.20}%
\end{equation}
where the dotless product of two vectors (e.g.,
$\boldsymbol{\boldsymbol{\nabla}}\mathbf{v}$) produces a second-order tensor,
and the superscript $^{tr}$ denotes transposition. The diffusion fluxes
$\mathbf{J}_{i}$ and the heat flux $\mathbf{Q}$ are related to the forces
$\mathbf{F}_{i}$, temperature $T$, and chemical potentials $G_{i}$ by%
\begin{equation}
\mathbf{J}_{i}=\sum_{j}D_{ij}\left[  \mathbf{F}_{j}-T\mathbf{\nabla}\left(
\frac{G_{j}}{T}\right)  \right]  -\frac{\zeta_{i}}{T}\mathbf{\nabla}T,
\label{2.21}%
\end{equation}%
\begin{equation}
\mathbf{Q}=\sum_{j}\zeta_{j}\left[  \mathbf{F}_{j}-T\mathbf{\nabla}\left(
\frac{G_{j}}{T}\right)  \right]  -\frac{\kappa}{T}\mathbf{\nabla}T,
\label{2.22}%
\end{equation}
where $D_{ij}$, $\zeta_{i}$, and $\kappa$ are the transport coefficients (all
three depend generally on $\rho_{i}$ and $T$).

To understand the physical meaning of expressions (\ref{2.21})--(\ref{2.22}),
rewrite them in the form%
\begin{equation}
\mathbf{J}_{i}=\sum_{j}D_{ij}\mathbf{F}_{j}-~\underset{\text{diffusion}%
}{\underbrace{\sum_{j}D_{ij}^{\prime}\mathbf{\nabla}\rho_{j}}}%
~-~\underset{\text{Soret effect}}{\underbrace{\zeta_{i}^{\prime}%
\rho\mathbf{\nabla}T}}, \label{2.23}%
\end{equation}%
\begin{equation}
\mathbf{Q}=\sum_{j}\zeta_{j}\mathbf{F}_{j}-~\underset{\text{Dufour
effect}}{\underbrace{\sum_{j}\zeta_{j}\sum_{k}\frac{\partial G_{j}}%
{\partial\rho_{k}}\mathbf{\nabla}\rho_{k}}~}-~\underset{\text{heat
conduction}}{\underbrace{\kappa^{\prime}\mathbf{\nabla}T}}, \label{2.24}%
\end{equation}
where%
\[
D_{ik}^{\prime}=\sum_{j}D_{ij}\frac{\partial G_{j}}{\partial\rho_{k}}%
,\qquad\zeta_{i}^{\prime}=\frac{\zeta_{i}}{T\rho}+\sum_{j}\frac{D_{ij}}{\rho
}\left(  \frac{\partial G_{j}}{\partial T}-\frac{G_{j}}{T}\right)  ,
\]%
\[
\kappa^{\prime}=\frac{\kappa}{T}+\sum_{j}\zeta_{j}\left(  \frac{\partial
G_{j}}{\partial T}-\frac{G_{j}}{T}\right)
\]
are the standard diffusivities, thermodiffusivities, and thermal conductivity,
respectively. The second term in expression (\ref{2.23}) corresponds to the
classical Fick Law (the fluxes depend linearly on the density gradients), and
the last term in expression (\ref{2.24}) characterises heat conduction
described by the Fourier Law. The last term in (\ref{2.23}) describes the
Soret effect ($\mathbf{\nabla}T$ gives rise to diffusion) and the second term
in (\ref{2.24}), the Dufour effect ($\mathbf{\nabla}\rho_{j}$ gives rise to
heat conduction).

The same four effects -- diffusion, heat conduction, the Soret and Dufour
effects -- are described, obviously, by the original expressions
(\ref{2.21})--(\ref{2.22}), albeit in a form where the terms cannot be matched
to a single effect each.

One might think that representing the fluxes in terms of $\mathbf{\nabla}%
\rho_{j}$ would be more natural than using $\mathbf{\nabla}\left(
G_{j}/T\right)  $. Observe, however, that the coefficient of $\left(
\mathbf{\nabla}T\right)  /T$ in expression (\ref{2.21}) coincides with the
coefficient of $\left[  \mathbf{F}_{j}-T\mathbf{\nabla}\left(  G_{j}/T\right)
\right]  $ in (\ref{2.22}). This symmetry reflects the so-called Onsager
reciprocal relations \citep{FerzigerKaper72}, which also imply%
\[
D_{ij}=D_{ji},
\]
i.e., the diffusion of an $i$-th species in a $j$-th species occurs the same
way as that of the $j$-th species in the $i$-th species.

It should also be assumed that the extended transport matrix,%
\[
D_{ij}^{(ext)}=%
\begin{bmatrix}
&  &  & \zeta_{1}\\
& D_{ij} &  & \vdots\\
&  &  & \zeta_{N}\\
\zeta_{1} & \cdots & \zeta_{N} & \kappa
\end{bmatrix}
\]
is positive semidefinite ($D_{ij}^{(ext)}\succeq0$) -- i.e.,
\[
\sum_{i=1}^{N+1}\sum_{j=1}^{N+1}d_{i}\,D_{ji}^{(ext)}d_{j}\geq0,
\]
for all $\left(  N+1\right)  $-dimensional arrays $d_{i}$. As seen later, this
property is essential for the entropy principle to hold.

Furthermore, the transport coefficients should be such that%
\begin{equation}
\sum_{i}D_{ij}=0,\qquad\sum_{i}\zeta_{i}=0. \label{2.25}%
\end{equation}
As a result, the density equation (\ref{2.17}) and expressions (\ref{2.21})
for the diffusion fluxes imply that%
\begin{equation}
\frac{\partial\rho}{\partial t}+\mathbf{\nabla}\cdot\left(  \rho
\mathbf{v}\right)  =0, \label{2.26}%
\end{equation}
where $\rho$ is the total density given by (\ref{2.2}). Observe that, for a
pure fluid, restrictions (\ref{2.25}) can be satisfied only if $D_{11}=0$ and
$\zeta_{1}=0$, which means that pure fluids neither diffuse nor thermodiffuse.

Equations (\ref{2.17})--(\ref{2.22}) have been first derived from the
thermodynamics of irreversible processes \citep{Meixner41}. They were also
derived from statistical mechanics \citep{BearmanKirkwood58,Mori58} and
non-equilibrium statistical thermodynamics \citep{Keizer87} -- for more
references, see \citep{Giovangigli99}. A derivation of the small-density
version of (\ref{2.17})--(\ref{2.22}) from the Boltzmann kinetic equation can
be found in any textbook on kinetic theory \citep[e.g.][]{FerzigerKaper72}.

In all these cases, the derived expressions for the transport coefficients
automatically satisfy the Onsager relations and the rest of the properties
listed above.

Note also that a reduction of the above equations for a binary fluid with no
Soret and Dufour effects ($N=2$, $\zeta_{i}=0$) was used by
\cite{LiuAmbergDoquang16} to show that such a model is able to describe the
phase equilibrium for a real binary mixture of $\mathrm{CO}_{2}$ and ethanol.

\subsection{Alternative forms of the momentum and energy
equations\label{Sec 2.4}}

Since the transport fluxes (\ref{2.21})--(\ref{2.22}) are expressed in terms
of $G_{i}$ and $T$ (not $\rho_{i}$ and $T$), it is convenient to do the same
for the pressure gradient in the momentum equation (\ref{2.18}). Recalling
identities (\ref{2.6})--(\ref{2.5}) which imply%
\[
\mathbf{\nabla}p=\sum_{i}\rho_{i}\mathbf{\nabla}G_{i}+\rho s\mathbf{\nabla}T,
\]
and using equation (\ref{2.26}) to simply the left-hand side of the momentum
equation (\ref{2.18}), one reduces it to%
\begin{equation}
\rho\left[  \frac{\partial\mathbf{v}}{\partial t}+\left(  \mathbf{v}%
\cdot\mathbf{\nabla}\right)  \mathbf{v}\right]  =\mathbf{\nabla}%
\cdot\boldsymbol{\Pi}+\sum_{i}\rho_{i}\left(  \mathbf{F}_{i}-\mathbf{\nabla
}G_{i}\right)  -\rho s\mathbf{\nabla}T. \label{2.27}%
\end{equation}
The energy equation, in turn, can be conveniently rewritten in terms of the
temperature (which is a measurable quantity, unlike the internal energy $e$).
Replacing, thus, in Eq. (\ref{2.19}),%
\[
\frac{\partial e}{\partial t}=\sum_{i}\frac{\partial e}{\partial\rho_{i}}%
\frac{\partial\rho_{i}}{\partial t}+\frac{\partial e}{\partial T}%
\frac{\partial T}{\partial t},
\]
one should use the density equation to eliminate $\partial\rho_{i}/\partial
t$. Using then equations (\ref{2.17}) and (\ref{2.27}), and recalling
identities (\ref{2.8})--(\ref{2.9}) and (\ref{2.11}), one obtains%
\begin{equation}
\rho c\left(  \frac{\partial T}{\partial t}+\mathbf{v}\cdot\mathbf{\nabla
}T\right)  +\mathbf{\nabla}\cdot\mathbf{Q}=\boldsymbol{\Pi}:\left(
\mathbf{\nabla v}\right)  +B\mathbf{\nabla}\cdot\mathbf{v}+\sum_{i}\left(
\mathbf{F}_{i}-\rho a_{i}\mathbf{\nabla}\right)  \cdot\mathbf{J}_{i},
\label{2.28}%
\end{equation}
where the symbol \textquotedblleft$:$\textquotedblright\ denotes double scalar
product of two tensors.

The first term on the right-hand side of equation (\ref{2.28}) describes heat
production by viscosity and the second term, that by fluid compression or expansion.

\subsection{The van der Waals force\label{Sec 2.5}}

Assume that a molecule of a $j$-th species exerts on a molecule of an $i$-th
species an isotropic force with a potential $\Phi_{ij}(r)$ where $r=\left(
x^{2}+y^{2}+z^{2}\right)  ^{1/2}$. Assuming for simplicity that the fluid is
unbounded, one can write the mass-averaged force affecting the $i$-th species
in the hydrodynamic equations (\ref{2.18})--(\ref{2.19}) in the form%
\begin{equation}
\mathbf{F}_{i}(\mathbf{r},t)=\mathbf{\nabla}\sum_{j}\int\frac{\rho
_{j}(\mathbf{r}^{\prime},t)}{m_{i}m_{j}}\Phi_{ij}(|\mathbf{r}^{\prime
}-\mathbf{r|})\,\mathrm{d}^{3}\mathbf{r}^{\prime}, \label{2.29}%
\end{equation}
where $m_{i}$ is the molecular mass an, the integration is implied to be over
the whole space. To guarantee the convergence of the integral in (\ref{2.29})
and those arising later, the potential $\Phi_{ij}(r)$ is assumed to decay
exponentially as $r\rightarrow\infty$.

Next, let the spatial scale of $\rho(\mathbf{r},t)$ be much larger than that
of $\Phi_{ij}(r)$, in which case expression (\ref{2.29}) can be simplified
asymptotically. To do so, change in it $\mathbf{r}^{\prime}\rightarrow
\mathbf{r}^{\prime}+\mathbf{r}$ and then expand $\rho_{j}(\mathbf{r}^{\prime
}+\mathbf{r},t)$ about $\mathbf{r}^{\prime}$, which yields%
\[
\mathbf{F}_{i}(\mathbf{r},t)=\mathbf{\nabla}\sum_{j}\int\left[  \rho
_{j}(\mathbf{r},t)+\mathbf{r}^{\prime}\cdot\mathbf{\nabla}\rho_{j}%
(\mathbf{r},t)+\frac{1}{2}\mathbf{r}^{\prime}\mathbf{r}^{\prime}%
:\mathbf{\nabla\nabla}\rho_{j}(\mathbf{r},t)+\cdots\right]  \frac{\Phi
_{ij}(r^{\prime})}{m_{i}m_{j}}\mathrm{d}^{3}\mathbf{r}^{\prime}.
\]
Given the isotropy of $\Phi_{ij}(r^{\prime})$, the second integral in the
above expansion vanishes, and one obtains%
\begin{equation}
\mathbf{F}_{i}=\sum_{j}W_{ij}\mathbf{\nabla}\rho_{j}+\sum_{j}K_{ij}%
\mathbf{\nabla}\nabla^{2}\rho_{j}+\cdots, \label{2.30}%
\end{equation}
where%
\[
W_{ij}=\int\frac{\Phi_{ij}(r^{\prime})}{m_{i}m_{j}}\mathrm{d}^{3}%
\mathbf{r}^{\prime},\qquad K_{ij}=\int\frac{r^{\prime2}}{2}\frac{\Phi
_{ij}(r^{\prime})}{m_{i}m_{j}}\mathrm{d}^{3}\mathbf{r}^{\prime}.
\]
Since Newton's Third Law implies that $\Phi_{ij}=\Phi_{ji}$, the matrices
$W_{ij}$ and $K_{ij}$ are symmetric.

Once expansion (\ref{2.30}) is substituted into the hydrodynamic equations
(\ref{2.18})--(\ref{2.19}), its first term can be absorbed into the internal
energy -- i.e., eliminated by the change%
\[
e\rightarrow e+\frac{1}{2\rho}\sum_{ij}W_{ij}\rho_{i}\rho_{j},\qquad
G_{i}\rightarrow G_{i}+\sum_{j}W_{ij}\rho_{j}.
\]
This does not come as a surprise, as the energy associated with potential
interactions of molecules can be viewed as a kind of internal energy -- in
fact, the second term of expansion (\ref{2.30}) could also be (and sometimes
is) absorbed into $e$. This is not done in this paper, however, as it would
make $e$ a functional (instead of a function), with the implication that all
the thermodynamic definitions and identities in Sect. \ref{Sec 2.1} would need
to be rewritten in terms of functional derivatives.

Thus, without loss of generality, one can set in expression (\ref{2.30})
$W_{ij}=0$. Omitting also the small terms hidden in \textquotedblleft$\cdots
$\textquotedblright, one obtains%
\begin{equation}
\mathbf{F}_{i}=\sum_{j}K_{ij}\mathbf{\nabla}\nabla^{2}\rho_{j}, \label{2.31}%
\end{equation}
which is the multicomponent extension of the standard DIM formula for the van
der Waals force \citep[e.g.,][]{PismenPomeau00}. The matrix $K_{ij}$ is the
extension of the so-called Korteweg parameter for pure fluids, and it will be
referred to as the Korteweg matrix. It should be positive definite,
$K_{ij}\succ0$, as the van der Waals force should be attractive, not repulsive.

Since the original representation (\ref{2.29}) was a phenomenological model
and, thus, the pair-wise potential $\Phi_{ij}$ cannot be measured, the
Korteweg matrix should be viewed as a set of adjustable parameters. As seen
later, they can be deduced from the measurements of the equation of state and
surface tension. One should keep in mind, however, that the Korteweg matrix
should not depend on the temperature. Such a dependence would be physically
unjustified, as the intermolecular attraction (characterised by $K_{ij}$)
should not depend on the molecules' velocities (characterised by $T$).

\subsection{Boundary conditions at a solid wall\label{Sec 2.6}}

Let the fluid occupy a domain $\mathcal{D}$, bounded by a smooth surface
$\partial\mathcal{D}$. For simplicity, the so-called Navier slip is disallowed
in this work, so that the boundary condition for the velocity is%
\begin{equation}
\mathbf{v}=\mathbf{0}\qquad\text{at}\qquad\mathbf{r}\in\partial\mathcal{D}.
\label{2.32}%
\end{equation}
To ensure mass conservation, one should require%
\begin{equation}
\mathbf{n\cdot J}_{i}=0\qquad\text{at}\qquad\mathbf{r}\in\partial\mathcal{D},
\label{2.33}%
\end{equation}
where $\mathbf{n}$ is the outward-pointing unit normal to $\partial
\mathcal{D}$.

The boundary condition for the temperature, in turn, depends on the problem at
hand. Since this paper is concerned, \emph{inter alia}, with the entropy
principle and energy conservation, it will be assumed that no heat escapes
through the boundary,%
\begin{equation}
\mathbf{n\cdot Q}=0\qquad\text{at}\qquad\mathbf{r}\in\partial\mathcal{D}.
\label{2.34}%
\end{equation}
Boundary conditions (\ref{2.32})--(\ref{2.34}) would be sufficient for the
standard compressible multi-component hydrodynamics, but the DIM requires an
extra condition [due to the presence of higher-order derivatives of the
density field in expression (\ref{2.31})].

The most common version of such condition -- prescribing a linear combination
of the boundary-value of the density and its normal gradient -- ascends to the
paper by \cite{CahnHilliard58}. In application to pure fluids, the Neumann
reduction of this condition was proposed by \cite{Seppecher96} and the
Dirichlet reduction, by \cite{PismenPomeau00}. As shown by \cite{Benilov20a},
the latter follows from the assumptions under which the whole DIM is derived
(pair-wise intermolecular interactions, slowly-varying density field), and so
it is used in the present paper. Thus, require that%
\begin{equation}
\rho_{i}=\rho_{0,i}\qquad\text{at}\qquad\mathbf{r}\in\partial\mathcal{D},
\label{2.35}%
\end{equation}
where the constant $\rho_{0,i}$ is specific to the fluid/solid combination
under consideration. The general version of the boundary condition for
$\rho_{i}$ is discussed briefly in Appendix \ref{Appendix B}.

To clarify the physical meaning of condition (\ref{2.35}), consider the van
der Waals forces acting on a fluid molecule located infinitesimally close to
the wall: the solid attracts it \emph{towards} the wall, while the other fluid
molecules pull it \emph{away}. The former force is fixed, whereas the latter
grows with the near-wall density -- so that the balance is achieved when the
density assumes a certain value -- namely, the parameter $\rho_{0,i}$ in
condition (\ref{2.35}). This argument suggests that a smaller value of
$\rho_{0,i}$ corresponds to a hydrophobic wall (characterised by a large
contact angle) and larger $\rho_{0,i}$, to a hydrophilic one.

According to its physical meaning, $\rho_{0,i}$ does not depend on the
temperature. As shown later, its value can be deduced from a measurement of
the contact angle.

\section{The entropy principle\label{Sec 3}}

\subsection{Conservation laws and the H-theorem\label{Sec 3.1}}

It can be verified that the governing equations and boundary conditions
introduced above conserve the mass of each species%
\begin{equation}
M_{i}=\int_{\mathcal{D}}\rho_{i}\mathrm{d}^{3}\mathbf{r}, \label{3.1}%
\end{equation}
and the total energy%
\begin{equation}
E=\int_{\mathcal{D}}\left[  \rho e+\frac{1}{2}\rho\left\vert \mathbf{v}%
\right\vert ^{2}+\frac{1}{2}\sum_{ij}K_{ij}\left(  \mathbf{\nabla}\rho
_{i}\right)  \cdot\left(  \mathbf{\nabla}\rho_{j}\right)  \right]
\mathrm{d}^{3}\mathbf{r}. \label{3.2}%
\end{equation}
The three terms in expression (\ref{3.2}) represent the internal, kinetic, and
van der Waals energies.

The governing equations and boundary conditions satisfy also an H-theorem,
reflecting the fact that the net entropy of a fluid in a thermally-insulated
container cannot decrease. To prove this, consider the following combination
of the governing equations:%
\[
\left(  \ref{2.28}\right)  +T\rho\sum_{i}\frac{\partial s}{\partial\rho_{i}%
}\times(\ref{2.17})+Ts\times\left(  \ref{2.26}\right)  .
\]
After straightforward algebra involving the use of the thermodynamic
identities presented in subsection \ref{Sec 2.1} and expressions
(\ref{2.21})--(\ref{2.22}) for the fluxes, one obtains%
\begin{multline}
\frac{\partial\left(  \rho s\right)  }{\partial t}+\mathbf{\nabla}\cdot\left(
\rho s\mathbf{v}-\sum_{i}\frac{G_{i}}{T}\mathbf{J}_{i}+\frac{\mathbf{Q}}%
{T}\right)  =\frac{\boldsymbol{\Pi}:\left(  \mathbf{\nabla v}\right)  }{T}\\
+\left\{  \sum_{ij}D_{ij}\left\vert \mathbf{F}_{j}-T\mathbf{\nabla}\left(
\frac{G_{j}}{T}\right)  \right\vert ^{2}+2\sum_{i}\zeta_{i}\left[
\mathbf{F}_{i}-T\mathbf{\nabla}\left(  \frac{G_{i}}{T}\right)  \right]
\cdot\left(  -\frac{\mathbf{\nabla}T}{T}\right)
\phantom{\left\vert -\frac{\mathbf{\nabla}T}{T}\right\vert ^{2}}\right. \\
+\left.
\vphantom{\sum_{ij}\left\vert \left( \frac{G_{j}}{T}\right) \right\vert ^{2}}\kappa
T\left\vert -\frac{\mathbf{\nabla}T}{T}\right\vert ^{2}\right\}  . \label{3.3}%
\end{multline}
The first term on the right-hand side of this equation is non-negative due to
the following (easily verifiable) identity,%
\begin{equation}
\boldsymbol{\Pi}:\left(  \mathbf{\nabla v}\right)  =\frac{1}{2}\mu
_{s}\left\vert \left(  \mathbf{\nabla v}\right)  +\left(  \mathbf{\nabla
v}\right)  ^{tr}-\frac{2}{3}\mathbf{I\nabla}\cdot\mathbf{v}\right\vert
^{2}+\mu_{b}\left(  \mathbf{\nabla}\cdot\mathbf{v}\right)  ^{2}, \label{3.4}%
\end{equation}
whereas the expression in curly brackets is non-negative because the extended
transport matrix is positive semidefinite (see subsection \ref{Sec 2.3}).
Thus, integrating equation (\ref{3.3}) over $\mathcal{D}$, and using boundary
conditions (\ref{2.32})--(\ref{2.34}), one obtains%
\begin{equation}
\frac{\mathrm{d}S}{\mathrm{d}t}\geq0, \label{3.5}%
\end{equation}
where%
\begin{equation}
S=\int_{\mathcal{D}}\rho s\,\mathrm{d}^{3}\mathbf{r}. \label{3.6}%
\end{equation}
Inequality (\ref{3.5}) is the desired H-theorem.

It follows from (\ref{3.3})--(\ref{3.4}) that the exact equality in
(\ref{3.5}) can only hold if the velocity field is spatially uniform; together
with the no-flow boundary conditions, this requirement amounts to
$\mathbf{v}=\mathbf{0}$ (i.e., the fluid is static).

\subsection{Stability via the entropy principle\label{Sec 3.2}}

The most common way to examine the stability of a steady solution of a set of
equations consist in linearizing these equations with respect to a small
perturbation, assuming the harmonic dependence of the perturbation on $t$, and
solving the resulting eigenvalue problem. In the problem at hand, however, it
is much simpler to examine stability using the entropy principle.

If, at a certain steady state, the total entropy $S$ has a local maximum
constrained by the conditions of fixed energy $E$ and mass $M_{i}$, this state
is stable. The inverse is also true: if $S$ does \emph{not} have a maximum,
the corresponding state is \emph{un}stable -- because a perturbed solution
with a higher entropy cannot evolve `back'. Neutrally stable oscillations are
also prohibited by the entropy principle -- hence, the system can only evolve
further away, towards a steady state where the entropy does have a maximum.

Let a fluid be enclosed in a container (which can be later assumed to be
infinitely large, if need be) and seek a maximum of $S$, constrained by the
conditions of fixed $M_{i}$ and $E$. This problem amounts to finding the
stationary points of the functional%
\begin{equation}
H[\rho_{1}...\rho_{N},T,\mathbf{v}]=S+\sum_{i}\eta_{i}M_{i}+\lambda E,
\label{3.7}%
\end{equation}
where $\lambda$ and $\mu_{i}$ are the Lagrange multipliers, and $S$, $E$, and
$M_{i}$ are given by (\ref{3.6}), (\ref{3.2}), and (\ref{3.1}), respectively.
Varying $H$ with respect to $\mathbf{v}$ and equating the variation to zero,
one obtains%
\begin{equation}
\mathbf{v}=\mathbf{0}, \label{3.8}%
\end{equation}
i.e., a steady state must be (unsurprisingly) static. Next, varying $H$ with
respect to $T$, one obtains%
\[
\frac{\partial s}{\partial T}+\lambda\frac{\partial e}{\partial T}=0.
\]
Comparison of this equality with the Gibbs relation (\ref{2.4}) yields%
\begin{equation}
\lambda=-\frac{1}{T}. \label{3.9}%
\end{equation}
Since $\lambda$ is a constant, equation (\ref{3.9}) implies that $T$ is
spatially uniform -- i.e., a steady state must be isothermal.

Finally, varying $H$ with respect to $\rho_{i}$, recalling expressions
(\ref{3.8})--(\ref{3.9}) for $\mathbf{v}$ and $\lambda$, and keeping in mind
definition (\ref{2.3}) of the chemical potential $G_{i}$, one obtains
\begin{equation}
\frac{1}{T}\left(  G_{i}-\sum_{j}K_{ij}\nabla^{2}\rho_{j}\right)  +\eta_{i}=0.
\label{3.10}%
\end{equation}
This equation describes all steady states of the governing equations, and it
will be extensively used in the rest of this paper. The temperature $T$ in
equation (\ref{3.10}) should be treated as a known parameter, whereas the
Lagrange multiplier $\eta_{i}$ is to be deduced from the boundary conditions.
The latter will be illustrated for the case of an infinite domain, plus an
assumption that the fluid at infinity be spatially uniform and characterised
by a coordinate-independent chemical potential $G_{i}=G_{\infty,i}$. In this
case equation (\ref{3.10}) yields%
\begin{equation}
\eta_{i}=-\frac{G_{\infty,i}}{T}, \label{3.11}%
\end{equation}
as required.

Note also that equation (\ref{3.10}) can also be recovered by adapting the
governing equations for the state of equilibrium. To do so, one should set
$\mathbf{v}=\mathbf{0}$, $T=\operatorname{const}$, and $\partial\rho
_{i}/\partial t=0$ in equations (\ref{2.17})--(\ref{2.18}) and, recalling
expression (\ref{2.21}) for the diffusion fluxes, obtain (\ref{3.10}), as required.

To examine a solution of equation (\ref{3.10}) for stability, one needs to
calculate the second variation of $H$. Omitting the algebra (involving the use
of identities (\ref{2.3})--(\ref{2.4}), definition (\ref{2.8}) of the heat
capacity $c$, and expression (\ref{3.9}) for $\lambda$), one obtains%
\[
\delta^{2}H=\frac{1}{T}\int_{\mathcal{D}}\left\{  \sum_{ij}\left[
-\frac{\partial G_{i}}{\partial\rho_{j}}\left(  \delta\rho_{i}\right)  \left(
\delta\rho_{j}\right)  -K_{ij}\left(  \mathbf{\nabla}\delta\rho_{i}\right)
\cdot\left(  \mathbf{\nabla}\delta\rho_{j}\right)  \right]  -\frac{\rho c}%
{T}\left(  \delta T\right)  ^{2}-\rho\left\vert \delta\mathbf{v}\right\vert
^{2}\right\}  \mathrm{d}^{3}\mathbf{r}.
\]
Evidently, perturbations of the temperature and velocity are negative and can
only lower the total entropy -- hence, the type of the stationary point
(maximum vs saddle) is fully determined by the variation of the density.
Setting, thus, $\delta T=0$ and $\delta\mathbf{v}=\mathbf{0}$, one obtains%
\begin{equation}
\delta^{2}H=\frac{1}{T}\int_{\mathcal{D}}\sum_{ij}\left[  -\frac{\partial
G_{i}}{\partial\rho_{j}}\left(  \delta\rho_{i}\right)  \left(  \delta\rho
_{j}\right)  -K_{ij}\left(  \mathbf{\nabla}\delta\rho_{i}\right)  \cdot\left(
\mathbf{\nabla}\delta\rho_{j}\right)  \right]  \mathrm{d}^{3}\mathbf{r}.
\label{3.12}%
\end{equation}
Expression (\ref{3.12}) is the main tool used in this paper for studying the
stability properties of steady states described by equation (\ref{3.10}).

\section{Nondimensionalization and the governing parameters\label{Sec 4}}

Introduce a characteristic density scale $\varrho$, a pressure scale $P$, a
temperature scale $T_{0}$, and a typical value $R$ of the specific gas
constant $R_{i}$ introduced in subsection \ref{Sec 2.2}. These scales allow
one to nondimensionalise all thermodynamics variables and functions introduced
in subsection \ref{Sec 2.1},%
\[
\left(  \rho_{i}\right)  _{nd}=\frac{\rho_{i}}{\varrho},\qquad\rho_{nd}%
=\frac{\rho}{\varrho},\qquad T_{nd}=\frac{T}{T_{0}},\qquad e_{nd}%
=\frac{\varrho}{P}e,\qquad s_{nd}=\frac{s}{R},
\]%
\[
p_{nd}=\frac{p}{P},\qquad\left(  G_{i}\right)  _{nd}=\frac{\rho}{P}%
G_{i},\qquad c_{nd}=\frac{c}{R},\qquad a_{nd}=\frac{\varrho^{2}}{P}a,\qquad
B_{nd}=\frac{B}{P}.
\]
It is also convenient to nondimensionalise the coefficients which appear in
the governing equations. Using their respective scales, one obtains%
\[
\left(  K_{ij}\right)  _{nd}=\frac{K_{ij}}{K},\qquad\left(  \mu_{s}\right)
_{nd}=\frac{\mu_{s}}{\mu},\qquad\left(  \mu_{b}\right)  _{nd}=\frac{\mu_{b}%
}{\mu},
\]%
\[
\left(  R_{i}\right)  _{nd}=\frac{R_{i}}{R},\qquad\kappa_{nd}=\frac{\kappa
}{\varkappa},\qquad\left(  D_{ij}\right)  _{nd}=\frac{D_{ij}}{D},\qquad\left(
\zeta_{i}\right)  _{nd}=\frac{\zeta_{i}}{\zeta}.
\]
Finally, introduce%
\[
\mathbf{r}_{nd}=\frac{\mathbf{r}}{l},\qquad t_{nd}=\frac{Vt}{l},\qquad
\mathbf{v}_{nd}=\frac{\mathbf{v}}{V},
\]%
\[
\boldsymbol{\Pi}_{nd}=\frac{l}{\mu V}\boldsymbol{\Pi},\qquad\left(
\mathbf{F}_{i}\right)  _{nd}=\frac{l^{3}}{K\varrho}\mathbf{F}_{i}%
,\qquad\left(  \mathbf{J}_{i}\right)  _{nd}=\frac{\mathbf{J}_{i}}{\varrho
V},\qquad\mathbf{Q}_{nd}=\frac{\mathbf{Q}}{PV},
\]
where%
\[
l=\sqrt{\frac{\varrho^{2}K}{P}},
\]
is the characteristic interfacial thickness, and%
\[
V=\frac{Pl}{\mu}%
\]
is the velocity scale reflecting the three-way balance of the pressure
gradient, viscous stress, and van der Waals force \citep{Benilov20b}. Note
also that the characteristic interfacial thickness $l$ is on a nano scale \citep{MagalettiGalloMarinoCasciola16,GalloMagalettiCoccoCasciola20,Benilov20a}.

Rewriting equations (\ref{2.17}), (\ref{2.27})--(\ref{2.28}), (\ref{2.21}%
)--(\ref{2.22}), and (\ref{2.31}), and omitting the subscript $_{nd}$, one
obtains%
\begin{equation}
\frac{\partial\rho_{i}}{\partial t}+\mathbf{\nabla}\cdot\left(  \rho
_{i}\mathbf{v}+\mathbf{J}_{i}\right)  =0, \label{4.1}%
\end{equation}%
\begin{equation}
\fbox{$\alpha$}\rho\left[  \frac{\partial\mathbf{v}}{\partial t}+\left(
\mathbf{v}\cdot\mathbf{\nabla}\right)  \mathbf{v}\right]  =\mathbf{\nabla
}\cdot\boldsymbol{\Pi}+\sum_{i}\rho_{i}\left(  \mathbf{F}_{i}-\mathbf{\nabla
}G_{i}\right)  -\rho s\mathbf{\nabla}T, \label{4.2}%
\end{equation}%
\begin{equation}
\fbox{$\tau$}\rho c\left(  \frac{\partial T}{\partial t}+\mathbf{v}%
\cdot\mathbf{\nabla}T\right)  +\mathbf{\nabla}\cdot\mathbf{Q}=\boldsymbol{\Pi
}:\left(  \mathbf{\nabla v}\right)  -B\mathbf{\nabla}\cdot\mathbf{v}+\sum
_{i}\left(  \mathbf{F}_{i}+\rho a_{i}\mathbf{\nabla}\right)  \cdot
\mathbf{J}_{i}, \label{4.3}%
\end{equation}%
\begin{equation}
\boldsymbol{\Pi}=\mu_{s}\left[  \mathbf{\nabla v}+\left(  \mathbf{\nabla
v}\right)  ^{tr}-\frac{2}{3}\left(  \mathbf{\nabla}\cdot\mathbf{v}\right)
\mathbf{I}\right]  +\mu_{b}\left(  \mathbf{\nabla}\cdot\mathbf{v}\right)
\mathbf{I}, \label{4.4}%
\end{equation}%
\begin{equation}
\fbox{$\delta$}\mathbf{J}_{i}=\sum_{j}D_{ij}\left[  \mathbf{F}_{j}%
-T\mathbf{\nabla}\left(  \frac{G_{j}}{T}\right)  \right]  -\fbox{$\dfrac
{\nu\delta}{\beta}$}\frac{\zeta_{i}}{T}\mathbf{\nabla}T, \label{4.5}%
\end{equation}%
\begin{equation}
\fbox{$\beta$}\mathbf{Q}=\fbox{$\nu$}\sum_{j}\zeta_{j}\left[  \mathbf{F}%
_{j}-T\mathbf{\nabla}\left(  \frac{G_{j}}{T}\right)  \right]  -\kappa
\mathbf{\nabla}T, \label{4.6}%
\end{equation}%
\begin{equation}
\mathbf{F}_{i}=\mathbf{\nabla}\sum_{j}K_{ij}\nabla^{2}\rho_{j}, \label{4.7}%
\end{equation}
where the nondimensional parameters%
\begin{equation}
\alpha=\frac{K\varrho^{3}}{\mu^{2}},\qquad\tau=\frac{\varrho RT_{0}}{P}%
,\qquad\beta=\frac{KP\varrho^{2}}{\mu\varkappa T_{0}},\qquad\nu=\frac{\zeta
P}{\varkappa T_{0}\varrho},\qquad\delta=\frac{K\varrho^{4}}{D\mu P},
\label{4.8}%
\end{equation}
are placed for better `visibility' in boxes.

Since $\alpha$ appears in front of the material derivative in equation
(\ref{4.1}), it should be interpreted as the microscopic Reynolds number
(associated with the flow near the interface, not the global flow). $\tau$ is
the nondimensional temperature. $\beta$ is the isothermality parameter
\citep{Benilov20b}: if it is small, the temperature field is close to being
spatially uniform (isothermal). The Nusselt number $\nu$ characterises the
importance of heat diffusion relative to heat conduction -- see expression
(\ref{4.6}) for the heat flux. Finally, the position of $\delta$ in equation
(\ref{4.5}) suggests that this parameter characterises advection relative to
diffusion (recall that the flux $\mathbf{J}_{i}$ was nondimensionalised using
the advection scale $\varrho V$).

One can also introduce the Prandtl and Schmidt numbers,%
\[
Pr=\frac{\beta\tau}{\alpha}=\frac{\mu R}{\varkappa},\qquad Sc=\frac{\delta
}{\alpha}=\frac{\mu\varrho}{DP},
\]
characterizing viscosity relative to heat conduction and diffusion, respectively.

The nondimensional versions of the boundary conditions look exactly the same
as their dimensional counterparts, (\ref{2.32})--(\ref{2.35}) -- as do the
nondimensional versions of the thermodynamic identities of subsection
\ref{Sec 2.1} except definition (\ref{2.8}) of the heat capacity, which
becomes%
\[
c=\frac{1}{\tau}\frac{\partial e}{\partial T}.
\]
This paper does not aim to present a comprehensive classification of
asymptotic regimes of the multicomponent DIM
\citep[for the pure-fluid DIM, see][]{Benilov20b}. Only the simplest regime
will be described and used later as a qualitative illustration of
theoretically-predicted behaviors. It corresponds to the following
assumptions:%
\[
\alpha\ll1,\qquad\beta\ll1,\qquad\nu\ll1.
\]
The smallness of $\alpha$ allows one to take advantage of the slow-flow
approximation -- whereas the other two assumptions and equations (\ref{4.3})
and (\ref{4.6}) imply that%
\[
T=1+\mathcal{O}(\beta),
\]
i.e., the fluid is almost isothermal. Setting, thus, $T=\operatorname{const}$
in expression (\ref{4.5}) for the diffusion flux and substituting it into the
density equation (\ref{4.1}), one obtains%
\begin{equation}
\frac{\partial\rho_{i}}{\partial t}+\mathbf{\nabla}\cdot\left[  \rho
_{i}\mathbf{v}+\sum_{j}D_{ij}\mathbf{\nabla}\left(  \sum_{n}K_{jn}\nabla
^{2}\rho_{n}-G_{j}\right)  \right]  =0, \label{4.9}%
\end{equation}
where it was assumed, without loss of generality, that $\delta=1$. Simplifying
similarly equation (\ref{4.2}) and substituting into it expression (\ref{4.7})
for $\mathbf{F}_{i}$, one obtains%
\begin{equation}
\mathbf{\nabla}\cdot\boldsymbol{\Pi}+\sum_{i}\rho_{i}\mathbf{\nabla}\left(
\sum_{j}K_{ij}\nabla^{2}\rho_{j}-G_{i}\right)  =0. \label{4.10}%
\end{equation}
Observe that equations (\ref{4.9})--(\ref{4.10}) do not involve the (small)
temperature variations -- hence, the temperature equation (\ref{4.3}) can be
simply omitted.

Equations (\ref{4.9})--(\ref{4.10}) and expression (\ref{4.4}) for the viscous
stress form a closed set of approximate equations for the unknowns $\rho_{i}$
and $\mathbf{v}$. The chemical potential $G_{i}$ in these equations should be
treated as a known function of $\rho_{1}...\rho_{N}$, and the temperature $T$,
as a known parameter.

Unlike the exact set -- which describes fast acoustic waves and slow
interfacial flow -- the approximate equations describe only the latter. This
is a clear advantage: in a numerical simulation, for example, waves
necessitate a small timestep and, thus, dramatically slow down the
computation. At the same time, the two sets of equations have very similar
properties: they share the same steady solutions, both conserve mass and
energy, and satisfy the H-theorem.

\section{Basic solutions and their stability\label{Sec 5}}

\subsection{Spatially uniform states\label{Sec 5.1}}

Consider a uniform fluid where there is no flow and all species are in vapour
phase. If the temperature drops, one of the species may become overcooled,
giving rise to rapid condensation. A similar instability may occur when all or
some of the species are in \emph{liquid} phase, and the temperature
\emph{increases} -- giving rise to rapid \emph{evaporation}.

To determine which exactly states are unstable (thus, do not occur in real
world), one could perform the usual linear analysis. For a pure fluid ($N=1$),
this is a straightforward task yielding the following stability criterion:%
\begin{equation}
\frac{\partial p}{\partial\rho}>0. \label{5.1}%
\end{equation}
Thus, instability occurs if an increase in density \emph{lowers} the pressure,
so that the flow generated by the pressure gradient brings even more fluid to
this region.

For an arbitrary $N$, however, the analysis of linearised equations is
extremely cumbersome. Instead, it will be examined via the entropy
principle\ -- i.e., using expression (\ref{3.12}). Taking the limit
$\mathcal{D}\rightarrow%
%TCIMACRO{\U{211d} }%
%BeginExpansion
\mathbb{R}
%EndExpansion
^{3}$ (unbounded fluid), assuming that $\rho_{i}$ is spatially uniform, and
recalling that $K_{ij}\succ0$ and $c>0$, one can deduce from (\ref{3.12}) that
$H$ has a maximum if%
\begin{equation}
\frac{\partial G_{i}}{\partial\rho_{j}}\succ0. \label{5.2}%
\end{equation}
This is the standard stability criterion for a spatially uniform state of a
multicomponent fluid \citep{GlansdorffPrigogine71}.

Four comments are in order:

\begin{itemize}
\item For a physically meaningful $G_{i}(\rho_{1}...\rho_{N},T)$, condition
(\ref{5.2}) holds for a sufficiently rarefied vapour or a sufficiently dense liquid.

\item The states with marginal stability are sometimes referred to as
\textquotedblleft spinodal points\textquotedblright\ and stable vapour, as
\textquotedblleft subspinodal vapour\textquotedblright.

\item Interestingly, the viscosity and transport coefficients do not appear in
criterion (\ref{5.2}). The corresponding effects can only slow the instability
down, but not eliminate it.

\item To reconcile criterion (\ref{5.2}) with its pure-fluid counterpart
(\ref{5.1}), note that for $N=1$, $\partial G/\partial\rho$ and $\partial
p/\partial\rho$ have the same sign [as implied by identity (\ref{2.6})].
\end{itemize}

\subsection{Two-phase saturated states\label{Sec 5.2}}

Consider an interface separating liquid and vapour of the same pure fluid. If
they are in equilibrium, their temperatures are equal, and the rest of the
parameters satisfy the so-called Maxwell construction \citep{Maxwell75},%
\begin{equation}
G(\rho^{(l)},T)=G(\rho^{(v)},T),\qquad p(\rho^{(l)},T)=p(\rho^{(v)},T),
\label{5.3}%
\end{equation}
where the superscripts $^{(l)}$ and $^{(v)}$ mark the parameters of the liquid
and vapour, respectively. The former and latter equalities in (\ref{5.3})
guarantee thermodynamic and mechanical equilibria of the interface, respectively.

Assume also that both phases are stable,%
\[
\left(  \frac{\partial G}{\partial\rho}\right)  _{\rho=\rho^{(l)}}%
>0,\qquad\left(  \frac{\partial G}{\partial\rho}\right)  _{\rho=\rho^{(v)}%
}>0,
\]
and the density of the liquid exceeds that of the vapour,%
\[
\rho^{(l)}>\rho^{(v)}.
\]
Subject to these conditions, the Maxwell construction (\ref{5.3}) uniquely
determines the \emph{saturated} densities $\rho^{(l)}$ and $\rho^{(v)}$ as
functions of $T$.

In what follows, the multicomponent version of the Maxwell construction will
be shown to follow from the DIM's entropy principle.

Consider an insulated container with fluid subdivided between two states,
liquid and vapour. If the liquid/vapour and fluid/wall interfaces are
sufficiently thin, the corresponding full entropy, mass, and energy can be
approximately written in the form%
\[
S=V^{(l)}\rho^{(l)}\,s(\rho_{1}^{(l)}...\rho_{N}^{(l)},T^{(l)})+V^{(v)}%
\rho^{(v)}\,s(\rho_{1}^{(v)}...\rho_{N}^{(v)},T^{(v)}),
\]%
\begin{equation}
M_{i}=V^{(l)}\rho_{i}^{(l)}+V^{(v)}\rho_{i}^{(v)}, \label{5.4}%
\end{equation}%
\begin{equation}
E=V^{(l)}\rho^{(l)}\,e(\rho_{1}^{(l)}...\rho_{N}^{(l)},T^{(l)})+V^{(v)}%
\rho^{(v)}\,e(\rho_{1}^{(v)}...\rho_{N}^{(v)},T^{(v)}), \label{5.5}%
\end{equation}
where $V^{(l)}$ and $V^{(v)}$ are the volumes of the liquid and vapour phases,
respectively. Introduce also the full volume of the container,%
\begin{equation}
V=V^{(l)}+V^{(v)}. \label{5.6}%
\end{equation}
The Maxwell construction can be derived by maximizing $S$ subject to the
constraints of fixed $M_{i}$, $E$, and $V$ (which are now \emph{functions}, as
opposed to being \emph{functionals} in the previous subsection).
Straightforward calculations show that the maximum of entropy is achieved if
$T^{(l)}=T^{(v)}$ (isothermality) and%
\begin{equation}
G_{i}(\rho_{1}^{(l)}...\rho_{N}^{(l)},T)=G_{i}(\rho_{1}^{(v)}...\rho_{N}%
^{(v)},T), \label{5.7}%
\end{equation}%
\begin{equation}
p(\rho_{1}^{(l)}...\rho_{N}^{(l)},T)=p(\rho_{1}^{(v)}...\rho_{N}^{(v)},T).
\label{5.8}%
\end{equation}
Four comments are in order:

\begin{itemize}
\item Since the solution describing coexistence of two phases satisfies the
entropy principle, it is automatically stable.

\item The Maxwell construction can sometimes yield a solution with negative
$V^{(l)}$or $V^{(v)}$. In such cases, the two-phase equilibrium is irrelevant,
and the fluid evolves towards the one-phase state with the same masses of the
species and total energy.

\item The multicomponent Maxwell construction (\ref{5.7})--(\ref{5.8})
comprises $N+1$ equations for $2N$ unknowns -- hence, does not uniquely fix
the liquid and vapour densities.\\*\hspace*{0.6cm}To close set (\ref{5.7}%
)--(\ref{5.8}), one should assume the masses $M_{i}$ and the total energy $E$
to be known and view equalities (\ref{5.4})--(\ref{5.6}) as additional
equations. They bring the total number of equations to $2N+3$ (making the
`expanded' Maxwell construction appear overdetermined) -- but in this
formulation, $V^{(l)}$, $V^{(v)}$, and $T$ should also be viewed as
unknowns.\\*\hspace*{0.6cm}Physically, if a certain amount of fluid, with a
certain amount of energy, is placed in a box, the entropy principle uniquely
determines the final temperature and the proportion in which the box is
subdivided between the liquid and vapour phases.

\item The closure of the Maxwell construction described in the previous bullet
is, obviously, inapplicable to containers of infinite volume. To understand
how conditions (\ref{5.7})--(\ref{5.8}) should be closed in this case,
consider the interface between the Earth's atmosphere and ocean. For this
setting, one should prescribe the (atmospheric) pressure and composition of
dry air above the ocean's surface. With these parameters given, (\ref{5.7}%
)--(\ref{5.8}) yield the saturated moisture content of the air and the
saturated amounts of nitrogen, oxygen, \emph{etc.} dissolved in the oceanic water.
\end{itemize}

\begin{figure}\begin{centering}
\includegraphics[width=129mm]{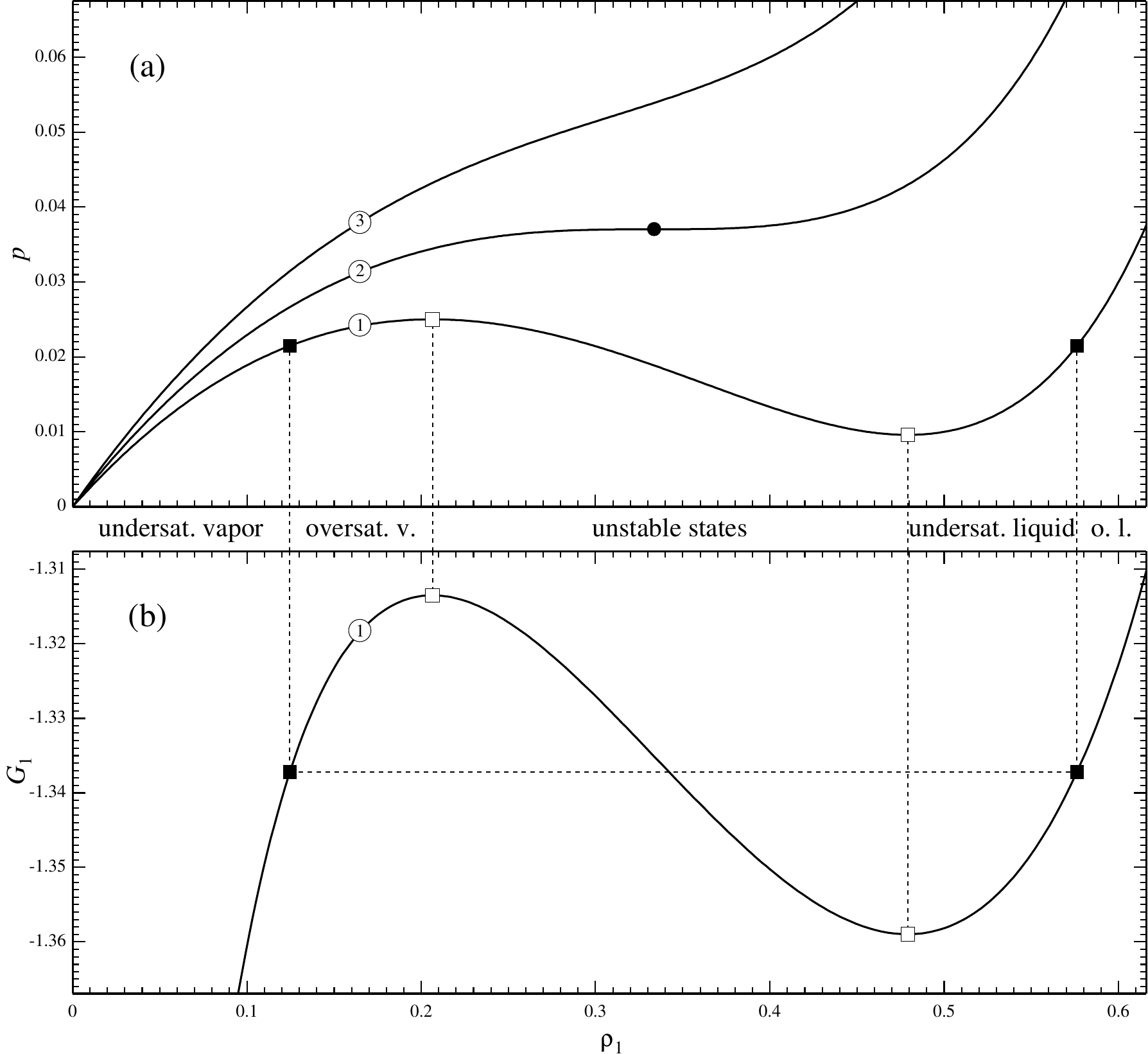}
\caption{The pressure [panel (a)] and chemical potential [panel (b)] of a pure van der Waals fluid as functions of the density, for three values of the temperature: (1) $T=0.26$ (subcritical); (2) $T=8/27$ (critical), (3) $T=0.33$ (supercritical). The critical point on curve (2) is marked with a filled circle; the saturation and spinodal points on curve (1) are shown by filled and empty squares, respectively. The labels \textquotedblleft undersat[urated] vapour\textquotedblright, \textquotedblleft oversat[urated]
v[apour]\textquotedblright, \emph{etc.} apply only to curve (1).}
\label{fig1}
\end{centering}\end{figure}

To illustrate the use of the Maxwell construction, consider a van der Waals
fluid, whose pressure and chemical potential are described by expressions
(\ref{2.15})--(\ref{2.16}). Assume for simplicity that the fluid is pure
($N=1$) and monatomic ($c_{i}=3R_{i}/2$) -- and let the scales $\varrho$, $P$,
and $R$ used for nondimensionalization be such that%
\begin{equation}
a_{11}=1,\qquad b_{1}=1,\qquad R_{1}=1. \label{5.9}%
\end{equation}
If $T<8/27$ (which is a \emph{subcritical} temperature of the van der Waals
fluid), two states exist representing the liquid and vapour phases, with some
(but not all) of the states in between being unstable (see figure \ref{fig1}).
If the temperature is \emph{supercritical}, only one phase exists and
interfaces do not.

\begin{figure}\begin{centering}
\includegraphics[width=91mm]{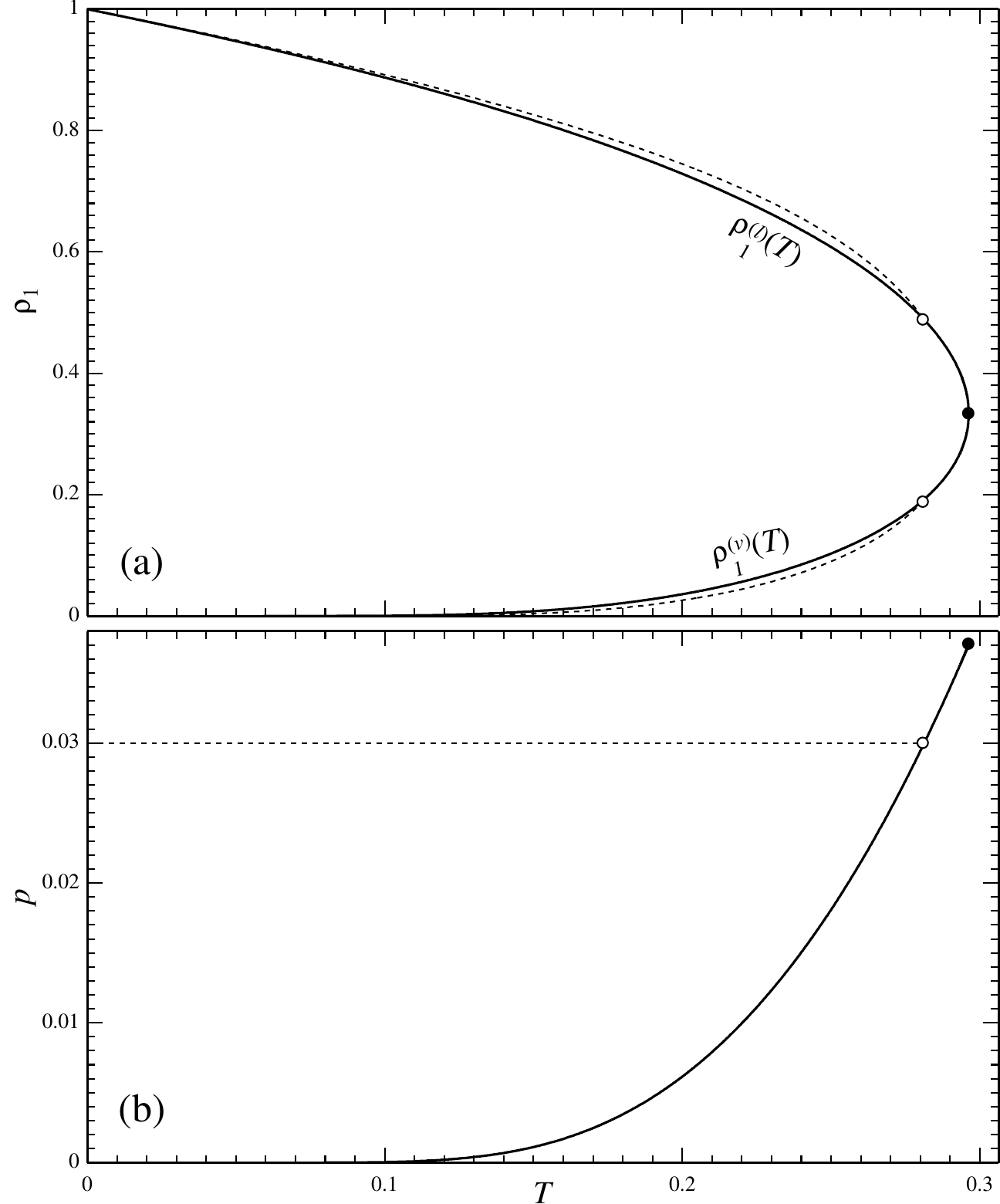}
\caption{The saturated densities [panel (a)] and pressure [panel (b)] vs the temperature, for a pure (solid line) and two-component (dotted line) van der Waals fluids, with the parameters (\ref{5.9})--(\ref{5.10}). The empty circles mark the boiling point, the filled circle marks the critical point. The saturated densities of the second species (of the two-component fluid) are not shown.}
\label{fig2}
\end{centering}\end{figure}

The simplest model describing a water/air interface is that with $N=2$. The
Maxwell construction in this case should be complemented with one extra
condition, setting the pressure above the interface equal to its atmospheric
value,%
\[
p(\rho_{1}^{(v)},\rho_{2}^{(v)},T)=p_{A}.
\]
The difference between a pure and a multicomponent fluids is illustrated in
figure \ref{fig2} for parameters (\ref{5.9}) and%
\begin{equation}
a_{22}=0.2,\qquad a_{12}=0,\qquad b_{2}=1,\qquad R_{2}=0.6,\qquad p_{A}=0.03.
\label{5.10}%
\end{equation}
These values reflect a compromise between simplicity, illustrative purposes
(the curves with $N=1$ and $N=2$ should be visibly different), and an attempt
to loosely match the parameters of water and air (loosely, because the van der
Waals model includes few adjustable constants). In particular, parameters
(\ref{5.10}) make the critical temperature of the second species noticeably
smaller than that of the first species. As a result, a temperature range
exists where the former is definitely vapour, whereas the latter can be either
vapour or liquid (as is indeed the case with water under normal conditions).

The most important difference between a multicomponent fluid at fixed pressure
and a pure fluid is that the former exhibits a boiling point. It occurs when
the saturated pressure of the liquid-phase species matches the applied
(atmospheric) pressure: as a result, the second species is \emph{completely}
replaced by the vapour of the first species -- and so the fluid becomes pure.
If it is heated beyond the boiling point, the pressure can no longer be kept
fixed, but ought to increase (to match the saturated pressure of the first
species, which grows with $T$).

One should keep in mind that the small-$T$ part of figure \ref{fig2} is
physically meaningless due to freezing (which is not described by the DIM).

\section{One-dimensional steady states\label{Sec 6}}

\subsection{Flat liquid/vapour interfaces\label{Sec 6.1}}

Consider an unbounded fluid involving liquid and vapour phases in equilibrium,
separated by a static flat interface. Its spatial structure is described by
equation (\ref{3.10}) derived from the entropy principle. To adapt it
specifically for a flat interface, let $\rho_{i}$ depend on a single
coordinate -- say, $z$ -- so that the interface is horizontal. Let the fluid
below the interface be liquid and above, vapour,%
\begin{equation}
\rho_{i}\rightarrow\rho_{i}^{(l)}\qquad\,\text{as}\qquad z\rightarrow-\infty,
\label{6.1}%
\end{equation}%
\begin{equation}
\rho_{i}\rightarrow\rho_{i}^{(v)}\qquad\text{as}\qquad z\rightarrow+\infty.
\label{6.2}%
\end{equation}
Taking in equation (\ref{3.10}) the limit $z\rightarrow+\infty$, one can
determine the constant $\eta_{i}$ and rewrite (\ref{3.10}) in the form%
\begin{equation}
\sum_{j}K_{ij}\frac{\mathrm{d}^{2}\rho_{j}}{\mathrm{d}z^{2}}=G_{i}-G_{i}%
(\rho_{1}^{(v)}...\rho_{N}^{(v)},T). \label{6.3}%
\end{equation}
Boundary-value problem (\ref{6.1})--(\ref{6.3}) is invariant with respect to
the change $z\rightarrow z+\operatorname{const}$ -- hence, its solution is not
unique. To make it such, an extra condition is needed -- say,%
\begin{equation}
\rho_{1}=\frac{1}{2}\left(  \rho_{1}^{(l)}+\rho_{1}^{(v)}\right)
\qquad\text{at}\qquad z=0. \label{6.4}%
\end{equation}
As shown before, the parameters $\rho_{i}^{(l)}$ and $\rho_{i}^{(v)}$ are not
arbitrary but have to satisfy the Maxwell construction, i.e., conditions
(\ref{5.7})--(\ref{5.8}). Thus, they must be intrinsic to boundary-value
problem (\ref{6.1})--(\ref{6.4}) and can be derived directly from it, as a
condition for existence of solution.

Indeed, consider equation (\ref{6.3}) in the limit $z\rightarrow-\infty$ and
recall (\ref{6.1}) -- which immediately yields the first half of the Maxwell
construction, equation (\ref{5.7}). To derive the second half, consider%
\[
\sum_{i}\int_{-\infty}^{\infty}\left(  \ref{6.3}\right)  \times\frac
{\mathrm{d}\rho_{i}}{\mathrm{d}z}\mathrm{d}z.
\]
The integral in this equation can be evaluated via identity (\ref{2.7}), and
the constant of integration can be fixed via boundary condition (\ref{6.2}).
After straightforward algebra, one obtains%
\begin{equation}
\frac{1}{2}\sum_{ij}K_{ij}\frac{\mathrm{d}\rho_{i}}{\mathrm{d}z}%
\frac{\mathrm{d}\rho_{j}}{\mathrm{d}z}=\sum_{i}\rho_{i}\left[  G_{i}%
-G_{i}(\rho_{1}^{(v)}...\rho_{N}^{(v)},T)\right]  -p+p(\rho_{1}^{(v)}%
...\rho_{N}^{(v)},T). \label{6.5}%
\end{equation}
Taking in this equation the limit $z\rightarrow-\infty$ and recalling boundary
condition (\ref{6.1}), one recovers the second part of the Maxwell
construction, equation (\ref{5.8}), as required.

\begin{figure}\begin{centering}
\includegraphics[width=129mm]{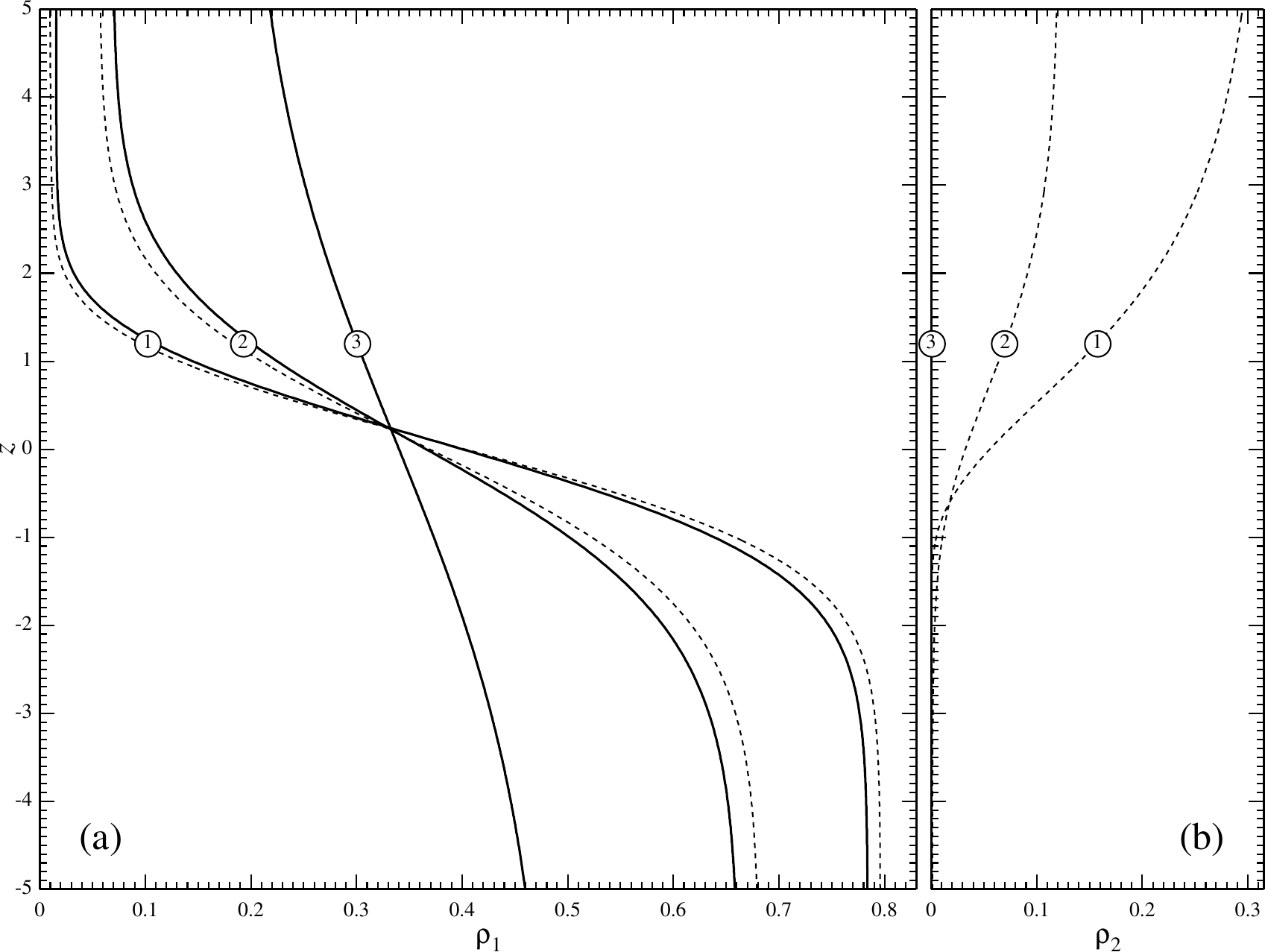}
\caption{Comparison between liquid/vapour interfaces in pure (solid line) and two-component (dotted line) van der Waals fluids with parameters (\ref{5.9})--(\ref{5.10}) and (\ref{6.6}). Curves (1)--(3) correspond to $T=0.17,~0.23,~0.28132$. The last of these values is the (approximate) boiling point, so that the solid and doted curves virtually coincide and $\rho_{2}$ is virtually zero.}
\label{fig3}
\end{centering}\end{figure}

Boundary-value problem (\ref{6.1})--(\ref{6.4}) was solved numerically for a
two-component van der Waals fluid, using the MATLAB function BVP4c based on
the three-stage Lobatto IIIa formula \citep{KierzenkaShampine01}. Examples of
numerical solutions with parameters (\ref{5.9})--(\ref{5.10}) and%
\begin{equation}
K_{11}=K_{22}=1,\qquad K_{12}=0, \label{6.6}%
\end{equation}
are shown in figure \ref{fig3}. Observe that an increase in temperature makes
the interface thicker, and it becomes infinitely thick when the critical
temperature is reached.

As shown in Appendix \ref{Appendix C.1}, all solutions describing
liquid/vapour interfaces, are stable as long as $\mathrm{d}\rho_{i}%
/\mathrm{d}z$ never vanishes ($\rho_{i}(z)$ is strictly monotonic) for all $i$.

\subsection{One-dimensional drops and bubbles}

Apart from liquid/vapour interfaces, equation (\ref{3.10}) admits spatially
localised solutions, describing drops or bubbles floating in vapour or liquid,
respectively. In this subsection, the simplest -- one-dimensional -- solutions
of this kind are discussed, describing a flat layer of increased or decreased
density. They both correspond to the following boundary condition%
\begin{equation}
\rho_{i}\rightarrow\rho_{\infty,i}\qquad\text{as}\qquad z\rightarrow\pm\infty,
\label{6.7}%
\end{equation}
where $\rho_{\infty,i}$ is the density of the vapour (liquid). Unlike the
interface solutions examined above, the fluid outside the drop (bubble) does
not have to be saturated vapour (liquid); it can actually be any stable or
metastable state.

The 1D reduction of the steady-state equation (\ref{3.10}) and expression
(\ref{3.11}) for $\eta_{i}$ can be written in the form%
\begin{equation}
\sum_{j}K_{ij}\frac{\mathrm{d}^{2}\rho_{j}}{\mathrm{d}z^{2}}=G_{i}%
-G_{\infty,i}, \label{6.8}%
\end{equation}
where $G_{\infty,i}=G_{i}(\rho_{\infty,1}...\rho_{\infty,N},T)$ is the
chemical potential at infinity.

Figure \ref{fig4} illustrates typical solutions of boundary-value problem
(\ref{6.7})--(\ref{6.8}) computed for a pure fluid. The following features can
be observed in panel (a) depicting some 1D drop solutions:

\begin{figure}\begin{centering}
\includegraphics[width=129mm]{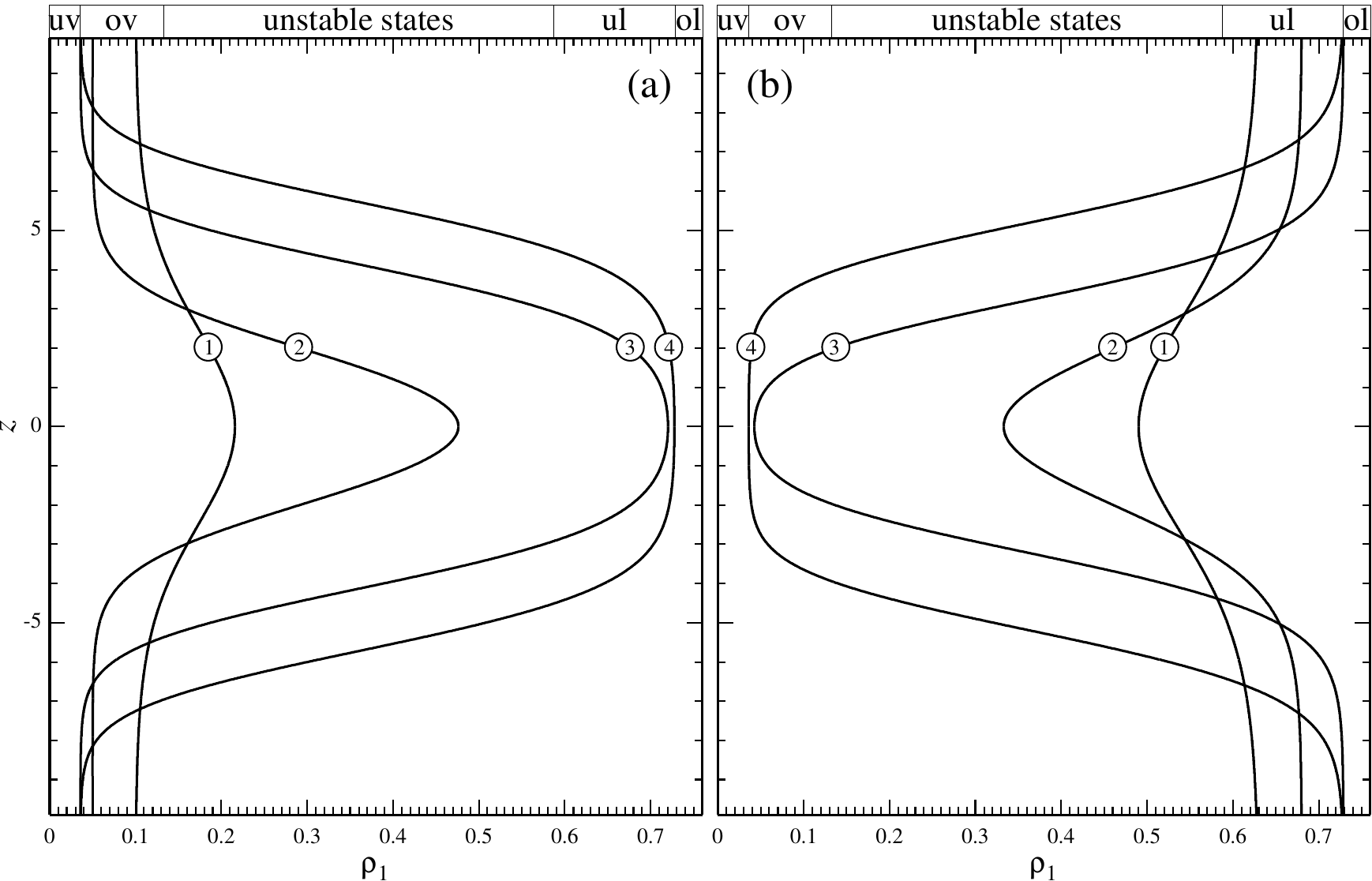}
\caption{Solutions of boundary-value problem (\ref{6.7})--(\ref{6.8}) for a pure van der Waals fluid with $T=0.2$. (a) Drop solutions with $\rho_{\infty,1}=0.1,~0.05,~0.03587,~0.0358485$ (curves (1)--(4), respectively). (b) Bubble solutions with $\rho_{\infty,1} =0.63,~0.68,~0.7286,~0.728671$ (curves (1)--(4), respectively). The labels \textquotedblleft u[ndersaturated] v[apour]\textquotedblright, \textquotedblleft o[versaturated] v[apour]\textquotedblright, \emph{etc.}
characterise $\rho_{\infty,1}$.}
\label{fig4}
\end{centering}\end{figure}

\begin{itemize}
\item Drop solutions exist only if $\rho_{\infty,1}$ corresponds to
oversaturated vapour.

\item As $\rho_{\infty,1}\rightarrow\rho_{1}^{(v)}+0$, the drop becomes
increasingly thick.

\item Once $\rho_{\infty,1}$ passes $\rho_{1}^{(v)}$, drop solutions cease to
exist. Physically, this is because liquid drops surrounded by undersaturated
vapour evaporate.
\end{itemize}

\noindent Similar tendencies have been observed for 1D bubble solutions,
illustrated in figure \ref{fig4}(b).

Most importantly, all drops (bubbles) in oversaturated vapour (undersaturated
liquid) are likely to be unstable. For pure fluids ($N=1$), the instability
can be proven rigorously: as shown in Appendix \ref{Appendix C.2}, the entropy
has a saddle (not maximum) on these solutions.

To understand how slightly perturbed drops and bubbles evolve, they were
simulated numerically using the simplest asymptotic version of the DIM -- the
one based on equations (\ref{4.9})--(\ref{4.10}) and (\ref{4.4}). It is
adapted for $N=1$ and a single spatial coordinate in Appendix \ref{Appendix D}%
, which also outlines the numerical method used.

\begin{figure}\begin{centering}
\includegraphics[width=129mm]{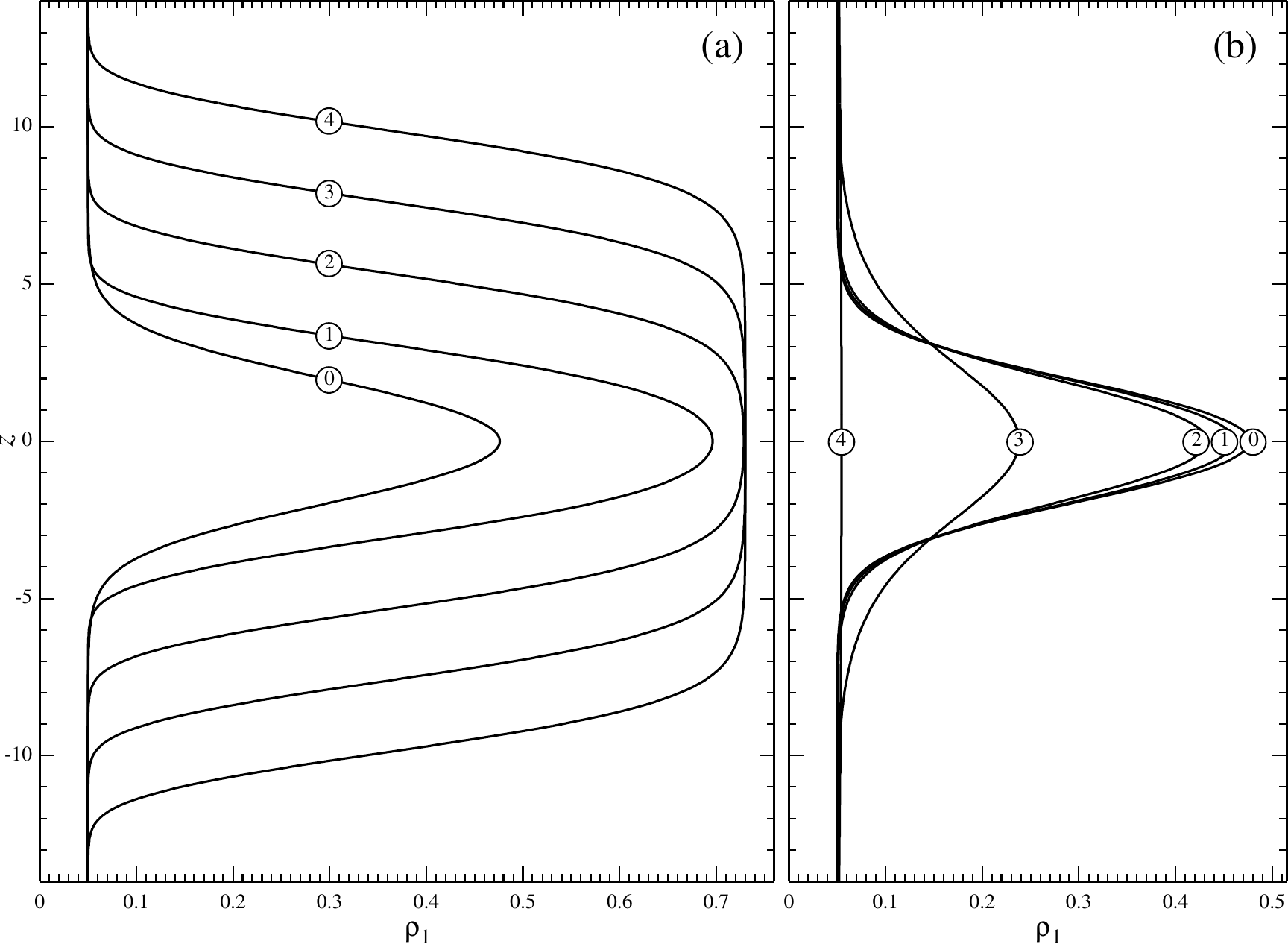}
\caption{Evolution of a perturbed drop, described by equations (\ref{D.3})--(\ref{D.4}) and boundary conditions (\ref{D.5})--(\ref{D.6}), for a pure van der Waals fluid and parameters (\ref{6.9}). (a) Initial condition (\ref{6.10}) (the steady solution plus extra fluid); the time $t$ of a snapshot and the corresponding curve number $n$ are inter-related via $t=5000\,n$. (b) Initial condition (\ref{6.11}) (the steady solution minus extra fluid); $t=500\,n$.}
\label{fig5}
\end{centering}\end{figure}

Various initial conditions have been simulated for the van der Waals pure
fluid, with only two patterns of evolution observed. If the initial condition
includes a steady-drop solution $\rho_{1}^{(sd)}(z)$ \emph{plus some extra
fluid}, spontaneous condensation is typically triggered off, giving rise to
two shock waves propagating away from the drop's center. This behavior is
illustrated in figure \ref{fig5}(a) for%
\begin{equation}
T=0.2,\qquad\rho_{\infty,1}=0.05, \label{6.9}%
\end{equation}
and the initial condition%
\begin{equation}
\rho_{1}=\rho_{1}^{(sd)}(1.01\,z),\qquad w=0\qquad\text{at}\qquad t=0,
\label{6.10}%
\end{equation}
where $w$ is the vertical velocity (the other two velocity components are zero
due to the 1D nature of the flow).

If, however, the initial condition includes \emph{less} fluid than $\rho
_{1}^{(nm)}(z)$, the drop would typically evaporate. This pattern is
illustrated in figure \ref{fig5}(b) for parameters (\ref{6.9}) and the
following initial condition,%
\begin{equation}
\rho_{1}=\rho_{1}^{(sd)}(0.99\,z),\qquad w=0\qquad\text{at}\qquad t=0.
\label{6.11}%
\end{equation}
The two patterns can be understood physically on the basis of the fact that
the steady-drop solutions exist only for drops surrounded by oversaturated
metastable vapour -- which is stable with respect to infinitesimally
small\emph{ }perturbations, but may be unstable with respect to finite ones.
One can thus assume that the drop solution provides a lower bound for the mass
of perturbations capable of triggering off instability of the surrounding vapour.

It is also interesting to see how a drop would evolve if it is surrounded by
\emph{saturated} (not \emph{over}saturated) vapour.

The fact that the thickness of a steady drop becomes infinite as $\rho
_{\infty,1}\rightarrow\rho_{1}^{(v)}$ suggests that a \emph{finite}-size
liquid layer, surrounded by saturated vapour, evaporates. The mechanism of
this evaporation, however, is not immediately clear. 2D and 3D drops, for
example, evaporate because the curvature of their boundary gives rise to a
liquid-to-vapour mass flux \citep{Benilov20c,Benilov21b,Benilov22a} -- but the
boundaries of 1D drops are flat. One can only assume that they evaporate due
to a long-range interaction of the drop's upper and lower interfaces.

To explore this hypothesis, 1D drops floating in saturated vapour were
simulated numerically (using the same model and numerical method as before).
It has turned out that these drops do evaporate, and their evolution can be
subdivided into three distinct stages:

\begin{enumerate}
\item[(i)] The boundaries of the drop rapidly assume the profile of a steady
liquid/vapour interface -- of the kind examined in the previous subsection.

\item[(ii)] The drop begins to get thinner, but the density of the drop's core
remains close to $\rho_{1}^{(l)}$.

\item[(iii)] Once the thickness of the drop becomes comparable to the
interfacial thickness, the density at the drop's center begins to rapidly
decrease, and the drop disappears.
\end{enumerate}

The main characteristic of such behavior is the evaporation time $t_{e}$,
which can be defined as the interval over which the density at the drop's
center falls by a factor of, say, $10$ -- i.e.,%
\[
\frac{\rho_{1}(0,t_{e})}{\rho_{1}(0,0)}=0.1.
\]
$t_{e}$ was computed as a function of the drop's initial size $W$, for the
following initial condition:%
\[
\rho_{1}=\rho_{1}^{(v)}+\frac{\rho_{1}^{(l)}-\rho_{1}^{(v)}}{1+\left(
\dfrac{2z}{W}\right)  ^{6}},\qquad w=0\qquad\text{at}\qquad t=0.
\]
It has turned out that, for $W$ changing from $0.2$ to $10$, the evaporation
time $t_{e}$ grows from $48.5$ to $10^{6}$ -- i.e., exponentially. This agrees
with the hypothesis that the drop evaporation is caused by long-range
interaction of the drop's boundaries and the fact that the density in
liquid/vapour interfaces tends to $\rho_{1}^{(v)}$ and $\rho_{1}^{(l)}$ (as
$z\rightarrow\pm\infty$) \emph{exponentially} quickly.

\subsection{Solid/fluid interfaces\label{Sec 6.2}}

Let the fluid be bounded below by a flat horizontal wall (substrate) located
at $z=0$, so that boundary condition (\ref{2.35}) reduces to%
\begin{equation}
\rho_{i}=\rho_{0,i}\qquad\text{at}\qquad z=0. \label{6.12}%
\end{equation}
Far above the substrate, the fluid is homogeneous,%
\begin{equation}
\rho_{i}\rightarrow\rho_{\infty,i}\qquad\text{as}\qquad z\rightarrow+\infty,
\label{6.13}%
\end{equation}
where $\rho_{\infty,i}$ corresponds to a stable or metastable state. Finally,
equation (\ref{6.8}) used in the previous subsection for 1D drops and bubbles
applies to the present case as well.

\begin{figure}\begin{centering}
\includegraphics[width=129mm]{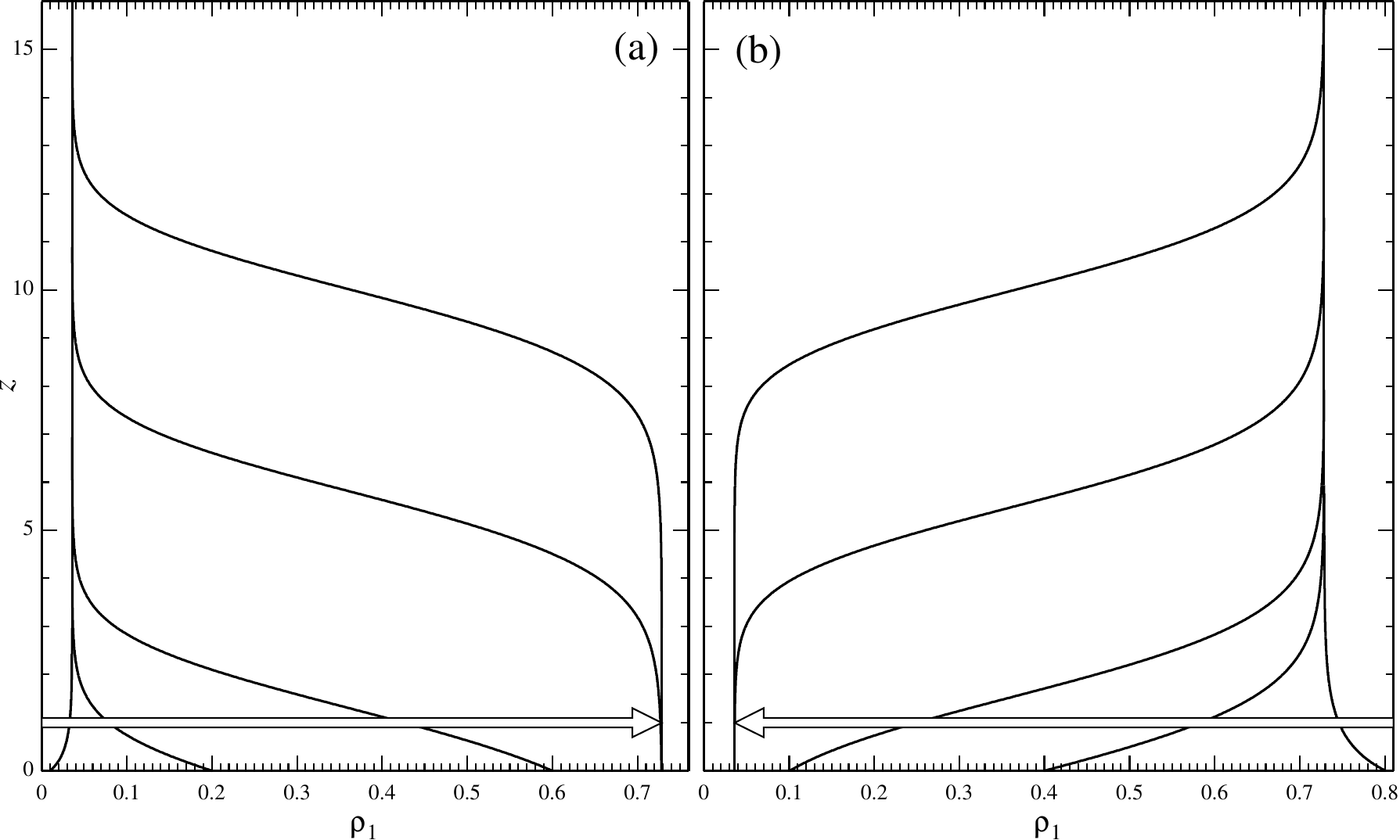}
\caption{Examples of solid/fluid interfaces for a pure van der Waals fluid with parameters (\ref{6.14}), for a given $\rho_{\infty,1}$: (a) $\rho_{\infty,1}=\rho_{1}^{(v)}$,
$\rho_{0,1}$ varies from $0$ to $\rho_{1}^{(l)}$; (b) $\rho_{\infty,1} =\rho_{1}^{(l)}$, $\rho_{0,1}$ varies from $1$ to $\rho_{1}^{(v)}$. The arrows show the direction towards the boundary of the pool of allowable values of $\rho_{0,1}$.}
\label{fig6}
\end{centering}\end{figure}

\begin{figure}\begin{centering}
\includegraphics[width=129mm]{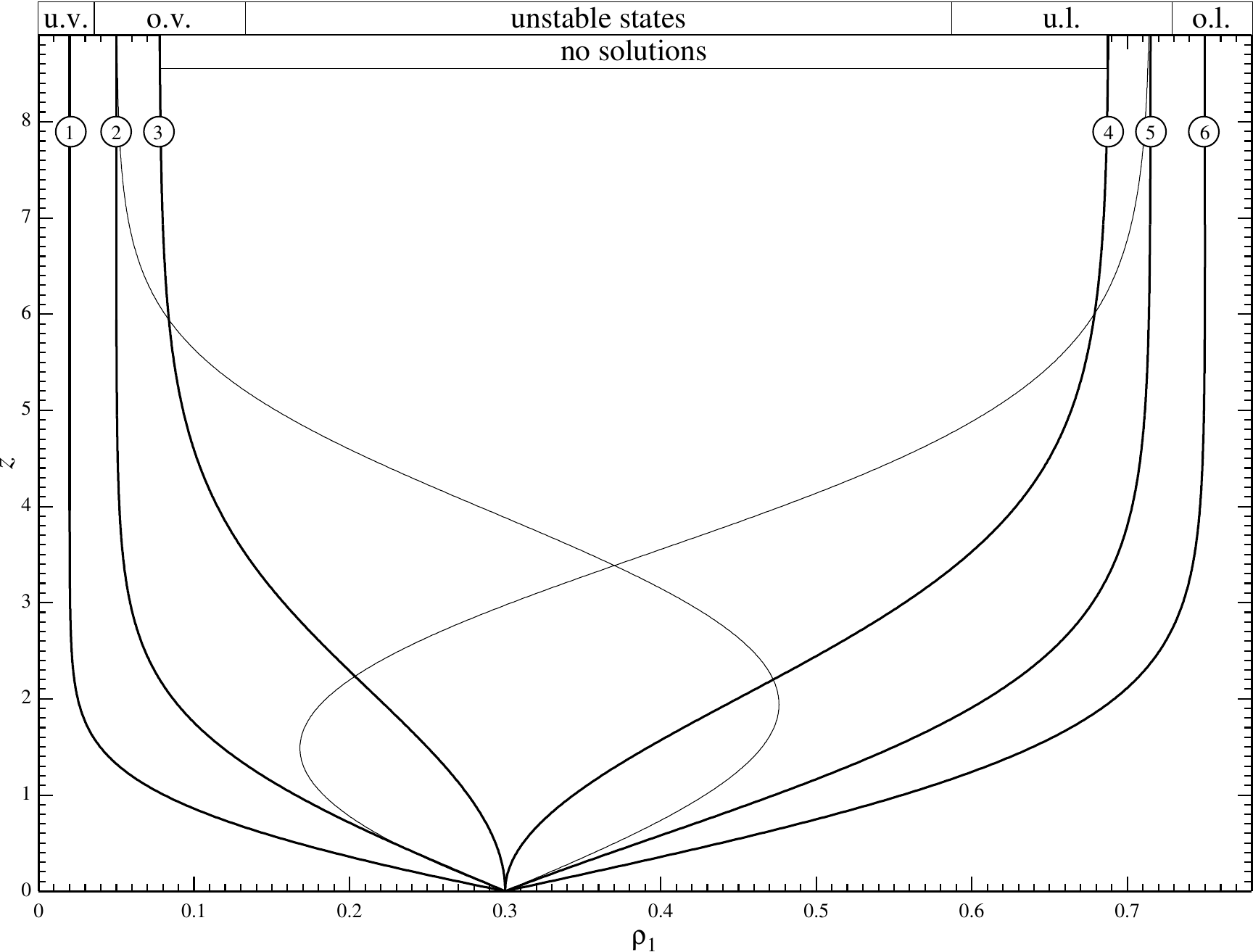}
\caption{Examples of solid/fluid interfaces for a pure van der Waals fluid with parameters (\ref{6.14}). The fluid at infinity is: (1) undersaturated vapour, (2) oversaturated vapour with two possible solutions (the unstable one is shown in thin line), (3) oversaturated vapour with a single solution, (4) undersaturated liquid with a single solution, \emph{etc.} Observe that there are no solutions between curves (3) and (4).}
\label{fig7}
\end{centering}\end{figure}

Boundary-value problem (\ref{6.8}), (\ref{6.12})--(\ref{6.13}) was solved
numerically for a van der Waals pure fluid with%
\begin{equation}
T=0.2,\qquad\rho_{0,1}=0.3. \label{6.14}%
\end{equation}
Its typical solutions are illustrated in figure \ref{fig6} ($\rho_{0,i}$
varies, $\rho_{\infty,1}$ is fixed) and in figure \ref{fig7} ($\rho_{0,i}$ is
fixed, $\rho_{\infty,1}$ varies). The following features can be
observed.$\smallskip$

(i) For a given $\rho_{\infty,1}$, there exists a certain pool of $\rho_{0,i}$
which can be `connected' to this $\rho_{\infty,1}$. If, for example,
$\rho_{\infty,1}$ equals the saturated vapour density (the case illustrated in
figure \ref{fig6}a), than $\rho_{0,1}$ must be smaller than the saturated
liquid density (and \emph{vice versa}: if $\rho_{\infty,1}=\rho_{1}^{(l)}$,
then $\rho_{0,i}>\rho_{1}^{(v)}$ -- see figure \ref{fig6}b). When
$\rho_{\infty,1}$ approaches the pool's boundary, a clearly visible
liquid/vapour interface emerges in the solution and is gradually moving away
from the substrate.$\smallskip$

(ii) It follows from the above that, for some pairs $\left(  \rho_{0,1}%
,~\rho_{\infty,1}\right)  $, no solution exists. As illustrated in figure
\ref{fig7}, some of such pairs involve unstable values of $\rho_{\infty,1}$
(and, thus, are unimportant), but there are also ones with a metastable
$\rho_{\infty,1}$.

To clarify what happens in such cases, the evolution was simulated numerically
(using the model and method used in the previous two subsections). It has
turned out that, if boundary-value problem (\ref{6.8}), (\ref{6.12}%
)--(\ref{6.13}) does not have a solution, the substrate triggers off a
spontaneous phase change. This result can be readily interpreted physically:
if a sufficiently \emph{hydrophilic} substrate (with a sufficiently small
$\rho_{0,1}$) touches \emph{oversaturated vapour}, it triggers off spontaneous
condensation, whereas a sufficiently \emph{hydrophobic} substrate touching
\emph{undersaturated liquid} triggers off evaporation. In either case, a shock
wave of phase change propagates away from the substrate, and the solid/fluid
interface cannot be stationary.\smallskip

(iii) For some pairs $\left(  \rho_{0,1},~\rho_{\infty,1}\right)  $, there are
two different solutions -- a monotonic and non-monotonic ones (see figure
\ref{fig7}). It is unclear which of the two occurs in reality.

It has turned out that only monotonic solutions of boundary-value problem
(\ref{6.12})--(\ref{6.13}), (\ref{6.8}) maximise the entropy, whereas
non-monotonic solutions do not (the former claim is proved in the general
case, but the latter, only for $N=1$ -- see Appendix \ref{Appendix C.3}). That
is, non-monotonic solutions cannot be ruled out with certainty for $N\geq2$ --
yet the mere fact that the general stability proof that works for monotonic
solutions cannot be extended to non-monotonic ones seems to resolve the
dilemma in favour of the former.\smallskip

(iv) Another physically important conclusion follows from the fact that
boundary-value problem (\ref{6.8}), (\ref{6.12})--(\ref{6.13}) has no more
than one stable solution -- and, thus, does \emph{not} admit solutions
describing a liquid layer of a finite thickness on a substrate, with saturated
vapour above it. All such layers, regardless of their thickness, evaporate --
and the Kelvin effect cannot be responsible for this effect (because the
liquid/vapour interface is flat). The evaporation in this case can only be
caused by long-range interaction between the interface and the
substrate.\smallskip

(v) It is interesting to compare the interfacial profiles shown in figures
\ref{fig6}--\ref{fig7} to those computed by \cite{EvansStewartWilding17} via
the density functional theory and Monte Carlo method for a Lennard-Jones fluid
bounded by a single wall or contained between two parallel walls. The
single-wall profiles of \cite{EvansStewartWilding17} (see their figure 3) are
qualitatively similar to those computed in this paper, but their two-wall
profiles are riddled with short-scale oscillations (see figures 6, 14, and
20). It is not clear whether the oscillations are caused by `interference' of
the walls: on the one hand, the distance between the walls exceeds the spatial
scale of wall--molecule interaction by a factor of $30$ (hence, the
`interference' should be weak) -- but, on the other hand, no oscillations
occur when this distance is infinite.

Whatever the nature of the oscillations is, one should not expect the DIM to
describe them due to the omission of the higher-order derivatives of $\rho
_{i}$ when obtaining expression (\ref{2.31}) for the van der Waals force.

\section{Surface tension and contact angle\label{Sec 7}}

\begin{figure}\begin{centering}
\includegraphics[width=90mm]{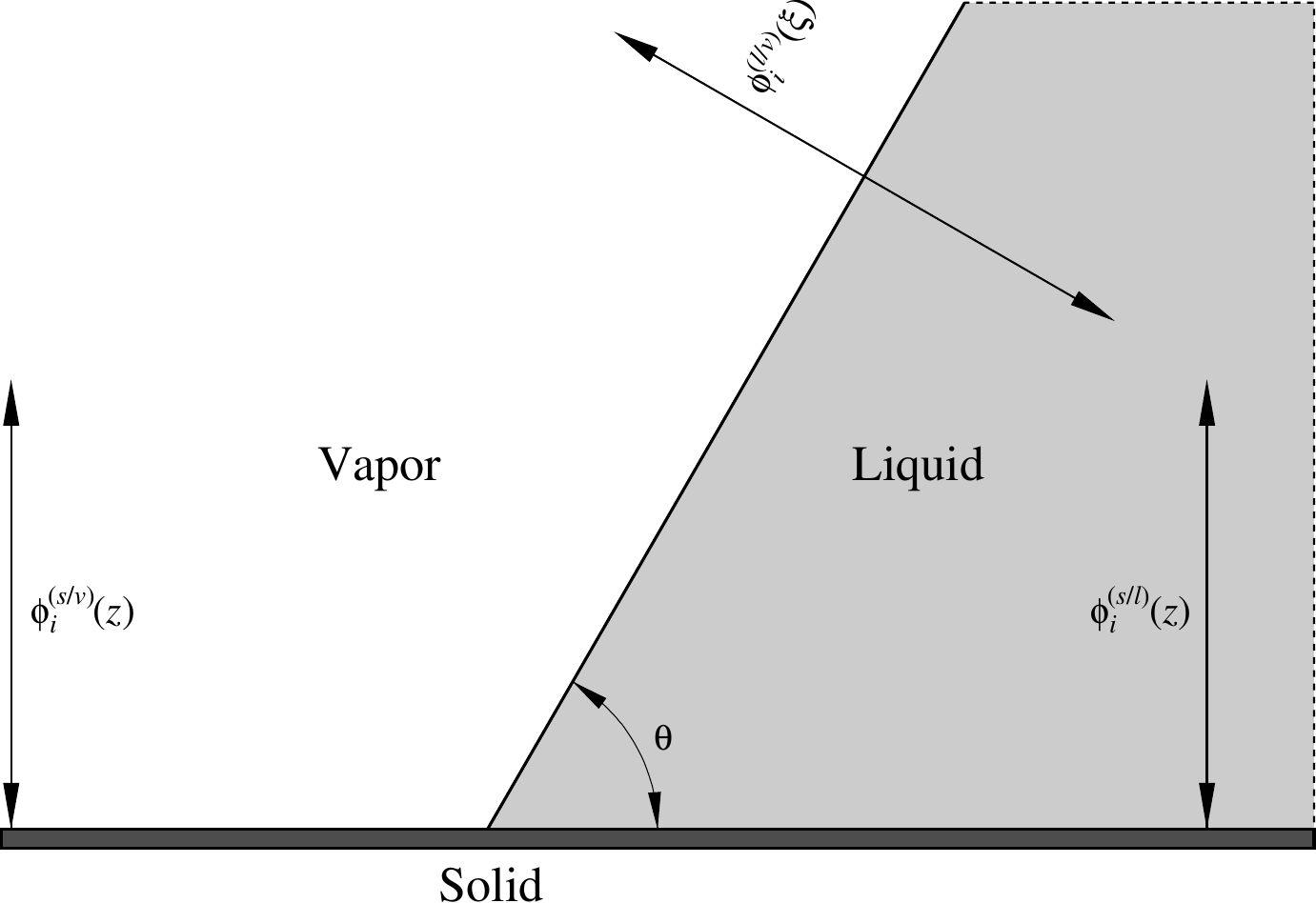}
\caption{A schematic illustrating boundary conditions (\ref{7.4})--(\ref{7.5}) for a static contact line. $\theta$ is the contact angle, $\phi^{(s/v)}(z)$ describes the \emph{s}olid/\emph{v}apour interface, \emph{etc.}}
\label{fig8}
\end{centering}\end{figure}

Consider a fluid bounded below by a horizontal substrate, and an oblique
liquid/vapour interface intersecting the substrate -- see figure \ref{fig8}.
Note that this figure depicts to a hydrophilic substrate, such that the
contact angle $\theta$ is smaller than $\pi/2$.

If in equilibrium, the setting outlined is described by the two-dimensional
reduction of equation (\ref{3.10}). Given expression (\ref{3.11}) for
$\eta_{i}$, one obtains%
\begin{equation}
G_{i}-\sum_{j}K_{ij}\left(  \frac{\partial^{2}\rho_{j}}{\partial x^{2}}%
+\frac{\partial^{2}\rho_{j}}{\partial z^{2}}\right)  +G_{\infty,i}=0.
\label{7.1}%
\end{equation}
Impose also the boundary condition%
\begin{equation}
\rho_{i}\rightarrow\rho_{i}^{(v)}\qquad\text{as}\qquad z\rightarrow+\infty,
\label{7.2}%
\end{equation}
which implies that the constant in (\ref{7.1}) is $G_{\infty,i}=G_{i}(\rho
_{1}^{(v)}...\rho_{N}^{(v)},T)$. At the substrate, the standard boundary
condition,%
\begin{equation}
\rho_{i}=\rho_{0,i}\qquad\text{at}\qquad z=0, \label{7.3}%
\end{equation}
is assumed.

To close boundary-value problem (\ref{7.1})--(\ref{7.3}), one should set
boundary conditions as $x\rightarrow\pm\infty$. The setting depicted in figure
\ref{fig8} corresponds to%
\begin{align}
\rho_{i}  &  \rightarrow\rho_{i}^{(s/v)}(z)\hspace{2.32cm}\text{as}\qquad
x\rightarrow-\infty,\label{7.4}\\
\rho_{i}  &  \rightarrow\rho_{i}^{(s/l)}(z)+\rho_{i}^{(l/v)}(\xi
)\qquad\text{as}\qquad x\rightarrow+\infty. \label{7.5}%
\end{align}
Here, the function $\rho_{i}^{(s/v)}(z)$ describes a solid/vapour interface
(i.e., satisfies boundary-value problem (\ref{6.8}), (\ref{6.12}%
)--(\ref{6.13}) with $\rho_{\infty,i}=\rho_{i}^{(v)}$); the function $\rho
_{i}^{(s/l)}(z)$ describes a solid/liquid interface; and $\rho_{i}^{(l/v)}%
(\xi)$ (where $\xi=z\cos\theta-x\sin\theta$) describes a liquid/vapour
interface tilted at an angle $\theta$ (it satisfies boundary-value problem
(\ref{6.1})--(\ref{6.3}) with $z$ changed to $\xi$).

In the framework of the DIM, the contact angle $\theta$ is not arbitrary, but
is fully determined by the fluid's thermodynamic properties, the Korteweg
matrix $K_{ij}$, and the near-wall density $\rho_{0,i}$. To derive an
expression for $\theta$ (similar to that derived by \cite{PismenPomeau00} for
pure fluids), consider%
\[
\sum_{i}\int_{0}^{\infty}\left(  \ref{7.1}\right)  \times\frac{\partial
\rho_{i}}{\partial x}\mathrm{d}z.
\]
After straightforward algebra involving integration by parts and the use of
boundary conditions (\ref{7.2})--(\ref{7.3}) and identity (\ref{2.7}), one
obtains%
\[
\frac{\partial}{\partial x}\int_{0}^{\infty}\left[  \sum_{j}\rho_{j}\left(
G_{j}-G_{j,\infty}\right)  -p+p_{\infty}+\frac{1}{2}\sum_{ij}K_{ij}\left(
\frac{\partial\rho_{i}}{\partial z}\frac{\partial\rho_{j}}{\partial z}%
-\frac{\partial\rho_{i}}{\partial x}\frac{\partial\rho_{j}}{\partial
x}\right)  \right]  \mathrm{d}z=0,
\]
where $p_{\infty}=p(\rho_{1}^{(v)}...\rho_{N}^{(v)},T)$. Finally, integrating
the above expression from $x=-\infty$ to $x=+\infty$ and recalling boundary
conditions (\ref{7.4})--(\ref{7.5}), one obtains%
\begin{multline*}
\int_{0}^{\infty}\left\{  \left[  \sum_{i}\rho_{i}\left(  G_{i}-G_{\infty
,i}\right)  -p+p_{\infty}\right]  _{\rho_{i}=\rho_{i}^{(s/v)}}+\frac{1}{2}%
\sum_{ij}K_{ij}\frac{\mathrm{d}\rho_{i}^{(s/v)}}{\mathrm{d}z}\frac
{\mathrm{d}\rho_{j}^{(s/v)}}{\mathrm{d}z}\right\}  \mathrm{d}z\\
=\int_{0}^{\infty}\left\{  \left[  \sum_{i}\rho_{i}\left(  G_{i}-G_{\infty
,i}\right)  -p+p_{\infty}\right]  _{\rho_{i}=\rho_{i}^{(s/l)}}+\frac{1}{2}%
\sum_{ij}K_{ij}\frac{\mathrm{d}\rho_{i}^{(s/l)}}{\mathrm{d}z}\frac
{\mathrm{d}\rho_{j}^{(s/l)}}{\mathrm{d}z}\right\}  \mathrm{d}z\\
+\int_{0}^{\infty}\left\{  \left[  \sum_{i}\rho_{i}\left(  G_{i}-G_{\infty
,i}\right)  -p+p_{\infty}\right]  _{\rho_{i}=\rho_{i}^{(l/v)}}%
\vphantom{\left[ \sum_{i}\right] _{\rho_{i}=\rho_{i}^{(l/v)}}\frac{\mathrm{d}\rho _{i}^{(l/v)}}{\mathrm{d}\xi}}\right.
\\
+\left.  \frac{1}{2}\left(  \cos^{2}\theta-\sin^{2}\theta\right)  \sum
_{ij}K_{ij}\frac{\mathrm{d}\rho_{i}^{(l/v)}}{\mathrm{d}\xi}\frac
{\mathrm{d}\rho_{i}^{(l/v)}}{\mathrm{d}\xi}%
\vphantom{\left[ \sum_{i}\right] _{\rho_{i}=\rho_{i}^{(l/v)}}\frac{\mathrm{d}\rho _{i}^{(l/v)}}{\mathrm{d}\xi}}\right\}
\frac{\mathrm{d}\xi}{\cos\theta}.
\end{multline*}
This equality can be simplified using identity (\ref{6.5}) (with $\rho_{i}$
changed to$\rho_{i}^{(l/v)}$), and similar identities for $\rho_{i}^{(s/v)}$
and $\rho_{i}^{(s/l)}$. After straightforward algebra involving re-denoting
$\xi\rightarrow z$, one obtains the DIM version of Young's formula,%
\begin{equation}
\cos\theta=\frac{\sigma^{(s/v)}-\sigma^{(s/l)}}{\sigma^{(l/v)}}, \label{7.6}%
\end{equation}
where%
\[
\sigma^{(s/v)}=\sum_{ij}K_{ij}%
%TCIMACRO{\dint _{0}^{\infty}}%
%BeginExpansion
{\displaystyle\int_{0}^{\infty}}
%EndExpansion
\dfrac{\mathrm{d}\rho_{i}^{(s/v)}}{\mathrm{d}z}\dfrac{\mathrm{d}\rho
_{j}^{(s/v)}}{\mathrm{d}z}\mathrm{d}z,\qquad\sigma^{(s/l)}=\sum_{ij}K_{ij}%
%TCIMACRO{\dint _{0}^{\infty}}%
%BeginExpansion
{\displaystyle\int_{0}^{\infty}}
%EndExpansion
\dfrac{\mathrm{d}\rho_{i}^{(s/l)}}{\mathrm{d}z}\dfrac{\mathrm{d}\rho
_{j}^{(s/l)}}{\mathrm{d}z}\mathrm{d}z,
\]%
\begin{equation}
\sigma^{(l/v)}=\sum_{ij}K_{ij}%
%TCIMACRO{\dint _{-\infty}^{\infty}}%
%BeginExpansion
{\displaystyle\int_{-\infty}^{\infty}}
%EndExpansion
\dfrac{\mathrm{d}\rho_{i}^{(l/v)}}{\mathrm{d}z}\dfrac{\mathrm{d}\rho
_{j}^{(l/v)}}{\mathrm{d}z}\mathrm{d}z \label{7.7}%
\end{equation}
are the surface tension coefficients of the solid/vapour, solid/liquid, and
liquid/vapour interfaces, respectively. Treating hydrophobic substrates
($\theta>\frac{1}{2}\pi$) in a similar fashion, one can show that formula
(\ref{7.6}) applies to that case as well.

To calculate the surface tension -- say, $\sigma^{(s/l)}$ -- one first needs
to solve boundary-value problem (\ref{6.1})--(\ref{6.4}) which determines the
function $\rho_{i}=\rho_{i}^{(l/v)}(z)$. For a pure fluid, however, the
expression for $\sigma^{(s/l)}$ can be rewritten as a closed-form integral. To
do so, let $N=1$ in equation (\ref{6.5}), which yields%

\[
\frac{\mathrm{d}\rho_{1}}{\mathrm{d}z}=-\sqrt{\frac{2}{K_{11}}\left\{
\rho_{1}\left[  G_{1}(\rho_{1},T)-G_{1}(\rho_{1}^{(v)},T)\right]  -p(\rho
_{1},T)+p(\rho_{1}^{(v)},T)\right\}  }.
\]
Using this equation to change the variable of integration in the expression
for $\sigma^{(l/v)}$ in (\ref{7.7}), one obtains%
\begin{equation}
\sigma^{(l/v)}=\sqrt{2K_{11}}%
%TCIMACRO{\dint _{\rho_{1}^{(v)}}^{\rho_{1}^{(l)}}}%
%BeginExpansion
{\displaystyle\int_{\rho_{1}^{(v)}}^{\rho_{1}^{(l)}}}
%EndExpansion
\sqrt{\rho_{1}\left[  G_{1}(\rho_{1},T)-G_{1}(\rho_{1}^{(v)},T)\right]
-p(\rho_{1},T)+p(\rho_{1}^{(v)},T)}\mathrm{d}\rho. \label{7.8}%
\end{equation}
Thus, to calculate $\sigma^{(l/v)}$, one should first use the given
$p(\rho_{1},T)$ and $G_{1}(\rho_{1},T)$ to find $\rho_{1}^{(l)}$ and $\rho
_{1}^{(v)}$ (via the Maxwell construction) and then evaluate integral
(\ref{7.8}).

\section{Parameterizing the diffuse--interface model for water/air
interfaces\label{Sec 8}}

To use the DIM in applications, one needs the following external parameters:
the fluid's thermodynamic properties, the dependence of the viscosity and
transport coefficients on $\left(  \rho_{1}...\rho_{N},T\right)  $, and the
Korteweg matrix. In this section, an approach to specifying these parameters
is described and applied to water/air interfaces.

For reasons described in subsection \ref{Sec 2.2}, the fluid's thermodynamic
properties will be approximated by the Enskog--Vlasov (EV) model. It involves
an undetermined matrix $a_{ij}$ and an undetermined function $\Theta(\rho
_{1}...\rho_{N})$, which should be fixed as the best fits of the empiric
characteristics of the fluid under consideration.

In subsections \ref{Sec 8.1}--\ref{Sec 8.2}, it will be explained how the
fitting should be carried out for a pure fluid (in application to water,
nitrogen, and oxygen). An approach to determining $K_{11}$ for a pure fluid
will be outlined in subsection \ref{Sec 8.3}. Subsection \ref{Sec 8.4}
describes how $a_{ij}$, $\Theta$, and $K_{ij}$ can be determined for
water--air mixture, and its viscosity and transport coefficients are dealt
with in subsection \ref{Sec 8.5}.

All these tasks will be carried out in terms of the original (dimensional) variables.

\subsection{The van der Waals parameter of pure water, nitrogen, and
oxygen\label{Sec 8.1}}

\begin{figure}\begin{centering}
\includegraphics[width=67mm]{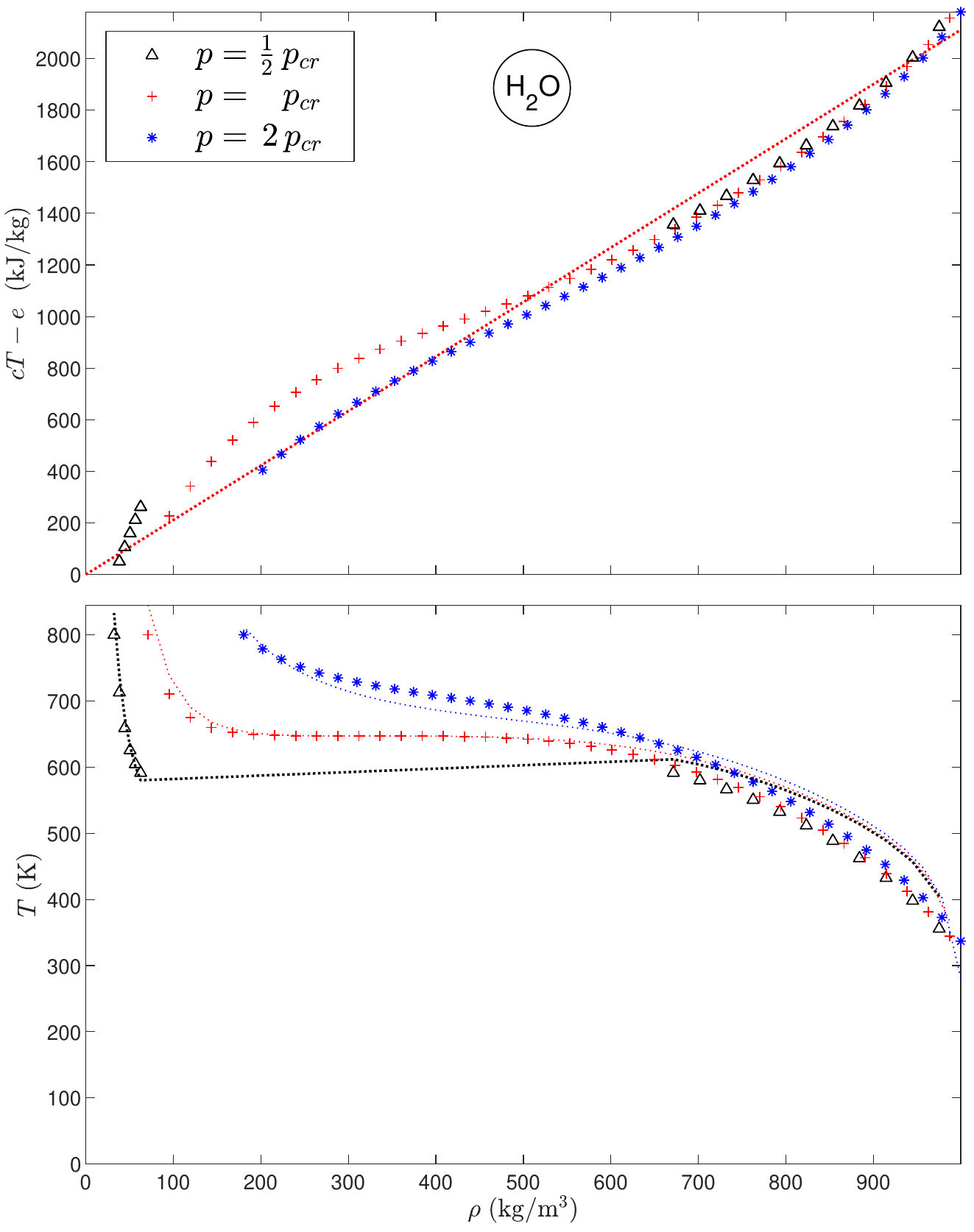} \includegraphics[width=66mm]{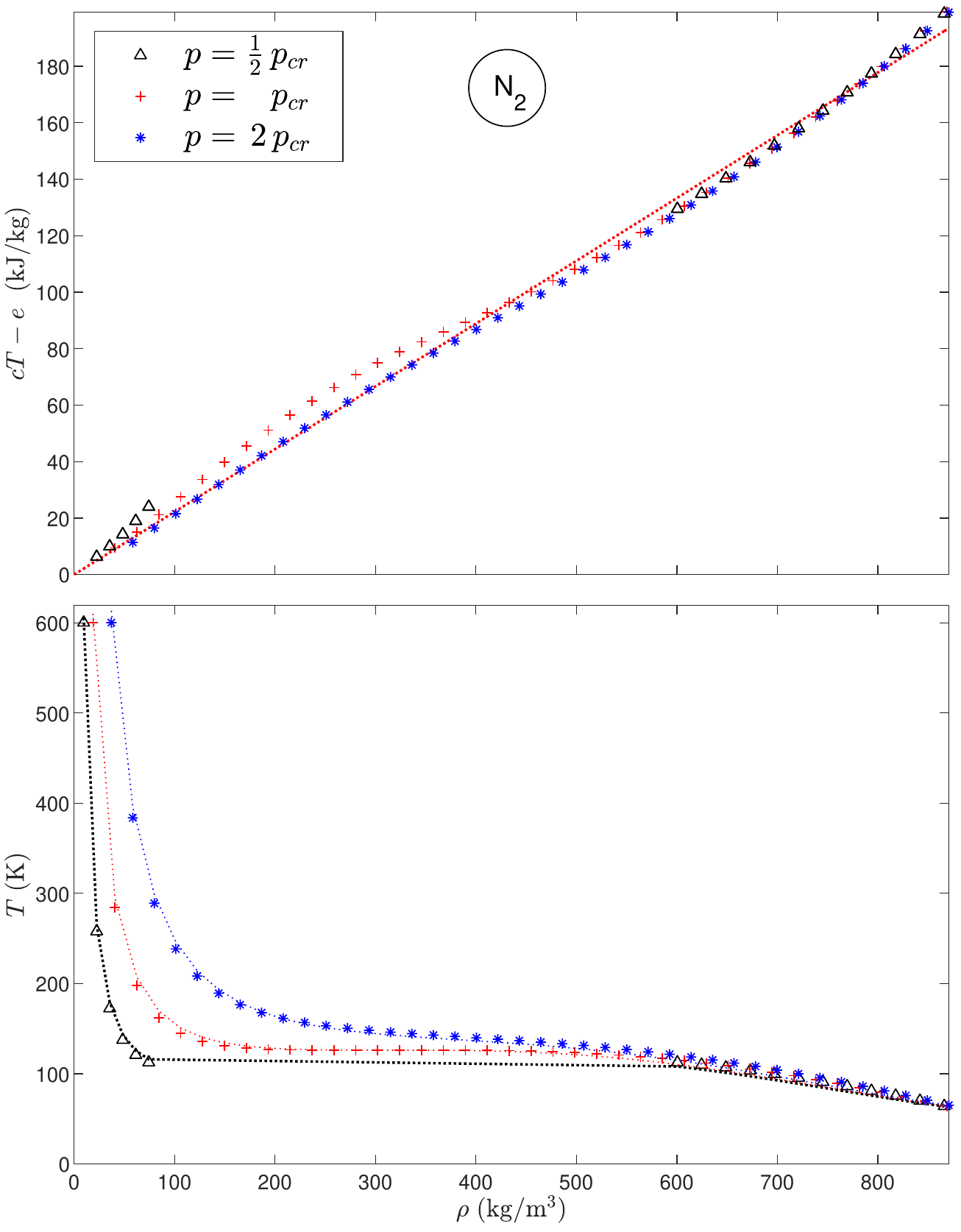}
\caption{Thermodynamic properties of $\mathrm{H}_{2}\mathrm{O}$ (left-hand panels) and $\mathrm{N}_{2}$ (right-hand panels). The non-connected symbols show the empiric data from \cite{LindstromMallard97} presented in isobaric form, for three values of the pressure $p$ relative to its critical value, $p_{cr}$ (see the legend). The gap between $\rho^{(v)}$ and $\rho ^{(l)}$ in the empiric data for $p=p_{cr}/2$ reflects the impossibility (difficulty) of measurements in an unstable (metastable) fluid. The dotted lines show the parameters calculated via the EV fluid model.}
\label{fig9}
\end{centering}\end{figure}

For a pure fluid, the EV expression (\ref{2.12}) for the internal energy
yields (the subscript $_{1}$ is omitted)%
\[
e=cT-a\rho,
\]
which suggests that the van der Waals parameter $a$ can be determined as the
\emph{linear} fit of the empiric dependence of $cT-e$ on $\rho$. The heat
capacity $c$ in this expression should be the same as that in the
Enskog--Vlasov kinetic theory -- i.e., $3R$ for water and $5R/2$ for nitrogen
and oxygen. For simplicity, the fitting was carried out using only the data on
the critical isobar $p=p_{cr}$ [as done by \cite{BenilovBenilov19a}], but the
resulting straight line fits the isobars $p=p_{cr}/2$ and $p=2p_{cr}$
reasonably well too (see the top panels of figure \ref{fig9}). The parameters
of the critical points for the fluids under consideration, as well as the
other parameters needed here and hereinafter, can be found in Tables
\ref{tab1}--\ref{tab2}.

\begin{table}
\begin{tabularx}{\textwidth}{@{}lYYYYYY@{}}
\hspace*{1mm}Fluid & $T_{tp}~(\mathrm{K})$ & $\rho_{tp}^{(l)}~(\mathrm{kg\,m}^{-3})$ & $\rho_{tr}^{(v)}~(\mathrm{kg\,m}^{-3})$ & $T_{cr}~(\mathrm{K})$ & $\rho
_{cr}~(\mathrm{kg\,m}^{-3})$ & $p_{cr}~(\mathrm{bar})$\\\hline
\hspace*{1mm}$\mathrm{H}_{2}\mathrm{O}$ & $273.160$ & $~999.79$ & $0.0048546$ & $647.096$ & $322.00$ & $220.64$\\\hline
\hspace*{1mm}$\mathrm{N}_{2}$ & $63.151$ & $~867.22$ & $0.6742700$ & $126.192$ & $313.30$ & $33.958$\\\hline
\hspace*{1mm}$\mathrm{O}_{2}$ & $54.361$ & $1306.10$ & $0.0103580$ & $154.581$ & $436.14$ & $50.430$\\\hline
\end{tabularx}
\caption{The parameters of the triple and critical points (subscripts $_{tr}$ and $_{cr}$, respectively) of $\mathrm{H}_{2}\mathrm{O}$, $\mathrm{N}_{2}$, and $\mathrm{O}_{2}$ \cite{LindstromMallard97}.}
\label{tab1}
\end{table}

\begin{table}
\begin{tabularx}{\textwidth}{@{}lYYYY@{}}
\hspace*{0.1cm}Fluid & $m~(10^{-26}\mathrm{kg})$ & $R~(\mathrm{m}^{2}\mathrm{s}^{-2}\mathrm{K}^{-1})$ & $a~(\mathrm{m}^{5}\mathrm{s}^{-2}\mathrm{kg}^{-1})$ & $K~(10^{-17}\mathrm{m}^{7}\mathrm{s}^{-2}\mathrm{kg}^{-1})$\\\hline
\hspace*{0.1cm}$\mathrm{H}_{2}\mathrm{O}$ & $2.9915$ & $461.53$ & $2112.1$ &
$1.87810$\\\hline
\hspace*{0.1cm}$\mathrm{N}_{2}$ & $4.6516$ & $296.81$ & $222.23$ & $1.50780$\\\hline
\hspace*{0.1cm}$\mathrm{O}_{2}$ & $5.3135$ & $259.84$ & $172.73$ & $0.84587$\\\hline
\hspace*{0.1cm}air & $4.7706$ & $289.41$ & $211.84$ & $1.36880$\\\hline
\end{tabularx}
\caption{The parameters of $\mathrm{H}_{2}\mathrm{O}$, $\mathrm{N}_{2}$, $\mathrm{O}_{2}$, and air: $m$ is the molecular mass, $R$ is the specific gas constant, $a$ is the van der Waals parameter, $K$ is the Korteweg constant. The parameters of air are calculated as the 79-21 weighted averages of the corresponding parameters of nitrogen and oxygen, respectively.}
\label{tab2}
\end{table}

The resulting values of the van der Waals parameter $a$ for $\mathrm{H}%
_{2}\mathrm{O}$, $\mathrm{N}_{2}$, and $\mathrm{O}_{2}$ are presented in Table
\ref{tab2}. Interestingly, $a$ of water exceeds significantly those of
nitrogen and oxygen. This is likely to be caused by the difference in the
molecular structure of these fluids: the water molecule has a non-zero dipolar
moment, whereas nitrogen and oxygen molecules are symmetric -- hence, do not.
Since the van der Waals force is of electric nature, one can assume that
dipolar molecules interact stronger than neutral ones.

\subsection{The equation of state of pure $\mathrm{H}_{2}\mathrm{O}$,
$\mathrm{N}_{2}$, and $\mathrm{O}_{2}$\label{Sec 8.2}}

Pure fluids will be described by the EV model (\ref{2.12})--(\ref{2.13}) with%
\begin{equation}
\Theta=R\left[  -q^{(0)}\ln\left(  1-0.99\frac{\rho}{\rho_{tp}}\right)
+\sum_{n=1}^{4}q^{(n)}\left(  \frac{\rho}{\rho_{tp}}\right)  ^{n}\right]  ,
\label{8.1}%
\end{equation}
where $q^{(0)}$... $q^{(4)}$ are undetermined coefficients and $\rho_{tp}$ is
the fluid's density at the triple point ($\rho_{tp}$ is just a convenient
density scale; the fact that, at the triple point, the liquid and vapour are
in equilibrium with the solid phase is irrelevant, as solids are not described
by the DIM). Note that the first term in expression (\ref{8.1}) sets the
maximum density at $\rho_{tp}/0.99$.

The coefficients $q^{(n)}$ were determined for $\mathrm{H}_{2}\mathrm{O}$,
$\mathrm{N}_{2}$, and $\mathrm{O}_{2}$ by ensuring that the expressions for
$p(\rho,T)$ and $G(\rho,T)$ corresponding to (\ref{8.1}) yield the `correct'
-- i.e., empiric -- values for the critical density, temperature, and
pressure, as well as the liquid and vapour densities at the triple-point
temperature $T=T_{tp}$ (five equations for the five unknown coefficients). All
the necessary empiric data can be found in Tables \ref{tab1}--\ref{tab2}, and
the computed values of $q^{(n)}$ are listed in Table \ref{tab3}. Such an
approach to calibrating the EV fluid model is a slight modification of that of
\cite{Benilov20a} -- which is, in turn, a modification of the approach of
\cite{BenilovBenilov18}.

\begin{table}
\begin{tabularx}{\textwidth}{@{}lYYYYY@{}}
\hspace*{0.1cm}Fluid & $q^{(0)}$ & $q^{(1)}$ & $q^{(2)}$ & $q^{(3)}$ & $q^{(4)}$\\\hline
\hspace*{0.1cm}$\mathrm{H}_{2}\mathrm{O}$ & $0.071894$ & $1.4139$ & $8.1126$ & $-8.3669$ & $4.0238$\\\hline
\hspace*{0.1cm}$\mathrm{N}_{2}$ & $-0.0013920$ & $0.72934$ & $6.4799$ & $-8.1143$ & $5.0186$\\\hline
\hspace*{0.1cm}$\mathrm{O}_{2}$ & $0.010770$ & $0.58901$ & $8.5357$ & $-12.034$ & $8.0872$\\\hline
\end{tabularx}
\caption{The coefficients of the equation of state (\ref{8.1}) for $\mathrm{H}_{2}\mathrm{O}$, $\mathrm{N}_{2}$, and $\mathrm{O}_{2}$.}
\label{tab3}
\end{table}

To illustrate how well the EV model describes real fluids, the isobaric (with
$p$ fixed) dependence of the temperature on the density is plotted for
$\mathrm{H}_{2}\mathrm{O}$ and $\mathrm{N}_{2}$ in the lower panels of figure
\ref{fig9}. The results for $\mathrm{O}_{2}$ (not presented) are similar to
those for $\mathrm{N}_{2}$.

Note that the DIM has been coupled with realistic equations of state before
\citep[e.g.][]{Caupin05,MagalettiGalloCasciola21}; for pure water, this was
typically done using the IAPWS-95 equation \citep{WagnerPruss02}. The EV model
used here is undoubtedly less accurate than the IAPWS-95 model -- but it
allows one to describe consistently all of the fluids involved ($\mathrm{H}%
_{2}\mathrm{O}$, $\mathrm{N}_{2}$, and $\mathrm{O}_{2}$).

\subsection{The Korteweg parameter of pure $\mathrm{H}_{2}\mathrm{O}$,
$\mathrm{N}_{2}$, and $\mathrm{O}_{2}$\label{Sec 8.3}}

\citep{Benilov20a} proposed to deduce $K$ from the requirement that the DIM
predict the correct value of the surface tension $\sigma^{(l/v)}$ of
liquid/vapour interface at the triple point. The same was done in the present
work: $\sigma^{(l/v)}$ was calculated via the DIM formula (\ref{7.8}) with $G$
and $p$ of the EV fluid, and the value of $K$ was chosen for which (\ref{7.8})
yields the same result as the empiric formula of \cite{Somayajulu88}. The
resulting $K$ for $\mathrm{H}_{2}\mathrm{O}$, $\mathrm{N}_{2}$, and
$\mathrm{O}_{2}$ can be found in Table \ref{tab2}.

It should be emphasised that the values of $K$ computed via the above approach
depend on the chosen fluid model. This explains the difference between the
result for $K$ in this paper and the one computed by \cite{Benilov20a}: the
former is based on the EV fluid model and the latter, on the van der Waals
model. The resulting 30\% difference in $\sigma^{(l/v)}$ reflects the fact
that the latter model is much less accurate.

\begin{figure}\begin{centering}
\includegraphics[width=83mm]{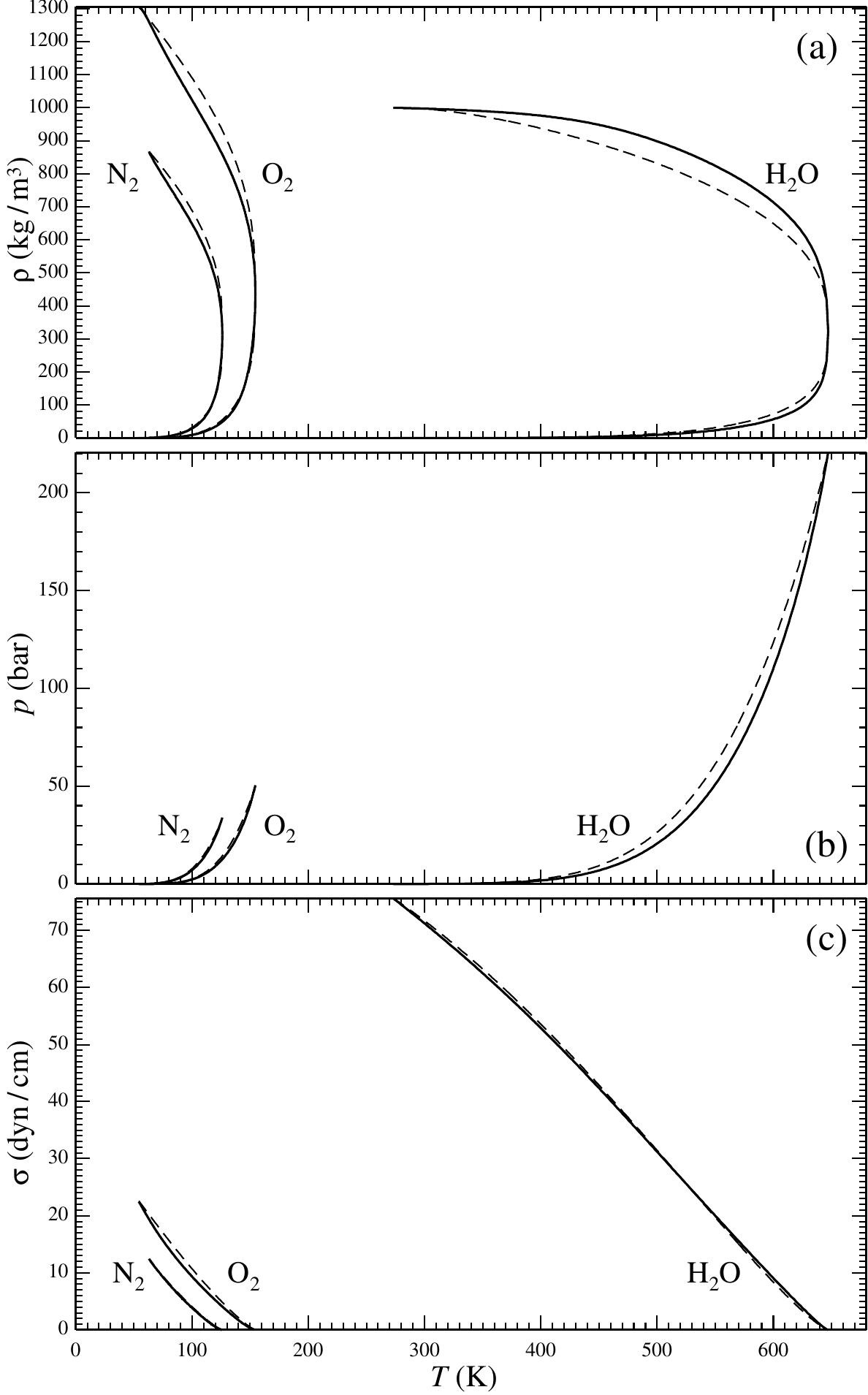}
\caption{Comparison of the results obtained via the DIM for an EV fluid (solid line) with the corresponding empiric data (dashed line): (a) the densities of saturated vapour and liquid, (b) saturated pressure, (c) surface tension of liquid/vapour interface. The empiric data in panels (a) and (b) are from \citep{LindstromMallard97}, and those in panel (c) are from \citep{Somayajulu88}.}
\label{fig10}
\end{centering}\end{figure}

As a test of the whole approach based on the DIM and EV models, the
theoretical values for the saturation characteristics and $\sigma^{(l/v)}$
have been compared to their empiric counterparts for the whole temperature
range where liquid/vapour interfaces exist, i.e., between the triple and
critical points. The results are shown in figure \ref{fig10}: evidently, the
theoretical prediction are reasonably accurate.

\subsection{Parameters of water--air interaction\label{Sec 8.4}}

Generally, the parameters of a fluid should be deduced from measurements of
its equation of state, surface tension, \emph{etc.} -- but these are rarely
known for \emph{multicomponent} fluids. The water--air mixture \emph{at normal
conditions} is an exception in this respect: the density of air is small in
this case, and its equation of state is close to that of an ideal gas. In
addition, air will be treated as a pure fluid with parameters equal to the
weighted averages of those of nitrogen and oxygen (see Table \ref{tab2}).

Thus, let the function $\Theta$ (the non-ideal part of the entropy of an EV
fluid) be independent of the air density $\rho_{2}$, so that expressions
(\ref{2.15})--(\ref{2.16}) with $N=2$ yield%
\begin{equation}
p=T\left[  R_{1}\rho_{1}+R_{2}\rho_{2}+\left(  \rho_{1}+\rho_{2}\right)
\frac{\mathrm{d}\Theta(\rho_{1})}{\mathrm{d}\rho_{1}}\right]  -a_{11}\rho
_{1}^{2}-2a_{12}\rho_{1}\rho_{2}, \label{8.2}%
\end{equation}%
\begin{equation}
G_{1}=T\left[  R_{1}\ln\rho_{1}+\Theta(\rho_{1})+\left(  \rho_{1}+\rho
_{2}\right)  \frac{\mathrm{d}\Theta(\rho_{1})}{\mathrm{d}\rho_{1}}\right]
-2\left(  a_{11}\rho_{1}+a_{12}\rho_{2}\right)  +T\left[  R_{1}+c_{1}\left(
1-\ln T\right)  \right]  , \label{8.3}%
\end{equation}%
\begin{equation}
G_{2}=T\left[  R_{2}\ln\rho_{2}+\Theta(\rho_{1})\right]  -2a_{12}\rho
_{1}+T\left[  R_{2}+c_{2}\left(  1-\ln T\right)  \right]  . \label{8.4}%
\end{equation}
Here, the function $\Theta(\rho_{1})$ is the one given by formula (\ref{8.1}),
with $\rho$ changed to $\rho_{1}$, and with its coefficients corresponding to water.

Before using expressions (\ref{8.2})--(\ref{8.4}), one needs to fix the
non-diagonal term $a_{12}$ of the matrix $a_{ij}$, responsible for the
interaction of water and air molecules. It can be deduced from $\rho_{2}%
^{(l)}$ (the amount of air dissolved in water): one should choose such
$a_{12}$ that the Maxwell construction based on (\ref{8.2})--(\ref{8.4})
predicts the correct value of $\rho_{2}^{(l)}$. Since $a_{12}$ is supposed to
not depend on $T$, such a calculation should be done for a single temperature
and the atmospheric pressure. For $T=25^{\circ}\mathrm{C}$ and
$p=1\,\mathrm{atm}$, for example, $\rho_{2}^{(l)}=0.0227\,\mathrm{kg\,m}^{-3}$
\citep{TheEngineeringToolbox-AirWaterSolubility}. It can then be deduced that the
Maxwell construction based on (\ref{8.2})--(\ref{8.4}) yields the correct
value of $\rho_{2}^{(l)}$ if%
\[
a_{12}=208\,\mathrm{m}^{5}\mathrm{s}^{-2}\mathrm{kg}^{-1},
\]
i.e., approximately equal to $a_{22}$ and ten times smaller than $a_{11}$.

The accuracy of expressions (\ref{8.2})--(\ref{8.4}) can be illustrated by the
corresponding value of the boiling point ($T\approx109^{\circ}\mathrm{C}$, as
opposed to the exact value of $T=100^{\circ}\mathrm{C}$). Furthermore, at the
\textquotedblleft room temperature\textquotedblright\ (say, $T=25^{\circ
}\mathrm{C}$), the error should be even smaller, because the room temperature
is close to the triple point of water ($T\approx0^{\circ}\mathrm{C}$) where
(\ref{8.2})--(\ref{8.4}) were calibrated.

It still remains to determine the non-diagonal term $K_{12}$ of the Korteweg
matrix $K_{ij}$.

Since $K_{ij}$ is supposed to be positive definite, $K_{12}$ should satisfy%
\begin{equation}
-1<\frac{K_{12}}{\left(  K_{11}K_{22}\right)  ^{1/2}}<1. \label{8.5}%
\end{equation}
One would think that $K_{12}$ could be determined by comparing the surface
tension of liquid-water/air interface to that of
liquid-water/vapour\emph{-water} interface. It turns out, however, that the
difference between these is so small that the existing experimental techniques
cannot detect it \citep[see][and references therein]{VargaftikVolkovVoljak83}.
This can be due to smallness of $K_{12}$ -- but more likely, because the
density of air is small. Either way, the determination of $K_{12}$ via
measurements of surface tension is impossible.

To at least confirm that air cannot affect significantly the surface tension
of water, boundary-value problem (\ref{6.1})--(\ref{6.4}) was solved for the
parameters of water and air at $T=25^{\circ}\mathrm{C}$, and $K_{12}$ varying
through range (\ref{8.5}). It has turned out that the resulting variation of
the surface tension is only $7\%$.

Even though the dependence of the surface tension (which is a \emph{global}
characteristic) on $K_{12}$ is weak, this parameter does influence the
\emph{local} structure of the interface.

\begin{figure}\begin{centering}
\includegraphics[width=108mm]{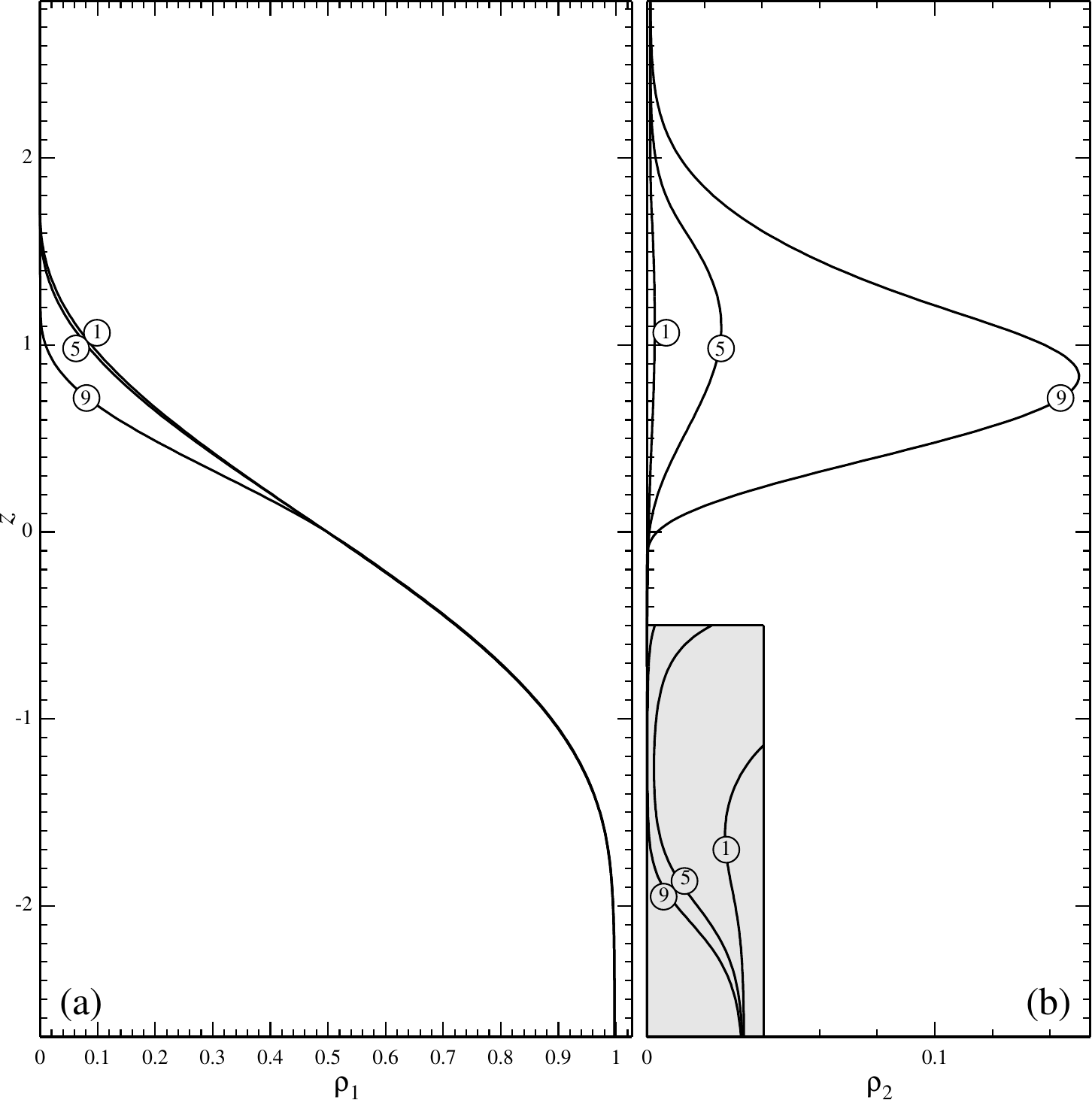}
\caption{The spatial structure of water/air interfaces, for $T=25^{\circ}\mathrm{C}$ and $K_{12}=\left(  K_{11}K_{22}\right)  ^{1/2}\times n/10$, where $n$ is the curve number: (a) density of water; (b) density of air. The horizontal scales of panel (a) and (b) differ by a factor of 5; the latter and the horisontal scale of the shaded inset in panel (b) differ by a factor of 1500. The inset shows that the concentration of the gas dissolved in liquid is non-zero.}
\label{fig11}
\end{centering}\end{figure}

Figure \ref{fig11} depicts the density profiles $\rho_{1}(z)$ and $\rho
_{2}(z)$ for the water/air interface at room temperature [i.e., the
above-mentioned solution of boundary-value problem (\ref{6.1})--(\ref{6.4})].
One can see that, as $K_{12}$ approaches the right endpoint of range
(\ref{8.5}), the profile of water density is getting steeper, and a layer of
high air density is developing inside the interface. For curve 5 of figure
\ref{fig11}(b) -- which is half-way through range (\ref{8.5}) -- the maximum
of $\rho_{2}(z)$ exceeds the atmospheric air density by a factor of
approximately $20$. Since the amount of air drawn into the interface grows
with $K_{12}$, one concludes that the air is `pulled in' by the van der Waals
attraction exerted by the bulk of the liquid.

This suggests a possibility of deducing $K_{12}$ from empiric data on
evaporation of drops (which depends on both global and local characteristics
of the interface \citep{Benilov22a}). Such an approach, however, requires an
extension of the evaporation model of \cite{Benilov22a} to multicomponent
fluids, which is currently in progress.

Observe that the interfaces depicted in figure \ref{fig11} are such that
$\rho_{2}(z)$ is not monotonic$\mathrm{\ }$-- hence, the sufficient stability
criterion from subsection \ref{Sec 6.1} and Appendix \ref{Appendix C.1} does
not apply. This does not, however, mean that this interface is unstable. It
is, in fact, difficult to imagine that a microscopic non-monotonicity of the
density of air dissolved in water could destabilise the whole interface -- but
this issue is still of interest theoretically and, thus, deserves further investigation.

\subsection{The viscosity and thermal conductivity of water--air
mixture\label{Sec 8.5}}

There is a large body of work on the viscosity of multicomponent fluids, with
even the simplest theoretical models -- e.g., that of Enskog--Chapman for
dilute gases \citep[e.g.,][]{FerzigerKaper72} -- yielding a fairly complicated
dependence of $\mu_{s}$ and $\mu_{b}$ on $\rho_{i}$. Phenomenological results,
on the other hand, tend to include many \emph{ad hoc} parameters specific to
the fluid under consideration
\citep[e.g.][and references therein]{Davidson93}. Overall, the simplest option
seems to be the expression for the shear viscosity proposed on a
phenomenological basis by \cite{HindMclaughlinUbbelohde60} and justified,
under certain approximations, via statistical mechanics by
\cite{BearmanJones60},%
\begin{equation}
\mu_{s}=\mu_{s,1}f_{1}^{2}+\left(  \mu_{s,1}+\mu_{s,2}\right)  f_{1}f_{2}%
+\mu_{s,2}f_{2}^{2}, \label{8.6}%
\end{equation}
where $\mu_{s,i}$ is the shear viscosity of the $i$-th species and%
\[
f_{i}=\frac{\rho_{i}/m_{i}}{\rho_{1}/m_{1}+\rho_{2}/m_{2}}%
\]
is its mole fraction. Expression (\ref{8.6}) does not include any
fluid-specific parameters, but is capable of predicting the shear viscosity
with a reasonable accuracy. If, for example, air is treated as a mixture of
$\mathrm{N}_{2}$ and $\mathrm{O}_{2}$, the error of (\ref{8.6}) for the range
$T=0-100^{\circ}\mathrm{C}$ is less than 1\%. A similar \textquotedblleft
mixture rule\textquotedblright\ can be assumed for the bulk viscosity and
thermal conductivity,%
\begin{equation}
\mu_{b}=\mu_{b,1}f_{1}^{2}+\left(  \mu_{b,1}+\mu_{b,2}\right)  f_{1}f_{2}%
+\mu_{b,2}f_{2}^{2}, \label{8.7}%
\end{equation}%
\begin{equation}
\kappa=\kappa_{1}f_{1}^{2}+\left(  \kappa_{1}+\kappa_{2}\right)  f_{1}%
f_{2}+\kappa_{2}f_{2}^{2}. \label{8.8}%
\end{equation}
Generally, various mechanical properties of a multicomponent fluid are often
described by the same mixture rule, in which case it is referred to as
\textquotedblleft generalised\textquotedblright.

It still remains to fix the dependence of $\mu_{s,i}$, $\mu_{b,i}$, and
$\kappa_{i}$ on $\left(  \rho_{1},\rho_{2},T\right)  $. In application to air
-- which can be treated as a dilute gas -- one can assume these parameter to
depend only on $T$ (as predicted by the kinetic theory of dilute gases), i.e.,%
\begin{equation}
\mu_{s,2}(\rho_{2},T)=\mu_{s,2}(0,T),\qquad\mu_{b,2}(\rho_{2},T)=\mu
_{b,2}(0,T),\qquad\kappa_{2}(\rho_{2},T)=\kappa_{2}(0,T), \label{8.9}%
\end{equation}
where $\mu_{s,2}(0,T)$, $\mu_{b,2}(0,T)$, and $\kappa_{2}(0,T)$ are the
small-density limiting values of the corresponding parameters.

\begin{figure}\begin{centering}
\includegraphics[width=76mm]{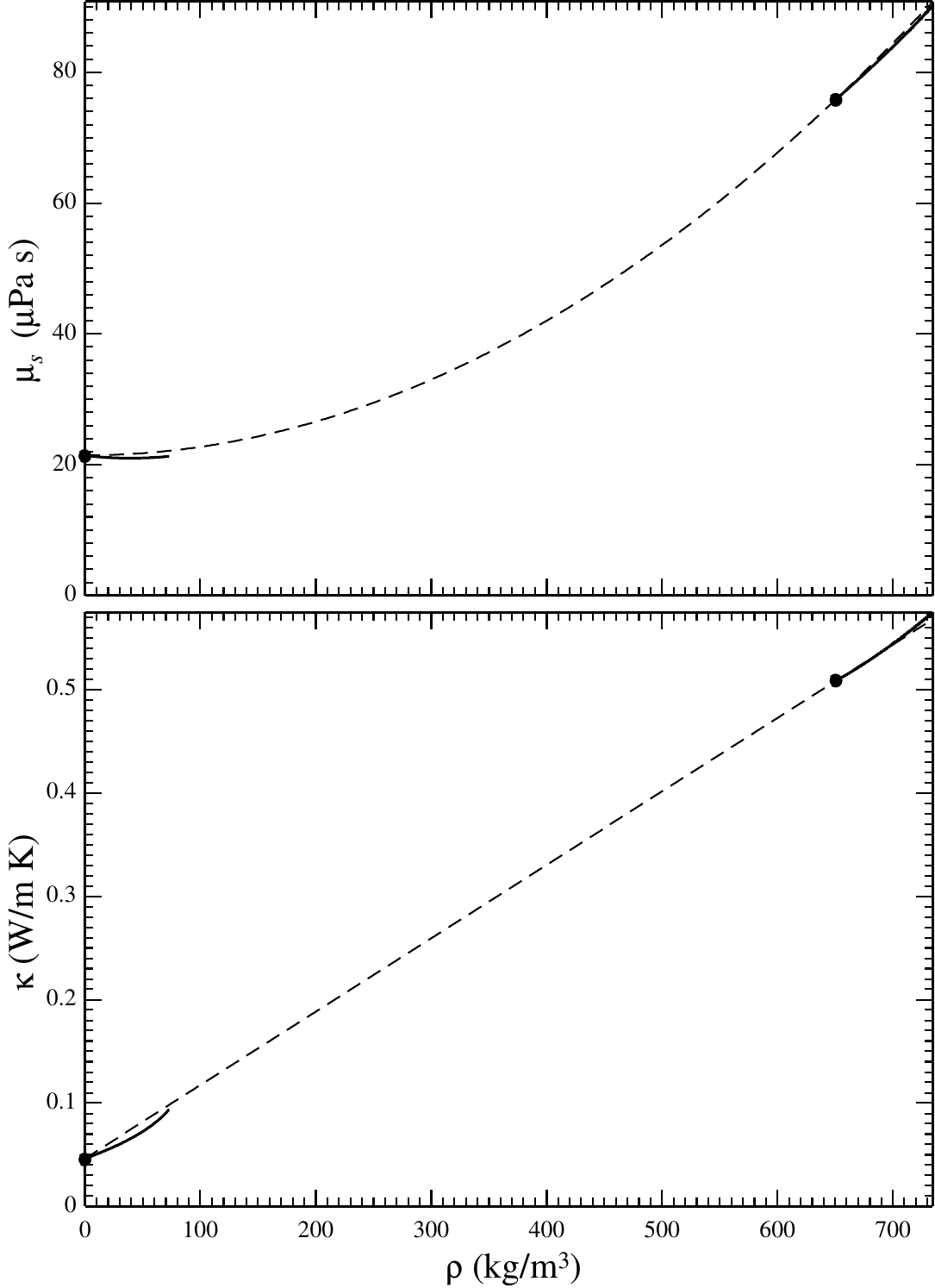}
\caption{The parameters of pure water at $T=327^{\circ}\mathrm{C}$: (a) shear viscosity, (b) thermal conductivity. The empiric data \citep{LindstromMallard97} are shown in solid line and approximations (\ref{8.10}) and (\ref{8.12}), in dashed line. The dots mark the reference points. The gap in the empiric data reflects the impossibility (difficulty) of
measurements in an unstable (metastable) fluid.}
\label{fig12}
\end{centering}\end{figure}

For water, whose liquid phase cannot be treated as a dilute gas, such an
approximation is inapplicable. Aiming again for simplicity, one can
approximate both viscosities by a parabola passing through two reference
points -- the zero-density limit and the saturated liquid state,%
\begin{equation}
\mu_{s,1}(\rho_{1},T)=\mu_{s,1}(0,T)+\left[  \mu_{s,1}(\rho_{1}^{(l)}%
,T)-\mu_{s,1}(0,T)\right]  \left[  \frac{\rho_{1}}{\rho_{1}^{(l)}(T)}\right]
^{2}, \label{8.10}%
\end{equation}%
\begin{equation}
\mu_{b,1}(\rho_{1},T)=\mu_{b,1}(0,T)+\left[  \mu_{b,1}(\rho_{1}^{(l)}%
,T)-\mu_{b,1}(0,T)\right]  \left[  \frac{\rho_{1}}{\rho_{1}^{(l)}(T)}\right]
^{2}. \label{8.11}%
\end{equation}
A similar approximation, but by a linear function, will be adopted for the
thermal conductivity,%
\begin{equation}
\kappa_{1}(\rho_{1},T)=\kappa_{1}(0,T)+\left[  \kappa_{1}(\rho_{1}%
^{(l)},T)-\kappa_{1}(0,T)\right]  \frac{\rho_{1}}{\rho_{1}^{(l)}(T)}.
\label{8.12}%
\end{equation}
The accuracy of the above approximations for $\mu_{s,1}$ and $\kappa_{1}$ is
illustrated in figure \ref{fig12}.

To use formulae (\ref{8.6})--(\ref{8.12}), one needs to know how the
viscosities and thermal conductivities of water and air depend on $T$. For
$\mu_{s}$ and $\kappa$, such data are widely available
\citep[e.g.,][]{LindstromMallard97,White05}. Measurements of $\mu_{b}$, on the
other hand, are scarce, but can still be found in \citep{HolmesParkerPovey11},
\citep{Cramer12}, and \citep{ShangWuWangYangYeHuTaoHe19} for liquid water,
vapour water, and air, respectively.

The empiric formulae proposed in these papers will not be discussed in detail.
The characteristic values they yield for normal conditions are listed in Table
\ref{tab4}.

\begin{table}
\begin{tabularx}{\textwidth}{@{}lYYY@{}}
Fluid & $\mu_{s}~(\mathrm{Pa\,s})$ & $\mu_{b}~(\mathrm{Pa\,s})$ & $\kappa~(\mathrm{W\,K}^{-1}\mathrm{m}^{-1})$\\\hline
liquid water & $0.890\times10^{-3}$\linebreak\citep{LindstromMallard97} & $2.47\times10^{-3}$\linebreak\citep{HolmesParkerPovey11} & $0.606460$\linebreak\citep{LindstromMallard97}\\\hline
vapour water & $0.971\times10^{-5}$\linebreak\citep{LindstromMallard97} & $7.20\times10^{-5}$\linebreak\citep{Cramer12} & $0.018433$\linebreak\citep{LindstromMallard97}\\\hline
air & $1.840\times10^{-5}$\linebreak\citep{White05} & $1.75\times 10^{-5}$\linebreak\citep{ShangWuWangYangYeHuTaoHe19} & $0.026089$ \linebreak\citep{White05}\\\hline
\end{tabularx}
\caption{The empiric viscosity and thermal conductivity of liquid water, water vapour, and air at $25^{\circ}\mathrm{C}$, and the corresponding references.}
\label{tab4}
\end{table}

\subsection{The transport coefficients\label{Sec 8.6}}

Recalling restrictions (\ref{2.25}), one can deduce that the extended
transport matrix for a two-component fluid is%
\[
D_{ij}^{(ext)}=%
\begin{bmatrix}
D_{11} & -D_{11} & \zeta_{1}\\
-D_{11} & D_{11} & -\zeta_{1}\\
\zeta_{1} & -\zeta_{1} & \kappa
\end{bmatrix}
.
\]
Evidently, it involves only three independent coefficients, one of which (the
thermal conductivity $\kappa$) has already been discussed in subsection
\ref{Sec 8.5}. The other two, $D_{11}$ and $\zeta_{1}$, are discussed below.

It can be safely assumed that diffusion is of importance only where the water
density is comparable to that of air. In the region where the former is high,
the latter is miniscule, and so is its influence on the global dynamics. This
effectively means that the diffusivities can be represented using the
Enskog--Chapman method: even though it does not apply to liquid water, the
error due to using it anyway is negligible.

According to the leading-order Enskog--Chapman formula
\citep[e.g.,][]{FerzigerKaper72},%
\begin{equation}
D_{11}=\frac{\rho_{1}\rho_{2}\left(  \rho_{1}m_{2}+\rho_{2}m_{1}\right)
}{\left(  \rho_{1}+\rho_{2}\right)  ^{2}k_{B}T}\mathfrak{D}(T), \label{8.13}%
\end{equation}
where $\mathfrak{D}(T)$ does not depend on $\rho_{1}$ and $\rho_{2}$. Its
dependence on $T$ can be found in
\citep{TheEngineeringToolbox-AirWaterDiffusion}, from which the following
characteristic value can be deduced,%
\begin{equation}
\mathfrak{D}(25^{\circ}\mathrm{C})=2.49\times10^{-5}\mathrm{m}^{2}%
\mathrm{s}^{-1}. \label{8.14}%
\end{equation}
The Enskog--Chapman expression for the thermodiffusivity is fairly bulky -- as
a result, one often uses the simpler formula of \cite{DegrootMazur1962},
\begin{equation}
\zeta_{1}=\left(  \rho_{1}+\rho_{2}\right)  \frac{\left(  \rho_{1}%
/m_{1}\right)  \left(  \rho_{2}/m_{2}\right)  }{\left(  \rho_{1}/m_{1}%
+\rho_{2}/m_{2}\right)  ^{2}}\mathfrak{U}(T). \label{8.15}%
\end{equation}
Unfortunately, no data on $\mathfrak{U}(T)$ for water--air mixture are
available in the literature; the author of this paper was able to find only an
estimate for a single temperature value \citep{LidonPerrotStroock21},%
\begin{equation}
\mathfrak{U}(21^{\circ}\mathrm{C})=-4.98\times10^{-6}\,\mathrm{m}%
^{2}\mathrm{s}^{-1}. \label{8.16}%
\end{equation}
It is shown later, however, that thermodiffusion in water-air interfaces is
negligible, so the lack of data on $\mathfrak{U}(T)$ is unimportant.

In the next subsection, characteristic values of the coefficients $D_{11}$ and
$\zeta_{1}$ will be needed. These will be estimated for the particular case
$\rho_{1}=\rho_{2}=1.17\,\mathrm{kg\,m}^{-3}$ -- i.e., when the water density
matches that of air at normal conditions. Substituting these values and the
molecular masses of water and air from Table \ref{tab2} into formulae
(\ref{8.13})--(\ref{8.16}), one obtains%
\begin{equation}
D_{11}\approx1.37\times10^{-10}\mathrm{m}^{3}\mathrm{s\,kg},\qquad\zeta
_{1}\approx-2.76\times10^{-6}\mathrm{m}^{-1}\mathrm{s}^{-1}\mathrm{kg}.
\label{8.17}%
\end{equation}
Even though these values apply to different temperatures ($T=25^{\circ
}\mathrm{C}$ and $T=21^{\circ}\mathrm{C}$), they will be used simultaneously
in the same qualitative estimate (under the implied assumption that the
four-degree difference does not matter).

\subsection{The nondimensional parameters}

To understand which effects are important for water/air interfaces under
normal conditions, one should estimate the nondimensional parameters
(\ref{4.8}). They will be calculated using the characteristics of water from
Tables \ref{tab2}--\ref{tab4} and estimates (\ref{8.17}). The characteristic
pressure scale will be assumed to be $P=a\varrho^{2}$ where $\varrho
=997\,\mathrm{kg\,m}^{-3}$ is the density of liquid water at $25^{\circ
}\mathrm{C}$ and $a$ equals the van der Waals parameter of water from Table
\ref{tab2}. The following expression for the viscosity scale will be used:%
\[
\mu=\frac{4}{3}\mu_{s}+\mu_{b},
\]
which arises naturally in problems involving liquid films
\citep{Benilov20d,Benilov22b} and drops \citep{Benilov22a}. In the estimates
for this paper, the viscosities $\mu_{s}$ and $\mu_{b}$ are those for water at
$25^{\circ}\mathrm{C}$.

The following values of parameters (\ref{4.8}) have been obtained:%
\begin{equation}
\alpha\approx0.00139,\qquad\tau\approx0.0653,\qquad\beta\approx0.0593,\qquad
\nu\approx0.0321,\qquad\delta\approx0.0176. \label{8.18}%
\end{equation}
The smallness of the microscopic Reynolds number $\alpha$ suggests that
inertia plays little role in the dynamics of water/air interfaces (hence, one
can use the Stokes approximation). The smallness of the nondimensional
temperature $\tau$ does not have physical implications, but enables one to use
asymptotic tools when calculating, say, the profile of the equilibrium
interface \citep{Benilov20d}. The smallness of the Nusselt number $\nu$
implies that the Soret and Dufour effects are negligible, and so the lack of
empiric data on them for water and air is unimportant. The smallness of
$\beta$, in turn, implies that the flow is almost isothermal, whereas the
small value of $\delta$ suggests that diffusion dominates advection.

Similar estimates have also been carried for the parameters of air at
$p=1\,\mathrm{atm}$ and $T=25^{\circ}\mathrm{C}$. It has turned out that
$\alpha$, $\beta$, $\nu$, and $\delta$ are even smaller than those for water,
but the nondimensional temperature $\tau$ for air is order-one. The latter is
not surprising, as the room temperature is much higher than the freezing
temperatures of nitrogen and oxygen, but is close to that for water.

All these observations should be helpful when using the DIM to examine
asymptotically the problems of evaporation of drops and moving contact lines.

One should keep in mind, however, that parameters (\ref{4.8}) are sensitive to
the choice of the viscosity scale $\mu$. \cite{Benilov20b}, for example, chose
$\mu$ equal the half-sum of the shear viscosities of liquid water and vapour
water -- as a result, $\alpha$ and $\beta$ were noticeably larger than those
calculated above. More generally, one should choose the viscosity scale, as
well as the other characteristic scales, to reflect the essentially physics of
the setting at hand.

Note also that estimates (\ref{8.18}) apply to water under \emph{normal}
conditions -- but in industrial applications, the governing nondimensional
parameters can be very different. In steam turbines, for example, the
temperature can be as high as $540^{\circ}\mathrm{C}$, and the pressure can
exceed $230\,\mathrm{atm}$ \citep{VassermanShutenko17}.

\section{Concluding remarks\label{Sec 9}}

The results obtained in this paper can be briefly summarised as follows:

\begin{enumerate}
\item The entropy principle and conservation laws of the multicomponent
diffuse-interface model have been used to examine the stability of
liquid/vapour interfaces. Several physically important results are reported:

\begin{itemize}
\item Flat liquid/vapour interfaces in an unbounded fluid are stable if the
density profiles of all species are monotonic.

\item There can exist up to two different solutions describing a solid/vapour
interface, one monotonic and one non-monotonic. The former is stable and the
latter is likely to be unstable (\emph{definitely} unstable for pure fluids).
Similar conclusions apply to solid/liquid interfaces.

\item For certain values of the near-substrate density (which is an external
parameter in DIM, linked to the contact angle of the substrate), no steady
solution exists describing a solid/vapour interfaces. Physically, such
substrates are too hydrophilic and, thus, trigger off spontaneous condensation
of adjacent vapour. Similarly, some substrates are too hydrophobic and trigger
off spontaneous evaporation of adjacent liquid.

\item A liquid layer between a flat substrate and a semi-space filled with
vapour can exist as a steady state only if the vapour is \emph{oversaturated}.
All such layers are unstable, however: depending on the perturbation, they
either fully evaporate or grow indefinitely due to condensation of vapour on
its upper boundary.\newline\hspace*{0.6cm}If the vapour above the layer is
\emph{saturated} or \emph{undersaturated}, the liquid evaporates and no steady
solution exists.\newline\hspace*{0.6cm}Similar conclusions apply to 1D vapour
layers between a flat substrate and a semi-space filled with liquid.

\item Similar conclusions to those in the previous bullet apply to a liquid
layer with vapour both above and below it, and a vapour layer with liquid
below and above it. Such solutions can be viewed as 1D drops and bubbles, respectively.
\end{itemize}

\item The multicomponent DIM has been fully parameterised for water/air
interfaces under normal conditions. It is shown that the Soret and Dufour
effects are weak in this case, which agrees with the results of
\cite{JiangStuderPodvin20}. It is also argued that the interfacial flow in
this case is isothermal.\newline\hspace*{0.6cm}These are of importance when
studying evaporation of water drops or the dynamics of their contact lines.
\end{enumerate}

\section*{Declaration of interests}

The author reports no conflict of interests.

\appendix

\section{The Gibbs relation\label{Appendix A}}

Introduce the volumetric densities of energy $\mathcal{U}(\rho_{1}...\rho
_{N},T)$, entropy $\mathcal{S}(\rho_{1}...\rho_{N},T)$, and partial chemical
potentials $\mathcal{G}_{i}(\rho_{1}...\rho_{N},T)$ -- related to the
corresponding specific quantities by%
\begin{equation}
\mathcal{U}=e\rho,\qquad\mathcal{S}=s\rho,\qquad\mathcal{G}_{i}=G_{i}.
\label{A1}%
\end{equation}
Now, definition (\ref{2.3}) of $G_{i}$ can be rewritten in the form%
\begin{equation}
\mathcal{G}_{i}=\frac{\partial\mathcal{U}}{\partial\rho_{i}}-T\frac
{\partial\mathcal{S}}{\partial\rho_{i}}. \label{A2}%
\end{equation}
The standard volumetric version of the Gibbs relation
\citep[e.g., Eq. (2.4) of][]{GiovangigliMatuszewski13} is%
\[
\mathrm{d}\mathcal{U}=T\,\mathrm{d}\mathcal{S}+\sum_{i}\mathcal{G}%
_{i}\,\mathrm{d}\rho_{i}.
\]
Substituting (\ref{A1})--(\ref{A2}) into this equality and rewriting it in
terms of $\mathrm{d}T$ and $\mathrm{d}\rho_{i}$ (instead of $\mathrm{d}%
\mathcal{S}$ and $\mathrm{d}\rho_{i}$), one obtains (\ref{2.4}) as required.

\section{The general boundary condition at a solid wall\label{Appendix B}}

Consider the following generalization of boundary condition (\ref{2.35}):%
\begin{equation}
\sum_{j}C_{ij}\left(  \mathbf{n}\cdot\mathbf{\nabla}\rho_{j}\right)  +\rho
_{i}=\rho_{0,i}\qquad\text{at}\qquad\mathbf{r}\in\partial\mathcal{D},
\label{B.1}%
\end{equation}
where $C_{ij}$ is a symmetric matrix, depending generally on $\rho_{1}%
...\rho_{N}$. (\ref{B.1}) is a multicomponent extension of a boundary
condition often used for pure fluids \citep[e.g.][]{MadrugaThiele09,GalloMagalettiCasciola21}.

To understand how the energy conservation law is affected by switching to a
different boundary condition, observe that the governing equations
(\ref{2.17})--(\ref{2.22}) and the other boundary conditions, (\ref{2.32}%
)--(\ref{2.34}), imply that%
\begin{multline}
\frac{\mathrm{d}}{\mathrm{d}t}\int_{\mathcal{D}}\left[  \rho e+\frac{1}{2}%
\rho\left\vert \mathbf{v}\right\vert ^{2}+\frac{1}{2}\sum_{ij}K_{ij}\left(
\mathbf{\nabla}\rho_{i}\right)  \cdot\left(  \mathbf{\nabla}\rho_{j}\right)
\right]  \mathrm{d}^{3}\mathbf{r}\\
+\int_{\partial\mathcal{D}}\sum_{i}\left(  \mathbf{n}\cdot\mathbf{\nabla}%
\rho_{i}\right)  \sum_{j}K_{ij}\frac{\partial\rho_{j}}{\partial t}%
\mathrm{d}A=0, \label{B.2}%
\end{multline}
where $\mathrm{d}A$ is the elemental aria on $\partial\mathcal{D}$. The `old'
boundary condition (\ref{2.35}) implies that the second integral in this
equality vanishes, making the integral in the first term be a conserved
quantity (the energy) in this case.

Next, assume that the `new' boundary condition (\ref{B.1}) is imposed.
Differentiate it with respect to $t$, change the indices -- first
$j\rightarrow k$, then $i\rightarrow j$ -- and use the resulting equality to
rearrange (\ref{B.2}) in the form%
\begin{multline}
\frac{\mathrm{d}}{\mathrm{d}t}\int_{\mathcal{D}}\left[  \rho e+\frac{1}{2}%
\rho\left\vert \mathbf{v}\right\vert ^{2}+\frac{1}{2}\sum_{ij}K_{ij}\left(
\mathbf{\nabla}\rho_{i}\right)  \cdot\left(  \mathbf{\nabla}\rho_{j}\right)
\right]  \mathrm{d}^{3}\mathbf{r}\\
-\int_{\partial\mathcal{D}}\sum_{ij}K_{ij}^{\prime}\left(  \mathbf{n}%
\cdot\mathbf{\nabla}\rho_{i}\right)  \frac{\partial\left(  \mathbf{n}%
\cdot\mathbf{\nabla}\rho_{j}\right)  }{\partial t}\mathrm{d}A=0, \label{B.3}%
\end{multline}
where%
\[
K_{ij}^{\prime}=\sum_{k}K_{ik}C_{kj}.
\]
Equation (\ref{B.3}) implies that the following quantity is conserved:%

\begin{multline*}
E=\int_{\mathcal{D}}\left[  \rho e+\frac{1}{2}\rho\left\vert \mathbf{v}%
\right\vert ^{2}+\frac{1}{2}\sum_{ij}K_{ij}\left(  \mathbf{\nabla}\rho
_{i}\right)  \cdot\left(  \mathbf{\nabla}\rho_{j}\right)  \right]
\mathrm{d}^{3}\mathbf{r}\\
-\frac{1}{2}\int_{\partial\mathcal{D}}\sum_{ij}K_{ij}^{\prime}\,\left(
\mathbf{n}\cdot\mathbf{\nabla}\rho_{i}\right)  \left(  \mathbf{n}%
\cdot\mathbf{\nabla}\rho_{j}\right)  \mathrm{d}A.
\end{multline*}
The second term in this expression is the surface energy corresponding to the
new boundary condition (\ref{B.1}).

By comparison to the Dirichlet boundary condition (\ref{2.35}), the mixed
condition (\ref{B.1}) includes additional $N\left(  N-1\right)  /2$ adjustable
parameters. Even though the extra parameters may enable the DIM to be more
accurate quantitatively, one should think that a physics-based model does not
need many adjustable parameters to capture the qualitative nature of the
phenomenon it is applied to.

\section{Stability of one-dimensional steady states\label{Appendix C}}

In this appendix, the stability of three families of 1D solutions are examined:

\begin{enumerate}
\item liquid/vapour interfaces,

\item 1D drops and bubbles,

\item solid/fluid interfaces.
\end{enumerate}

\subsection{Liquid/vapour interfaces\label{Appendix C.1}}

As mentioned before, equation (\ref{3.10}) describes all steady states in the
problem at hand, and expression (\ref{3.12}) describes the corresponding
second variation of the entropy. To adapt the latter for flat liquid/vapour
interfaces, one needs to replace the 3D integral over the domain $\mathcal{D}$
with a 1D integral over the domain $-\infty<z<\infty$,%
\[
\delta^{2}H=\frac{1}{T}\int_{-\infty}^{\infty}\sum_{ij}\left[  -\frac{\partial
G_{i}}{\partial\rho_{j}}\left(  \delta\rho_{i}\right)  \left(  \delta\rho
_{j}\right)  -K_{ij}\frac{\mathrm{d}\left(  \delta\rho_{i}\right)
}{\mathrm{d}z}\frac{\mathrm{d}\left(  \delta\rho_{j}\right)  }{\mathrm{d}%
z}\right]  \mathrm{d}z.
\]
This expression can be rewritten in the form%
\begin{equation}
\delta^{2}H=\frac{1}{T}\int_{-\infty}^{\infty}\sum_{ij}\delta\rho
_{i}\,\mathrm{\hat{O}}_{ij}\,\delta\rho_{j}\,\mathrm{d}z, \label{C.1}%
\end{equation}
where%
\begin{equation}
\mathrm{\hat{O}}_{ij}=-\frac{\partial G_{i}}{\partial\rho_{j}}+K_{ij}%
\frac{\mathrm{d}^{2}}{\mathrm{d}z^{2}} \label{C.2}%
\end{equation}
is a second-order differential operator. Since the matrices $\partial
G_{i}/\partial\rho_{j}$ and $K_{ij}$ are symmetric (see definition (\ref{2.3})
of $G_{i}$ and subsection \ref{Sec 2.5}, respectively), $\mathrm{\hat{O}}%
_{ij}$ is self-adjoint -- hence, its spectrum is real. Note also that
$\mathrm{\hat{O}}_{ij}$ depends on the steady state $\rho_{i}(z)$ via the
coefficient $\partial G_{i}/\partial\rho_{j}$.

It follows from equation (\ref{C.1}) that the functional $H$ has a maximum at
$\rho_{i}(z)$ if and only if the corresponding operator $\mathrm{\hat{O}}%
_{ij}$ is negative definite -- or, equivalently, its spectrum (the set of all
discrete and continuous eigenvalues) is negative.\medskip

Theorem 1. If the solution $\rho_{i}(z)$ of boundary-value problem
(\ref{6.1})--(\ref{6.3}) is monotonic for all $i$, the spectrum of the
corresponding operator $\mathrm{\hat{O}}_{ij}$ is negative.\medskip

Proof. Let $\Lambda$ be an eigenvalue of the \emph{discrete} spectrum (if any)
and $\psi_{i}$, the corresponding eigenfunction,
\begin{equation}
\sum_{j}\left(  -\frac{\partial G_{i}}{\partial\rho_{j}}\psi_{j}+K_{ij}%
\frac{\mathrm{d}^{2}\psi_{j}}{\mathrm{d}z^{2}}\right)  =\Lambda\psi_{i},
\label{C.3}%
\end{equation}%
\begin{equation}
\psi_{i}\rightarrow0\qquad\text{as}\qquad z\rightarrow\pm\infty. \label{C.4}%
\end{equation}
Introduce%
\begin{equation}
\phi_{i}=\left(  \frac{\mathrm{d}\rho_{i}}{\mathrm{d}z}\right)  ^{-1}\psi_{i},
\label{C.5}%
\end{equation}
and observe that, since $\rho_{i}(z)$ is monotonic, $\mathrm{d}\rho
_{i}/\mathrm{d}z$ never vanishes and $\phi_{i}(z)$ is non-singular.

Rewriting boundary-value problem (\ref{C.3})--(\ref{C.4}) in terms of
$\phi_{i}$ and keeping in mind that $\rho_{i}(z)$ satisfies equation
(\ref{6.3}), one obtains%
\begin{equation}
\sum_{j}K_{ij}\left(  2\frac{\mathrm{d}^{2}\rho_{j}}{\mathrm{d}z^{2}}%
\frac{\mathrm{d}\phi_{j}}{\mathrm{d}z}+\frac{\mathrm{d}\rho_{j}}{\mathrm{d}%
z}\frac{\mathrm{d}^{2}\phi_{j}}{\mathrm{d}z^{2}}\right)  =\Lambda
\frac{\mathrm{d}\rho_{i}}{\mathrm{d}z}\phi_{i}, \label{C.6}%
\end{equation}%
\begin{equation}
\frac{\mathrm{d}\rho_{i}}{\mathrm{d}z}\phi_{i}\rightarrow0\qquad
\text{as}\qquad z\rightarrow\pm\infty. \label{C.7}%
\end{equation}
Let $\left(  K_{ij}\right)  ^{-1}$ be the inverse matrix to $K_{ij}$ (the
latter is positive definite -- hence, invertible) and rewrite equation
(\ref{C.6}) in the form%
\[
\frac{\mathrm{d}}{\mathrm{d}z}\left[  \left(  \frac{\mathrm{d}\rho_{j}%
}{\mathrm{d}z}\right)  ^{2}\frac{\mathrm{d}\phi_{j}}{\mathrm{d}z}\right]
=\Lambda\sum_{i}\left(  K_{ij}\right)  ^{-1}\frac{\mathrm{d}\rho_{j}%
}{\mathrm{d}z}\frac{\mathrm{d}\rho_{i}}{\mathrm{d}z}\phi_{i}.
\]
Multiplying this equation by $\phi_{j}$, summing it with respect to $j$,
integrating from $z=0$ to $z=\infty$, integrating the left-hand side by parts,
and recalling boundary conditions (\ref{C.7}), one obtains%
\begin{equation}
-\int_{-\infty}^{\infty}\sum_{j}\left(  \frac{\mathrm{d}\rho_{j}}{\mathrm{d}%
z}\right)  ^{2}\left(  \frac{\mathrm{d}\phi_{j}}{\mathrm{d}z}\right)
^{2}\mathrm{d}z=\Lambda\int_{-\infty}^{\infty}\sum_{ij}\left(  K_{ij}\right)
^{-1}\left(  \phi_{i}\frac{\mathrm{d}\rho_{i}}{\mathrm{d}z}\right)  \left(
\phi_{j}\frac{\mathrm{d}\rho_{j}}{\mathrm{d}z}\right)  \mathrm{d}z.
\label{C.8}%
\end{equation}
It can be shown that the integrands of both integrals in equation (\ref{C.8})
decay exponentially as $z\rightarrow+\infty$ -- hence, the integrals converge.
Keeping also in mind that $K_{ij}$ is positive definite (hence, $\left(
K_{ij}\right)  ^{-1}$ is such), one concludes that equality (\ref{C.8})
implies that $\Lambda<0$, as required.

Next, let $\Lambda$ be an eigenvalue of the \emph{continuous} spectrum and
$\psi(z)$, the corresponding eigenfunction. They satisfy equation (\ref{C.3})
and the boundary conditions%
\begin{equation}
\psi_{i}\sim A_{i\pm}\operatorname{e}^{ik_{\pm}z}\qquad\text{as}\qquad
z\rightarrow\pm\infty, \label{C.9}%
\end{equation}
where $A_{i\pm}$ and $k_{\pm}$ are undetermined constants (the latter is
real). Substituting asymptotics (\ref{C.9}) into equation (\ref{C.3}), one
obtains%
\[
\sum_{j}\left[  -\left(  \frac{\partial G_{i}}{\partial\rho_{j}}\right)
_{\rho_{1}=\rho_{1}^{(l)}...\rho_{N}=\rho_{N}^{(l)}}-k_{-}^{2}K_{ij}\right]
A_{j-}=\Lambda A_{-},
\]%
\[
\sum_{j}\left[  -\left(  \frac{\partial G_{i}}{\partial\rho_{j}}\right)
_{\rho_{1}=\rho_{1}^{(v)}...\rho_{N}=\rho_{N}^{(v)}}-k_{+}^{2}K_{ij}\right]
A_{j+}=\Lambda A_{+},
\]
Evidently, $\Lambda$ is a common eigenvalue of the matrices%
\[
-G_{ij-}^{\prime}-k_{-}^{2}K_{ij}\qquad\text{and}\qquad-G_{ij+}^{\prime}%
-k_{+}^{2}K_{ij}%
\]
where%
\[
G_{ij-}^{\prime}=\left(  \frac{\partial G_{i}}{\partial\rho_{j}}\right)
_{\rho_{1}=\rho_{1}^{(l)}...\rho_{N}=\rho_{N}^{(l)}},\qquad G_{ij+}^{\prime
}=\left(  \frac{\partial G_{i}}{\partial\rho_{j}}\right)  _{\rho_{1}=\rho
_{1}^{(v)}...\rho_{N}=\rho_{N}^{(v)}}.
\]
Recall that, at infinity, the liquid and vapour are supposed to be stable,
which implies that $G_{ij\pm}^{\prime}$ are both positive definite. The
Korteweg matrix $K_{ij}$ is also positive definite -- hence, all eigenvalues
of $-G_{ij\pm}^{\prime}-k_{\pm}^{2}K_{ij}$ are negative, as required.

\subsection{1D drops and bubbles\label{Appendix C.2}}

The entropy maximization problem in this case can again be reduced to the
analysis of the operator $\mathrm{\hat{O}}_{ij}$. Furthermore, the second part
of the proof of Theorem 1 can be applied to 1D drops and bubbles without
modifications, and so the \emph{continuous} spectrum of $\mathrm{\hat{O}}%
_{ij}$ is, again, negative.

The \emph{discrete}-spectrum part of Theorem 1, however, cannot be generalised
for non-monotonic $\rho_{i}(z)$. Indeed, if $\mathrm{d}\rho_{i}/\mathrm{d}z$
vanishes somewhere, the function $\phi_{i}(z)$ defined by (\ref{C.5}) is
singular, and the integral on the left-hand side of (\ref{C.8}) diverges.
Thus, for 1D drops and bubbles, $\mathrm{\hat{O}}_{ij}$ may have positive
discrete eigenvalues, but proving that it definitely does is not easy. In what
follows, such a proof is presented only for $N=1$.\medskip

Theorem 2. The operator $\mathrm{\hat{O}}_{11}$ with $\rho_{1}(z)$ satisfying
boundary-value problem (\ref{6.7})--(\ref{6.8}) with $N=1$, has at least one
positive discrete eigenvalue (corresponding to an even eigenfunction).\medskip

Proof. For $N=1$, boundary-value problem (\ref{6.7})--(\ref{6.8}) becomes%
\begin{equation}
K_{11}\frac{\mathrm{d}^{2}\rho_{1}}{\mathrm{d}z^{2}}=G_{1}-G_{\infty,1},
\label{C.10}%
\end{equation}%
\begin{equation}
\rho_{1}\rightarrow\rho_{\infty,1}\qquad\text{as}\qquad z\rightarrow\pm\infty.
\label{C.11}%
\end{equation}
It can be readily shown that%
\begin{equation}
\rho_{1}(z)\sim\rho_{\infty,1}+A\operatorname{e}^{\sqrt{G^{\prime}}z}%
\qquad\text{as}\qquad z\rightarrow-\infty, \label{C.12}%
\end{equation}
where $G^{\prime}=\left(  \partial G_{11}/\partial\rho_{1}\right)  _{\rho
_{1}=\rho_{\infty,1}}$ and $A$ is a real constant (which can be expressed via
a certain integral, but its exact value is unimportant). Note that $A$ is
positive for drops [i.e., solutions of the kind illustrated in figure
\ref{fig4}(a)] and negative for bubbles [solutions illustrated in figure
\ref{fig4}(b)].

Next, eigenvalue problem (\ref{C.3})--(\ref{C.4}) with $N=1$ has the form%
\begin{equation}
-\frac{\partial G_{1}}{\partial\rho_{1}}\psi_{1}+K_{11}\frac{\mathrm{d}%
^{2}\psi_{1}}{\mathrm{d}z^{2}}=\Lambda\psi_{1}, \label{C.13}%
\end{equation}%
\begin{equation}
\psi_{1}\rightarrow0\qquad\text{as}\qquad z\rightarrow-\infty, \label{C.14}%
\end{equation}%
\begin{equation}
\psi_{1}\rightarrow0\qquad\text{as}\qquad z\rightarrow+\infty. \label{C.15}%
\end{equation}
Note that, for 1D drops an bubbles, $\rho_{1}(z)$ is even -- hence, so is
$\partial G_{1}/\partial\rho_{1}$ which appears in equation (\ref{C.13}).
Since all other coefficients in this (linear second-order) equation are
constants, one concludes that $\psi_{1}$ is either even or odd. Assuming the
latter, one can reduce the interval where equation (\ref{C.13}) is to be
solved to $\left(  0,\infty\right)  $ and replace boundary condition
(\ref{C.15}) with%
\begin{equation}
\frac{\mathrm{d}\psi_{1}}{\mathrm{d}z}=0\qquad\text{at}\qquad z=0.
\label{C.16}%
\end{equation}
Next, define an auxiliary function $\chi(z)$ which satisfies equation
(\ref{C.13}) and boundary condition (\ref{C.14}) -- but not necessarily
condition (\ref{C.16}). To still have two boundary conditions, require that%
\begin{equation}
\chi(z)\sim A\sqrt{G^{\prime}+\Lambda}\operatorname{e}^{\sqrt{G^{\prime
}+\Lambda}z}\qquad\text{as}\qquad z\rightarrow-\infty, \label{C.17}%
\end{equation}
where $A$ is the constant from asymptotics (\ref{C.10}). Note that $\chi(z)$
exists for \emph{any} value $\Lambda$ (depends on it as a parameter). In
particular,%
\begin{equation}
\chi(z)=\frac{\mathrm{d}\rho_{1}}{\mathrm{d}z}\qquad\text{for}\qquad\Lambda=0,
\label{C.18}%
\end{equation}
which can be verified by comparing equation (\ref{C.13}) with the derivative
of equation (\ref{C.10}) and asymptotics (\ref{C.17}) with the derivative of
asymptotics (\ref{C.12}). For a large $\Lambda$, on the other hand, it can be
deduced from equation (\ref{C.13}) and boundary condition (\ref{C.17}) that%
\begin{equation}
\chi(z)\sim A\sqrt{\Lambda}\operatorname{e}^{\sqrt{\Lambda}z}\qquad
\text{as}\qquad\Lambda\rightarrow+\infty. \label{C.19}%
\end{equation}
This asymptotics applies to all finite $z$.

Now, consider a drop solution [e.g., one of those illustrated in figure
\ref{fig4}(a). It is evident from the figure (and common sense) that $\left(
\mathrm{d}^{2}\rho_{1}/\mathrm{d}z^{2}\right)  _{z=0}<0$ and $A>0$ -- hence,
(\ref{C.18})--(\ref{C.19}) imply%
\[
\left(  \frac{\mathrm{d}\chi}{\mathrm{d}z}\right)  _{z=0}<0\qquad
\text{for}\qquad\Lambda=0,
\]%
\[
\left(  \frac{\mathrm{d}\chi}{\mathrm{d}z}\right)  _{z=0}>0\qquad
\text{as}\qquad\Lambda\rightarrow+\infty.
\]
These two inequalities indicate that there exists a $\Lambda\in\left(
0,+\infty\right)  $ such that $\left(  \mathrm{d}\chi/\mathrm{d}z\right)
_{z=0}=0$ -- hence, the corresponding function $\chi(z)$ satisfies boundary
condition (\ref{C.16}) [as well as condition (\ref{C.14}) and equation
(\ref{C.13})]. The corresponding (positive) $\Lambda$ is, obviously, an
eigenvalue, as required.

The case of bubble solutions is similar to that of drops (examined above) and
yields the same conclusion.

\subsection{Solid/fluid interfaces\label{Appendix C.3}}

In this case, $\rho_{i}(z)$ satisfies equation (\ref{6.8}) and boundary
conditions (\ref{6.12})--(\ref{6.13}).\medskip

Theorem 3. The spectrum of the operator $\mathrm{\hat{O}}_{ij}$ corresponding
to a \emph{monotonic} $\rho_{i}(z)$ is real and negative.\medskip

The proof of this theorem is similar to that of Theorem 1. The only difference
is in the boundary conditions: since the density is constrained to a fixed
value at the substrate, its variation is zero -- and the eigenfunctions of the
operator $\mathrm{\hat{O}}_{ij}$ should satisfy%
\[
\psi_{i}=0\qquad\text{at}\qquad z=0.
\]
This boundary condition should hold for both discrete and continuous
spectra.\medskip

Theorem 4. If $N=1$, the operator $\mathrm{\hat{O}}_{11}$ with non-monotonic
$\rho_{1}(z)$ has at least one positive discrete eigenvalue.\medskip

The proof of this theorem is similar to that of Theorem 2

\section{1D pure-fluid reduction of equations (\ref{4.9})--(\ref{4.10}) and
(\ref{4.4})\label{Appendix D}}

To adapt asymptotic equation (\ref{4.9})--(\ref{4.10}) for a pure fluid,
recall that the transport coefficients $D_{11}$ and $\zeta_{1}$ are zero in
this case (because this is the only possibility to satisfy restrictions
(\ref{2.25}) for $N=1$). Thus, one obtains%
\begin{equation}
\frac{\partial\rho_{1}}{\partial t}+\mathbf{\nabla}\cdot\left(  \rho
_{1}\mathbf{v}\right)  =0, \label{D.1}%
\end{equation}%
\begin{equation}
\mathbf{\nabla}\cdot\boldsymbol{\Pi}+\rho_{1}\mathbf{\nabla}\left(
K_{11}\nabla^{2}\rho_{1}-G_{1}\right)  =0. \label{D.2}%
\end{equation}
Let the flow be 1D, so that the unknowns depend only on $z$ and $t$, and the
velocity has only one component $\mathbf{v}=\left[  0,0,w\right]  $. Then,
equations (\ref{D.1})--(\ref{D.2}) and expression (\ref{4.4}) for the viscous
stress yield%
\begin{equation}
\frac{\partial\rho_{1}}{\partial t}+\frac{\partial\left(  \rho_{1}w\right)
}{\partial z}=0, \label{D.3}%
\end{equation}%
\begin{equation}
\frac{\partial}{\partial z}\left[  \left(  \frac{4}{3}\mu_{s}+\mu_{b}\right)
\frac{\partial w}{\partial z}\right]  +\rho_{1}\frac{\partial}{\partial
z}\left(  K_{11}\frac{\partial^{2}\rho_{1}}{\partial z^{2}}-G_{1}\right)  =0.
\label{D.4}%
\end{equation}
Let the substrate be located at $z=0$, so that boundary conditions
(\ref{2.32}) and (\ref{2.35}) yield%
\begin{equation}
w=0,\qquad\rho_{1}\rightarrow\rho_{0,1}\qquad\text{at}\qquad z=0. \label{D.5}%
\end{equation}
At infinity, the density tends to a fixed value and there should be no stress
-- hence,%
\begin{equation}
\rho_{1}\rightarrow\rho_{\infty,1},\qquad\frac{\partial w}{\partial
z}\rightarrow0\qquad\text{as}\qquad z\rightarrow+\infty. \label{D.6}%
\end{equation}
Using identity (\ref{2.6}), one can replace in equation (\ref{D.4}) $\rho
_{1}\partial G_{1}/\partial z\rightarrow\partial p/\partial z$ and then
integrate this equation with respect to $z$. Fixing the constant of
integration via boundary conditions (\ref{D.6})--(\ref{D.7}), one obtains%
\begin{equation}
\left(  \frac{4}{3}\mu_{s}+\mu_{b}\right)  \frac{\partial w}{\partial
z}+K_{11}\left[  \rho_{1}\frac{\partial^{2}\rho_{1}}{\partial z^{2}}-\frac
{1}{2}\left(  \frac{\partial\rho_{1}}{\partial z}\right)  ^{2}\right]  =0.
\label{D.7}%
\end{equation}
Given a suitable initial condition, equations (\ref{D.3}) and (\ref{D.7}),
with boundary conditions (\ref{D.5}) fully determine $\rho_{1}(z,t)$ and
$w(z,t)$.

For numerical simulations, equations (\ref{D.3}) and (\ref{D.7}) were
discretised in $z$ using central differences, and the resulting set of
ordinary differential equations were solved using the MATLAB function ODE23tb
(which can handle stiff problems). This approach is usually referred to as the
\textquotedblleft method of lines\textquotedblright\ \citep{Schiesser78}.

\bibliographystyle{jfm}
\bibliography{../../bib/refs}
%\bibliography{}

\end{document}